\documentclass[aps,prd,amsmath,floats,floatfix,superscriptaddress,nofootinbib,showpacs]{revtex4}

\usepackage[T1]{fontenc}
\usepackage[utf8]{inputenc}
\usepackage{lmodern}
\usepackage{xfrac}
\usepackage{verbatim}
\usepackage{footmisc}

\usepackage[dvipsnames, usenames]{xcolor}
\definecolor{linkcolor}{rgb}{0.0,0.3,0.5}
\usepackage[hypertexnames=false, unicode, colorlinks=true, linkcolor=linkcolor,
citecolor=linkcolor, filecolor=linkcolor,urlcolor=linkcolor,
pdfusetitle]{hyperref}

\usepackage[all]{hypcap}
\usepackage{graphicx}
\usepackage{xspace}
\usepackage{amssymb}
\usepackage[normalem]{ulem} 
\usepackage{bm} 

\usepackage{microtype}

\usepackage[english]{babel}
\usepackage{blindtext}

\graphicspath{%
  {figs/}%
}

\DeclareMathAlphabet{\mathpzc}{OT1}{pzc}{m}{it}

\newcommand{\mat}[1]{\begin{pmatrix}#1\end{pmatrix}}

\definecolor{verde}{rgb}{0,0.5,0}

\renewcommand{\vec}[1] {\boldsymbol{#1}} 
\newcommand*{\vq} {\vec{q}}
\newcommand*{\vp} {\vec{p}}

\newcommand*{\vk} {\vec{k}}
\newcommand*{\p}  {\partial}
\newcommand*{\df}  {\delta}

\newcommand*{\non}  {\nonumber}
\newcommand*{\lb}  {\left(}
\newcommand*{\rb}  {\right)}

\newcommand*{\la}  {\left\langle}
\newcommand*{\ra}  {\right\rangle}
\newcommand*{\dd}  {\mathrm{d}}

\newcommand{\ba}{\[\begin{aligned}}
\newcommand{\ea}{\end{aligned}\]}

\newcommand{\eq}[1]{\begin{align}#1\end{align}}
\newcommand{\eeq}[1]{\begin{equation}#1\end{equation}}

\definecolor{verde}{rgb}{0,0.5,0}

 \def\bea  {\begin{eqnarray}}   \def\eea  {\end{eqnarray}}
\def\noi{{\noindent}}

\begin{document}

\rightline{\scriptsize RBI-ThPhys-2022-24}
\title{
Perturbation theory of LSS 
in the $\Lambda$CDM Universe:\\
exact time evolution and the two-loop power spectrum
}

\newcommand\matteohomeone{\affiliation{Instituto de F\'{i}sica T\'{e}orica UAM/CSIC, calle Nicol\'{a}s Cabrera 13-15, Cantoblanco, 28049,
Madrid, Spain}}
\newcommand\matteohometwo{\affiliation{Institute of Cosmology  and Gravitation, University of Portsmouth, PO1 3FX, UK}}
\newcommand\tomoohomeone{\affiliation{Waseda Institute for Advanced Study, Waseda University,
1-6-1 Nishi-Waseda, Shinjuku, Tokyo 169-8050, Japan}}
\newcommand\tomoohometwo{\affiliation{Research Center for the Early Universe,  The University of Tokyo, Bunkyo, Tokyo 113-0033, Japan}}
\newcommand\zvonehomeone{\affiliation{Ru\dj er Bo\v{s}kovi\'c Institute, Bijeni\v{c}ka cesta 54, 10000 Zagreb, Croatia}}
\newcommand\zvonehometwo{\affiliation{Kavli Institute for Cosmology, University of Cambridge, Cambridge CB3 0HA, UK}}
\newcommand\zvonehomethree{\affiliation{Department of Applied Mathematics and Theoretical Physics, University of Cambridge, Cambridge CB3 0WA, UK }}

\author{Matteo Fasiello}
\email{matteo.fasiello@csic.es}
\matteohomeone
\matteohometwo

\author{Tomohiro Fujita}
\email{tomofuji@aoni.waseda.jp}
\tomoohomeone
\tomoohometwo

\author{Zvonimir Vlah}
\email{zvlah@irb.hr}
\zvonehomeone
\zvonehometwo
\zvonehomethree

\hypersetup{pdfauthor={Fasiello et al.}}


\begin{abstract}
\vspace{.5cm}
\noindent 
We derive exact analytic solutions for density and velocity fields to all orders in Eulerian  standard perturbation theory for $\Lambda$CDM cosmology. In particular, we show that density and velocity field kernels can be written in a separable form in time and momenta at each perturbative order. The kernel solutions are built from an analytic basis of momentum operators and their time-dependent coefficients, which solve a set of recursive differential equations.  We also provide an exact closed perturbative solution for such coefficients, expanding around the (quasi-)EdS approximation. We find that the perturbative solution rapidly converges towards the numerically obtained solutions and its leading order result suffices for any practical requirements. To illustrate our findings, we compute the exact two-loop dark matter density and velocity power spectra in  $\Lambda$CDM cosmology. We show that the difference  between the exact $\Lambda$CDM and the (quasi-)EdS approximated result can reach the level of several percent (at redshift zero, for wavenumbers $k<1h/$Mpc). This deviation can be partially mitigated by exploiting the degeneracy with the EFT counterterms. As an additional benefit of our algorithm for the solutions of time-dependent coefficients, the computational complexity of power spectra loops in $\Lambda$CDM is comparable to the EdS case.  In performing the two-loop computation, we devise an explicit method to implement the so-called IR cancellations, as well as the cancellations arising as a consequence of mass and momentum conservation. 

\end{abstract}

\maketitle
\section{Introduction} 
\label{sec:intro}

\noindent The large scale structure (LSS) is a repository of key information on our 
universe's origin and evolution, all the way to the current dark energy dominated era. 
Data on inflationary interactions is encoded in the initial conditions for structure formation 
while LSS dynamical evolution also depends on the presence of additional components 
that may drive late-time acceleration. Astronomical surveys of the galaxy distribution 
(e.g. Euclid, LSST, SKA) promise to soon cross the qualitative threshold on cosmological 
parameter, such as a percent level accuracy on the dark energy equation of state 
parameters \cite{Amendola:2016saw,Abell:2009aa,Bacon:2018dui}. 
Crucially, it is by going beyond the background cosmology that we will, for example, 
extract information on non-Gaussianities and identify different dark energy models 
that otherwise support the same expansion history.

LSS dynamics is amenable to a perturbative description for a limited range of wavenumbers: those for which the separation of scales underlying a consistent effective treatment can 
be advocated. The large hierarchy separating the size of the observable Universe $1/H_0$ 
and the onset of non-linearities $1/k_{\rm NL}$ in structure formation explains the success 
of linear perturbation theory in describing the essential features observed in galaxy surveys. 
At scales as large as 10 Mpc, non-linearities become relevant: different Fourier modes stop 
evolving independently showing hints of a UV/IR mixing typical of non-linear regimes. 

At the interface between the linear and highly non-linear regime are so-called 
quasi-(or mildly-non-) linear scales. Gaining perturbative control over the quasi-linear range 
significantly increases  the number of modes at our disposal ($N\propto k^3$). 
A plethora of distinct perturbative approaches have been put forward in this direction 
\cite{Goroff++:1986,Buchert+:1993, Jain+:1993, Valageas:2001,
Bernardeau:2001qr,Taruya+:2007,
McDonald:2009, Carlson:2009it, Baumann:2010tm, Carrasco:2012cv, 
Bertolini:2016bmt, Fujita:2016dne, Pajer:2013jj, McDonald+:2017,
Matsubara:2008wx, Matsubara:2007wj, Carlson:2012bu, Porto:2013qua,Bartelmann++:2014, Vlah:2014nta, Vlah:2015sea, Vlah+:2018}, 
a programme that has been altogether quite successful. The exact $k$-reach of the perturbative 
treatment in particular has been the subject of intense research activity, especially 
within the context of the EFT framework \cite{Baumann:2010tm, Carrasco:2012cv} 
(see also \cite{Desjacques++:2016, Cabass++:2022} for recent reviews). 
Even though several aspects need further development on the ``model building'' front, there are already definite predictions on given observables, consisting mainly of the one-loop power spectrum and the tree-level bispectrum,  that have already been employed in obtaining cosmological information from the LSS galaxy surveys \cite{Ivanov++:2019, DAmico++:2019,Ivanov:2021, Chen++:2021,Zhang+:2021,Philcox++:2021,Farren++:2021,Chen++:2022}.

Our work tackles the perturbative treatment of LSS in $\Lambda$CDM cosmology. 
In this context, striving for \textit{exact analytical} solutions serves multiple purposes. 
Besides being necessary for a view of the projected accuracy of soon-to-be operational probes, 
such solutions are also important to ensure that approximations do not get in the way 
(i.e. create degeneracies out) of otherwise distinct signatures. 
In this work, we present all order exact recursive solutions for perturbation theory kernels 
of the density and velocity fields (i.e. $F_n,G_n$) in $\Lambda$CDM cosmology. 
The need to go beyond the so-called extended (quasi-)EdS approximation\footnote{This method consists 
in handling the time dependence of kernels as in an Einstein-de Sitter universe (only matter content), 
where e.g. $\delta^{(n)}(a) \propto D^{n}(a)$ but with the added prescription to employ 
the linear growth rate $D$ of a $\Lambda$CDM universe. Henceforth, in order to adhere to common parlance, we refer to this approximation as EdS, rather than (quasi-)EdS. \label{fn1}} 
has long been recognised as a timely step (see, e.g. \cite{Bernardeau+:1993, Takahashi:2008, 
Carrasco:2012cv, Matsubara:2015, Rampf++:2015,Lewandowski++:2016,Schmidt:2020, Garny+:2020, Donath+:2020, 
Steele+:2020}), with our own previous work \cite{Fasiello+:2016} providing 
for the first time exact all order solutions for $\Lambda$CDM and beyond. In this manuscript 
we shall take \cite{Fasiello+:2016} as the starting point and report on the significant progress 
in manifold directions.

We are after separable solutions accounting for the time and momenta dependence of density 
and velocity kernels. We identify, for each order in perturbation theory (PT), a complete 
``basis'' of operators in a separable form that make up the solution for the $F,G$ kernels. 
We derive such basis recursively, i.e. by employing the results at lower perturbative orders 
as building blocks. By construction, the derivation of time-dependent coefficients needs no input 
from the momenta operators and vice-versa, greatly simplifying and  speeding up the calculation. 
We provide an algorithm that unambiguously couples time and momenta operators to give 
each basis element. Our  algorithm completely eliminates the need (still present in \cite{Fasiello+:2016}) 
for an ansatz to be put forward at every perturbative step to identify the solution. 
This is a striking improvement, especially relevant as the community has been increasingly 
recognising the importance of tackling higher orders in PT
\cite{Blas++:2013a,Carrasco++:2013a,Blas++:2013b,Carrasco++:2013b,Baldauf++:2015,Foreman++:2015,Konstandin++:2019,Baldauf++:2021,Taruya++:2021,Alkhanishvili++:2021}.
Moreover, we derive explicit perturbative solutions for time-dependent coefficients. By 
a suitable choice of time variable our perturbative solutions are valid 
for generic cosmological parameters within the $\Lambda$CDM cosmology. 

We also develop a systematic way to deal with IR (and UV) divergences in loop integrals. 
As is well-known (see e.g. \cite{Jain+:1993,Scoccimarro+:1995,Peloso+:2013a,Carrasco++:2013a}) the equivalence principle guarantees 
the cancellation of leading and sub-leading IR divergencies. The presence of several large 
IR contributions in the expression for higher order observables ahead of their overall 
cancellation hinder calculational efficiency. In addition to the cancellation of these 
IR divergences, mass and momentum conservation also plays a role in 
determining the scale dependence of loop contributions by imposing 
cancellations of large contributions sensitive to UV scales \cite{Peebles:1980,Goroff++:1986,Bernardeau:2001qr,DAmico++:2021}.
By introducing suitable window functions, we are able to renormalize correlation functions and 
make contact with so-called perturbation theory counterterms in the context of the EFT framework. 
\bigskip

This paper is organised as follows: in \textit{Section} \ref{sec:eom} 
we set the stage with the equations of motion for the $\Lambda$CDM system, 
we also briefly report on previous works on the subject. In \textit{Section} \ref{sec:PT_coeffs} 
we lay out our algorithm and derive recursive separable solutions for the kernels that may 
be used up to any order in perturbation theory. We further show how, starting from the 
Einstein-de Sitter approximation, one may derive solutions arbitrarily close to the exact result. 
In \textit{Section} \ref{sec:one_two_loops} we focus on one- and two-loop
results for the density and velocity cross and auto power spectra. 
We draw our conclusions in \textit{Section} \ref{sec:conclusion} and comment on future work. 
A significant fraction of our derivations has been delegated to the appendices.
Thus, in \textit{Appendix} \ref{sec:linear_growth} we review the linear growth equations and derive a new 
expansion of the specific form of the linear growth rate combination.  
In \textit{Appendix} \ref{sec:direct integral solution} we review the integral solutions 
for perturbation theory kernels. Based on these results in \textit{Appendix} 
\ref{sec:derivtion of kernels} we derive separable kernel form, for which we give 
the perturbative solution of the time-dependence in the \textit{Appendix} 
\ref{sec: perturbation of WU}. In \textit{Appendix} 
\ref{sec: IR and UV limits} we explore the various IR and UV limits of the 
newly obtained kernels, which we use in \textit{Appendix} 
\ref{app:two-loops} to explore the IR and UV properties of the two-loop power spectra.

Throughout the paper, we assume a Euclidean cosmology with $\Omega_m = 0.3$, 
$\sigma_8 = 0.8$ and $h = 0.7$ with the BBKS linear power spectrum. 
We work under the assumptions of adiabatic Gaussian perturbations and General Relativity. 
As mentioned in footnote \footref{fn1}, in the rest of the paper when we refer to the EdS solutions, 
we have in mind the usual (quasi-EdS) approximation of setting the $n$-th order growth factor $D_+^n$,
instead of the $a^n$ which would be the solution in the actual EdS Universe.
Our results for time coefficients and momentum kernels are provided in the \textit{Mathematica} notebook,
in the \textit{arXiv} source file of this paper.

\section{Dynamics in the $\Lambda$CDM Universe}
\label{sec:eom}

As is well known, we may describe the large-scale structure as a fluid in the non-relativistic 
limit obeying the following equations of motion for the fluctuations of the density contrast 
$\delta$ and the peculiar velocity $\theta\equiv \partial_iv^i$:
\eq{
\label{eq:eom_v1}
\frac{\p \delta_{\vk}}{\p \tau} + \theta_{\vk}
   &=-\int_{\vq_{1},\vq_{2}} \delta^{D}_{\vk-\vq_{12}} \alpha(\vq_1,\vq_2) \theta_{\vq_1}\delta_{\vq_2}\, , \\
\frac{\p \theta_{\vk}}{\p \tau} +\mathcal{H} \theta_{\vk}+\frac{3}{2}\Omega_m\mathcal{H}^2 \delta_{\vk} 
   &= -\int_{\vq_{1},\vq_{2}} \delta^{D}_{\vk-\vq_{12}}  \beta(\vq_1,\vq_2) \theta_{\vq_1}\theta_{\vq_2}\, , \non
}
where $\delta^D_{\vq}$
is the Dirac's delta function, $\vq_{12}\equiv \vq_{1}+\vq_{2}$, 
$\int_{\vq}\equiv \int \dd^3 q/(2\pi)^3$, and 
$\mathcal{H} = d \ln a/d \tau$. Here $a$ is the scale factor, and $\tau$ is conformal time. 
The kernels $\alpha,\beta$ are defined as 
 $\alpha(\vq_1,\vq_2) \equiv 1+({\vq}_1 \cdot {\vq}_2)/{\vq}_1^2$,\,
$\beta(\vq_1,\vq_2) \equiv ({\vq}_{12})^2 ({\vq}_1\cdot{\vq}_2) /2q_1^2 q_2^2$.
At linear order, assuming the growing mode initial conditions, 
the time and momentum dependent parts are clearly separable
\eq{
\delta^{(1)}_{\vk}(\tau)\equiv D_+(\tau)\delta^{\rm in}_{\vk}\; ,\; 
\quad 
\theta^{(1)}_{\vk}(\tau)\equiv -\mathcal{H}(\tau)f_+(\tau) D_+(\tau) \delta^{\rm in}_{\vk}\,,
}  
where $D_+$ is the linear growth factor, $f_+ \equiv d{\rm ln}\,D_\pm/d{\rm ln\,}a$ is the linear growth rate 
(see Appendix~\ref{sec:linear_growth} for a brief review of results in the linear regime).
In addition to the growth mode, we also have the decaying mode with linear decay factor $D_-$, 
and equivalently defined decay rate $f_-$. 
$\delta_{\vk}^{\rm in}$ represents the initial value of the density contrast. 
The growing and decaying factors $D_+$ and $D_-$ satisfy the differential equation,
\bea
\label{eq:Dpm EoM}
\frac{d^2 D(\tau)}{d\tau^2}+ \mathcal{H}(\tau)\frac{d D(\tau)}{d \tau} 
-\frac{3}{2}\Omega_m(\tau)\mathcal{H}^2(\tau) \,D(\tau) = 0\; .
\eea
In $\Lambda$CDM cosmology, the solutions for $D_\pm(\tau)$ can be expressed 
in a closed form (see Eq.~\eqref{eq:D+D- solutions}). 
In order to identify the solutions for density contrast and velocity beyond the linear order, 
we employ the following perturbative ansatz:
\eq{
\label{eq: sym kernels}
\delta_{\vk}(\tau)
&=\sum_{n=1}^{\infty} 
\delta^{D}_{\vk-\vq_{1n}} F^{s}_n({\vq}_1,..,{\vq}_n,\tau) D_+^{n}(\tau)
\delta_{\vq_1}^{\rm in}..\delta_{\vq_n}^{\rm in},\;\; \\
\theta_{\vk}(\tau)
&=\sum_{n=1}^{\infty} 
\delta^{D}_{\vk-\vq_{1n}} G^{s}_n({\vq}_1,..,{\vq}_n,\tau) D_+^{n}(\tau)
\delta_{\vq_1}^{\rm in}..\delta_{\vq_n}^{\rm in}, \non
}
where $\vq_{1n}\equiv \vq_1+\vq_2+\cdots\vq_n$. Henceforth we shall  not display the integration over $\vq_1\ldots \vq_n$ on the right-hand side, which is taken as granted. 
The kernel functions $F_n^s$, and $G_n^s$ are fully symmetrized with respect to the momenta in their argument. Hereafter all the kernels are to be understood as symmetrized and we omit the superscript ``s''.
Although the non-linear kernels $F_n, G_n$ are constant in time and more easily obtained 
in the EdS universe \cite{Goroff++:1986,Bernardeau:2001qr}, they become time-dependent functions 
in $\Lambda$CDM.
The standard approximation in the field is to keep the $\Lambda$CDM growth rate $D_+^{n}$
and keep the EdS, time-independent, solution for the $F_n, G_n$ kernels. 

Recently the full time-dependent solution for the kernels in $\Lambda$CDM has been found in \cite{Fasiello+:2016}.  
This solution has been derived in an integral recursive form which is somewhat 
impractical for direct use when computing correlators in perturbation theory. 
Here we will start from the results of \cite{Fasiello+:2016}, casting them in a slightly modified but equivalent form,  with the goal of expressing such solutions in an explicitly separable form, disentangling the time dependence
of the kernels from their momentum dependence.

We thus start from the full implicit $\Lambda$CDM solution of the kernels at $n$-th order 
\eq{
\label{Fn-tdep}
F_{n}(\vq_1,..,\vq_n,a)
&=\int_{0}^{a}   \frac{\dd\tilde{a}}{\tilde a} 
 \Big( w_\alpha^{(n)}(a, \tilde a) h^{(n)}_{\alpha} (\vq_1,..,\vq_n,\tilde a) 
 + w_\beta^{(n)}(a, \tilde a) h^{(n)}_{\beta}  (\vq_1,..,\vq_n,\tilde a) \Big),
\\
G_{n}(\vq_1,..,\vq_n,a)
&=\int_{0}^{a}   \frac{\dd\tilde{a}}{\tilde a} 
 \Big( u_\alpha^{(n)}(a, \tilde a) h^{(n)}_{\alpha} (\vq_1,..,\vq_n,\tilde a) 
 + u_\beta^{(n)}(a, \tilde a) h^{(n)}_{\beta}  (\vq_1,..,\vq_n,\tilde a) \Big),\non
}
where we use the scale factor $a$ as the time variable, and 
$w^{(n)}_{\alpha,\beta}, u^{(n)}_{\alpha,\beta}$ are the ``Green's functions'', 
given in the explicit form \eqref{omegaabn}. As clear by inspection, these are 
completely determined by the $D_\pm$ and $f_\pm$ functions. 
In addition to the purely time-dependent Green's functions, we have  
source terms $h_{\alpha,\beta}^{(n)}$, which also depend on time as well as the momenta. 
These source terms are recursively constructed from 
the lower order kernels $F_{n'}$ and $G_{n'}$, such that $n'<n$.
The explicit form of these source terms is given in \eqref{eq:hahb}. 
For the full derivation of this result see Appendix~\ref{sec:direct integral solution}.

As mentioned, in the integral solution in Eq.~(\ref{Fn-tdep}), the source functions 
$h_{\alpha,\beta}^{(n)}$ depend both on time and momenta, and a considerable calculational 
advantage would be achieved if one could provide solutions whose time and momenta dependent 
parts are separable. Furthermore, given the importance of higher order corrections, one should aim 
at recursive solutions, which would enable us to do without, for example, 
the order-specific ansatz used in \cite{Fasiello+:2016} to arrive at the exact analytical solution 
for the $n=3$ case. Here we present a systematic derivation of recursive separable 
functions that make up the kernels solution at any given order.

We start by suggesting the separable ansatze for the $\Lambda$CDM solutions in Eq.~\eqref{Fn-tdep},
\eq{
\label{eq:FGn}
F_n(\vq_1,..,\vq_n,a)
&= \sum_{\ell=1}^{N(n)} 
\lambda_n^{(\ell)}(a) H_n^{(\ell)}(\vq_1,..,\vq_n)= \boldsymbol{\lambda}_n(a) \cdot \vec H_n(\vq_1,..,\vq_n)\, , \\
G_n(\vq_1,..,\vq_n,a)
&= \sum_{\ell=1}^{N(n)} 
\kappa_n^{(\ell)}(a) H_n^{(\ell)}(\vq_1,..,\vq_n) =  \boldsymbol{\kappa}_n(a) \cdot \vec H_n(\vq_1,..,\vq_n) \, , \non
}
where the time dependent coefficients $\lambda_n^{(\ell)}, \kappa_n^{(\ell)}$ and the momentum operators part $H_n^{(\ell)}$ are explicitly separated.
The last equalities in Eqs.~\eqref{eq:FGn} are written with a more compact 
notation that we shall be using later in the text. For now, we keep the index ``$\ell$'' explicit  to make each manipulation of the operators as clear as possible. We stress that the same momentum 
operators $H_n^{(\ell)}$ are used for both $F_n$ and $G_n$,
while the time-dependent coefficients $\lambda_n^{(\ell)}$ and $\kappa_n^{(\ell)}$ are different. 
The number of terms in the sum $N(n)$ gives us the number of the basis elements at $n$-th perturbative order
which are, for the first few orders,
\eq{
\label{eq:N1-5}
N(1)=1, \quad N(2)=2, \quad N(3)=6, \quad N(4)=25, \quad N(5)=111.} 
In general, counting the number of terms generated by the recursive form of Eq. \eqref{Fn-tdep},
gives us the expression
\eeq{
\label{eq:N(n) exp}
N(n) = \df^K_{\frac{n}{2},\lfloor \frac{n}{2} \rfloor} 
\frac{1}{2} N \lb \tfrac{n}{2}\rb \big( 3 N \lb \tfrac{n}{2}\rb +  1 \big)
+ 3 \sum_{m=1}^{\left \lfloor (n-1)/2 \right \rfloor } N(m)N(n-m) \, .
}
Note that $N(n)$ provides a useful upper bound on the dimension of the basis operators at each given order so that  our basis may contain redundant elements. In order to obtain the minimal number of independent terms, one would need to employ relations such as the one in Eq.~(\ref{eq:wu relation}) as well as other physical 
constraints that arise from requirements such as the equivalence principle as well as mass and momentum 
conservation \cite{Fujita+:2020, DAmico++:2021}. We shall not linger on extracting all such relations at this stage but just point out that the solutions for the time coefficient we obtain should manifest all such properties, as we will show later on. 
Our task is thus split in two parts: determining the explicit form of the momentum operator basis $H_n^{(\ell)}$, as well as computing the time coefficients $\lambda_n^{(\ell)}$ and $\kappa_n^{(\ell)}$ at each  
perturbative order. For the detailed derivation of how the split of the momentum operators and the  time coefficients is performed, we refer the reader to Appendix~\ref{sec:derivtion of kernels}. Here we focus on presenting the main results.

The momentum operator basis $H_n^{(\ell)}$ is given by the recursive relation involving only the lower order basis operators. 
This is similar to the EdS solutions for the $F_n$ and $G_n$ kernels, 
although the expression for the $H_n^{(\ell)}$ contains more terms, and 
we have 
\eq{
\label{eq: H exp}
H_n^{(\ell)}(\vq_1,..,\vq_n) &= \df^K_{\frac{n}{2} , \lfloor \frac{n}{2} \rfloor } 
\sum_{i=1}^{N(n/2)} \Bigg[ \sum_{j=1}^{N(n/2)} [h_{\alpha}]_{\frac{n}{2},\frac{n}{2}}^{(ij)} \delta^K_{\ell, \phi_1}
+ \sum_{j=i}^{N(n/2)} \Big[ 2  -  \df^K_{ij} \Big] [h_{\beta}]_{\frac{n}{2},\frac{n}{2}}^{(ij)} \delta^K_{\ell, \phi_2} \Bigg]  \\
&\hspace{3.0cm} + \sum^{\left \lfloor (n-1)/2 \right \rfloor }_{m=1} \sum_{i=1}^{N(m)} \sum_{j=1}^{N(n-m)}
\Big( [h_{\alpha}]_{m,n-m}^{(ij)}\delta^K_{\ell, \phi_3}+ [h_{\alpha}]_{n-m, m}^{(ji)} \delta^K_{\ell, \phi_4} 
+ 2 [h_{\beta}]_{m,n-m}^{(ij)} \delta^K_{\ell, \phi_5} \Big) \, , \non
}
where the sourcing term $[h_{\alpha}]$ above is given by
\eq{
\label{eq:[h] def}
[h_{\alpha}]_{m,n-m}^{(ij)} (\vq_1,..,\vq_n) &=  \frac{m!(n-m)!}{n!}  \sum_{\pi- {\rm cross}} \alpha({\vq_m,\vq_{n-m}})
 H_m^{(i)} (\vec q_1, .., \vec q_m) H_{n-m}^{(j)} (\vec q_{m+1}, .., \vec q_n) \, ,
}
and the expression for $[h_{\beta}]_{m,n-m}^{(ij)}$ is obtained by simply replacing $\alpha$ with $\beta$ in Eq.(\ref{eq:[h] def}).
The Kronecker delta $\delta^K_{\ell,\phi_i}$ selects only one of the specific $[h_{\alpha,\beta}]_{m,n-m}^{(ij)}(\vq_1,..,\vq_n)$ 
operators and identifies it with $H_n^{(\ell)}$. The key to this counting are the bijective maps $\phi_i$, 
which depend on the indices $\{n,m,i,j\}$ and relate them to the set of numbers that go from 1 to $N(n)$. 
The explicit expressions for $\phi_i$ are provided in \eqref{eq:phiis}. 
At second order, one immediately recovers  $H_2^{(1)} = \alpha_s,\  H_2^{(2)} = \beta$, as expected.
Having obtained the expressions for the momentum operator basis, we now move on to determining the time dependent coefficients.

The expressions for the coefficients $\lambda_n^{(\ell)}$ (and similarly for $\kappa_n^{(\ell)}$) 
introduced in Eq.~\eqref{eq:FGn} give
\eq{
\label{eq:lambdakappa}
\lambda_n^{(\ell)}(a) &= \df^K_{\frac{n}{2} , \lfloor \frac{n}{2} \rfloor }  \sum_{i=1}^{N(n/2)}
\Bigg[ \sum_{j=1}^{N(n/2)}  
W^{(ij)}_{\alpha; n/2,n/2}\delta^K_{\ell, \phi_1}
+ \sum_{j=i}^{N(n/2)}  W^{(ij)}_{\beta; \frac{n}{2},\frac{n}{2}} \delta^K_{\ell, \phi_2} \Bigg]  \\
&\hspace{3.5cm} + \sum^{\left \lfloor (n-1)/2 \right \rfloor }_{m=1}
 \sum_{i=1}^{N(m)} \sum_{j=1}^{N(n-m)}\Big( W^{(ij)}_{\alpha; m,n-m} \delta^K_{\ell, \phi_3}
+ W^{(ji)}_{\alpha; n-m,m} \delta^K_{\ell, \phi_4}
+ W^{(ij)}_{\beta; m,n-m} \delta^K_{\ell, \phi_5} \Big) \, , \non
}
where the explicit time-integral representation for functions $W$ is given in Eq.~\eqref{eq:WU_defs}.  Analogously to what happens for the momentum basis ``vector'' $H_n^{(\ell)}$,
one of the $W_{\alpha; m,n-m}^{(ij)}(a)$ or $W_{\beta; m,n-m}^{(ij)}(a)$ functions 
with fixed indices is identified as $\lambda_n^{(\ell)}$. 
Equivalent expression holds for $\kappa_n$ as shown in Eq.~\eqref{eq:lambdakappa_II},
which identifies one of functions $U$ as $\kappa_n^{(\ell)}$.
Note that, 
the momenta operators $[h_{}]_{m,n-m}^{(ij)}$ 
and the time coefficients $W_{ m,n-m}^{(ij)}$ and $U_{ m,n-m}^{(ij)}$ 
share the same index structure. 
For $n=1$ one has $\lambda^{(1)}_1=\kappa^{(1)}_1=1$ and at $n=2$ one finds
\eq{
\label{lam2kap2}
\lambda_2^{(1)}(a) =  W^{\,2\,(1,1)}_{\alpha; 1,1} \, , 
\qquad
\lambda_2^{(2)}(a) =  W^{\,2\,(1,1)}_{\beta; 1,1} \, ,
\qquad
\kappa_2^{(1)}(a) =  U^{\,2\,(1,1)}_{\alpha; 1,1} \, ,
\qquad\ 
\kappa_2^{(2)}(a) =  U^{\,2\,(1,1)}_{\beta; 1,1} \, ,
}
which, as expected, agrees with the previous findings \cite{Sefusatti+:2011, Fasiello+:2016}.
Combined with those in Appendix~\ref{sec:derivtion of kernels}, the formulas in Eqs.~\eqref{eq: H exp} 
and \eqref{eq:lambdakappa} for $H_n^{(\ell)},\lambda_n^{(\ell)},\kappa_n^{(\ell)}$ have a close (recursive) 
structure allowing us to systematically compute kernels up to an arbitrary order $n$.
The operators $H_n^{(\ell)}$ are made up by a combination of the basic building blocks $\alpha$ and $\beta$ , 
making it straightforward to automatize the calculation with a computer program.

Although the time-dependent coefficients, $\lambda_n^{(\ell)},\kappa_n^{(\ell)}$,
can also be systematically written down, their expressions given in \eqref{eq:WU_defs} involve recursive time integrals.
This in itself is not a problem and these expressions can easily be used to obtain the numerical values for the 
time coefficients. However, instead of these integral representations, we can recast these expressions 
in the form of coupled differential equations
\eq{
\label{tk1}
\dot{W}^{\,n\,(ij)}_{\alpha; m_1,m_2} + n {W}^{\,n\,(ij)}_{\alpha; m_1,m_2} - {U}^{\,n\,(ij)}_{\alpha; m_1,m_2} 
& =\kappa^{(i)}_{m_1} \lambda^{(j)}_{m_2}\, , \\
\dot{W}^{\,n\,(ij)}_{\beta; m_1,m_2} + n {W}^{\,n\,(ij)}_{\beta; m_1,m_2} - {U}^{\,n\,(ij)}_{\beta; m_1,m_2} &= 0 \, , \non\\
\dot{U}^{\,n\,(ij)}_{\alpha; m_1,m_2} + (n-1) {U}^{\,n\,(ij)}_{\alpha; m_1,m_2} -\frac{f_{-}}{f_{+}^{2}} 
\Big[  {U}^{\,n\,(ij)}_{\alpha; m_1,m_2} - {W}^{\,n\,(ij)}_{\alpha; m_1,m_2}\Big] &= 0\, , \non\\
\dot{U}^{\,n\,(ij)}_{\beta; m_1,m_2} + (n-1) {U}^{\,n\,(ij)}_{\beta; m_1,m_2} -\frac{f_{-}}{f_{+}^{2}} 
\Big[  {U}^{\,n\,(ij)}_{\beta; m_1,m_2} - {W}^{\,n\,(ij)}_{\beta; m_1,m_2}\Big] &= \kappa^{(i)}_{m_1} \kappa^{(j)}_{m_2}\, , \non
}
where one may identify  $m_1=m,\; m_2=n-m$. The time variable is $\eta \equiv \ln D_+$, 
and a dot denotes a derivative w.r.t. to $\eta$, that is $\,\dot\,\equiv \dd/\dd\eta$.
According to Eq.~\eqref{eq:lambdakappa}, one selects time coefficients $\lambda_{n}^{(\ell)}$ 
and $\kappa_{n}^{(\ell)}$ from the functions $W$ and $U$, respectively. 
One can recursively solve Eq.~\eqref{tk1} with the initial conditions 
$\lambda_1^{(1)}=\kappa_1^{(1)}=1$ and obtain the time-dependent coefficients 
up to the desired order.

The differential equation  for the time coefficients of the kernels is amenable to the direct numerical treatment, and indeed we will use this approach  to obtain our main reference results further on. In addition, the differential equation representation is particularly useful in formulating the analytical, perturbative solution which we discuss in the next section. 

\begin{figure}[t!]
\hspace{-2cm}
\includegraphics[width=0.65\linewidth]{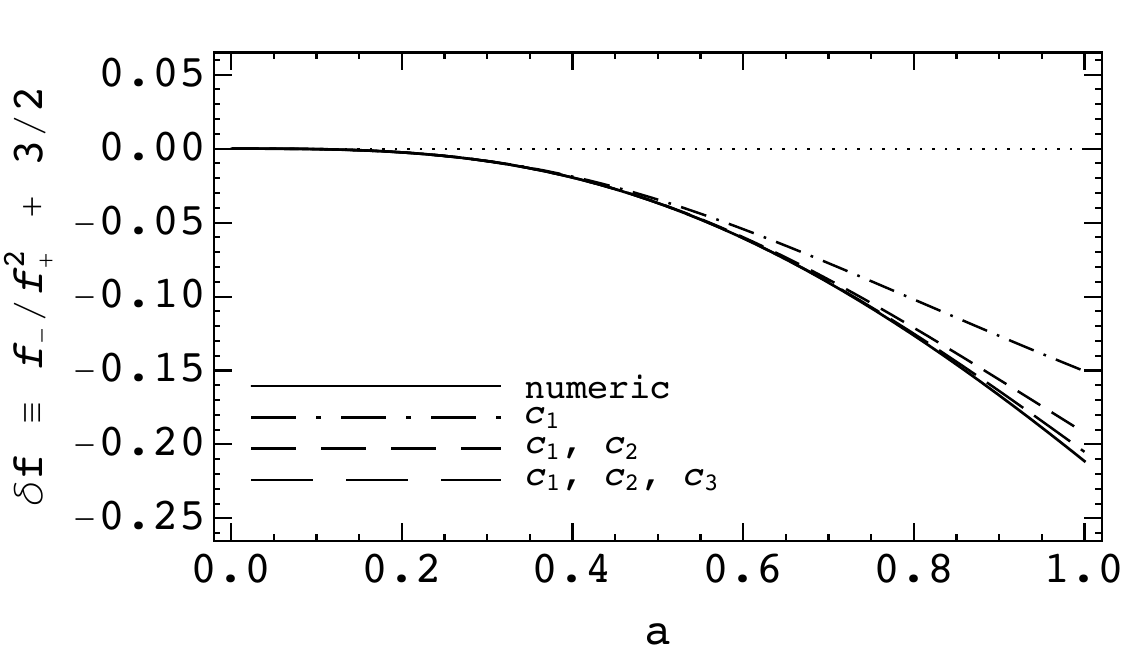}
\caption{$\delta f \equiv f_-/f_+^2 + 3/2$ is shown as a function of the scale factor. 
Numerical results (solid black line) are compared to the perturbative 
expansion in powers of $\zeta=\Omega_{\Lambda 0}/\Omega_{m0} e^{3\eta}$.
The expansion up to the first (dot-dashed lines), second (dashed) and third (long-dashed) order 
is shown using the $c_1$, $c_2$ and $c_3$ coefficients given in Eq.~\eqref{df ansatz}.
In EdS approximation this quantity vanishes identically, while beyond EdS the deviations from zero source the time dependence   of all the $\lambda_n^{(\ell)}$ and 
$\kappa_n^{(\ell)}$ coefficients.
}
\label{fig:fmfp2}
\end{figure}

Before we continue towards the solution of these equations, we stress here an interesting and practical point about the dependence of Eqs.~\eqref{tk1}  on cosmological parameters. 
The only dependence on the  cosmological parameters $\Omega_{m0}$ and $\Omega_{\Lambda0}$ ($z=0$ values) comes 
from the $f_{-}/f_{+}^{2}$ factor in the equations for $U_\alpha$ and $U_\beta$.
Moreover, in Appendix \ref{sec:linear_growth} we show that the functional dependence 
of $f_{-}/f_{+}^{2}$ can be written in the form of the single variable $\Omega_{\Lambda 0}/\Omega_{m0} e^{3\eta}$,
which captures the full dependence on the cosmological parameters. In other 
words, in Eq.~\eqref{eq:df_taylor} we show that we can write 
\eeq{
\frac{f_{-}}{f_{+}^{2}} = -\frac{3}{2}+ 
c_1 \lb \frac{\Omega_{\Lambda 0}}{\Omega_{m0}} e^{3\eta} \rb
+ c_2 \lb \frac{\Omega_{\Lambda 0}}{\Omega_{m0}} e^{3\eta} \rb^2 
+ c_3 \lb \frac{\Omega_{\Lambda 0}}{\Omega_{m0}} e^{3\eta} \rb^3  + \ldots ,
}
with some numerical coefficients $c_i$, fixed within the $\Lambda$CDM paradigm.
Figure~\ref{fig:fmfp2} shows the convergence of this expansion. 
This implies that in Eqs.~\eqref{tk1} we can change the variable to 
$\zeta \equiv \Omega_{\Lambda 0}/\Omega_{m0} e^{3\eta}$,
 which would alter only the first derivative terms with 
$\partial_\eta = 3\zeta \partial_\zeta$.
Then a change of the cosmological parameters merely results in a shift of time $\zeta$.\footnote{This fact has subsequently been also observed, at the level of one-loop  results, in the reference \cite{Rampf:2022tpg}.}

Thus, our equations are independent 
from cosmological parameters $\Omega_{m0}$ and $\Omega_{\Lambda0}$
and once solved, the solutions are valid for all choices of cosmological parameters. 

\section{Numerical and Perturbative solutions of the kernel time dependence}
\label{sec:PT_coeffs}

\begin{figure}[t!]
\centering
\includegraphics[width=0.95\linewidth]{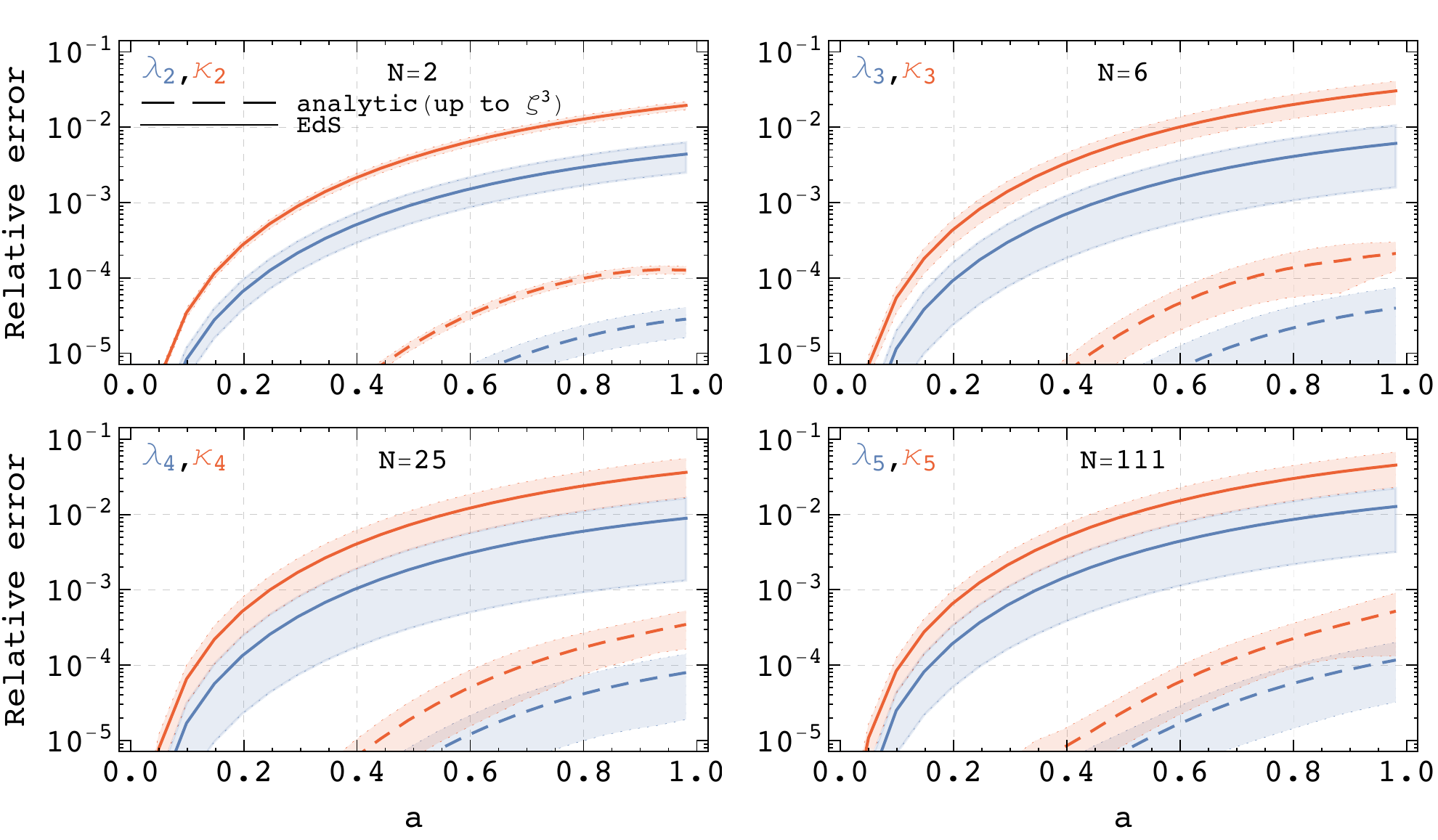}
\caption{
Relative error showing the deviation of the analytic (dashed) and EdS (solid) 
$\lambda_n$ and $\kappa_n$ coefficients from the numerical calculations.
The error is defined as $X/X_\mathrm{num}-1$ for $X=\lambda_n^{(\ell)}$ (blue) 
and $\kappa_n^{(\ell)}$ (red). 
The above four panels illustrate the typical time evolution of the relative deviations for 
$n=$2 (top left), 3 (top right), 4 (bottom left) and 5 (bottom right). 
Analytic results correspond to the perturbative calculations up to the third order in $\zeta$, 
see Eqs.~\eqref{eq:lamdakappa2} and ~\eqref{eq:lambda3}.
The central lines are the average of $N(n)$ lines for each order, while the 
coloured bands indicate the spread of all of the coefficients in $\ell$ (i.e. 
one standard deviation around the mean of all coefficients in $\ell$ is used).
As can be seen in the figure, the analytic results generally agree with the 
full numerical solution to $0.1\%$ accuracy while the deviation associated 
to the EdS results can reach up $10\%$ at late times.  
}
\label{fig:kappa_lambda_I}
\end{figure}

As we have anticipated in the previous section, the $\lambda_n^{(\ell)}$ and 
$\kappa_n^{(\ell)}$ solutions can be obtained either by using the 
explicit integral solutions given in Eqs.~\eqref{eq:WU_defs}, 
or alternatively by numerically solving the differential Eqs.~\eqref{tk1}
and using the correspondence in Eq.~\eqref{eq:lambdakappa}. 
Either one is a viable option, although given the plethora of existing tools 
for solving coupled differential equations, the path via differential equations
seems the most practical and efficient. We have used this method to obtain the results in Figure~\ref{fig:kappa_lambda_I}. 
Solid lines denote the relative deviations of $\lambda_n^{(\ell)}$ and $\kappa_n^{(\ell)}$ 
obtained in the EdS limit from the exact numerical results.
Since the $\Lambda$CDM universe matches the EdS universe at early times, 
the deviations vanish at $a=0$. As expected, the deviations grow with time
in all the panels of Figure~\ref{fig:kappa_lambda_I},
and can reach values barely shy of ten percent. Note in particular that
the typical deviation at redshift $z=1$ is a factor of a few smaller than its $z=0$ counterpart. One expects this difference to propagate all the way to correlators.

\begin{figure}[t!]
\centering
\includegraphics[width=0.9\linewidth]{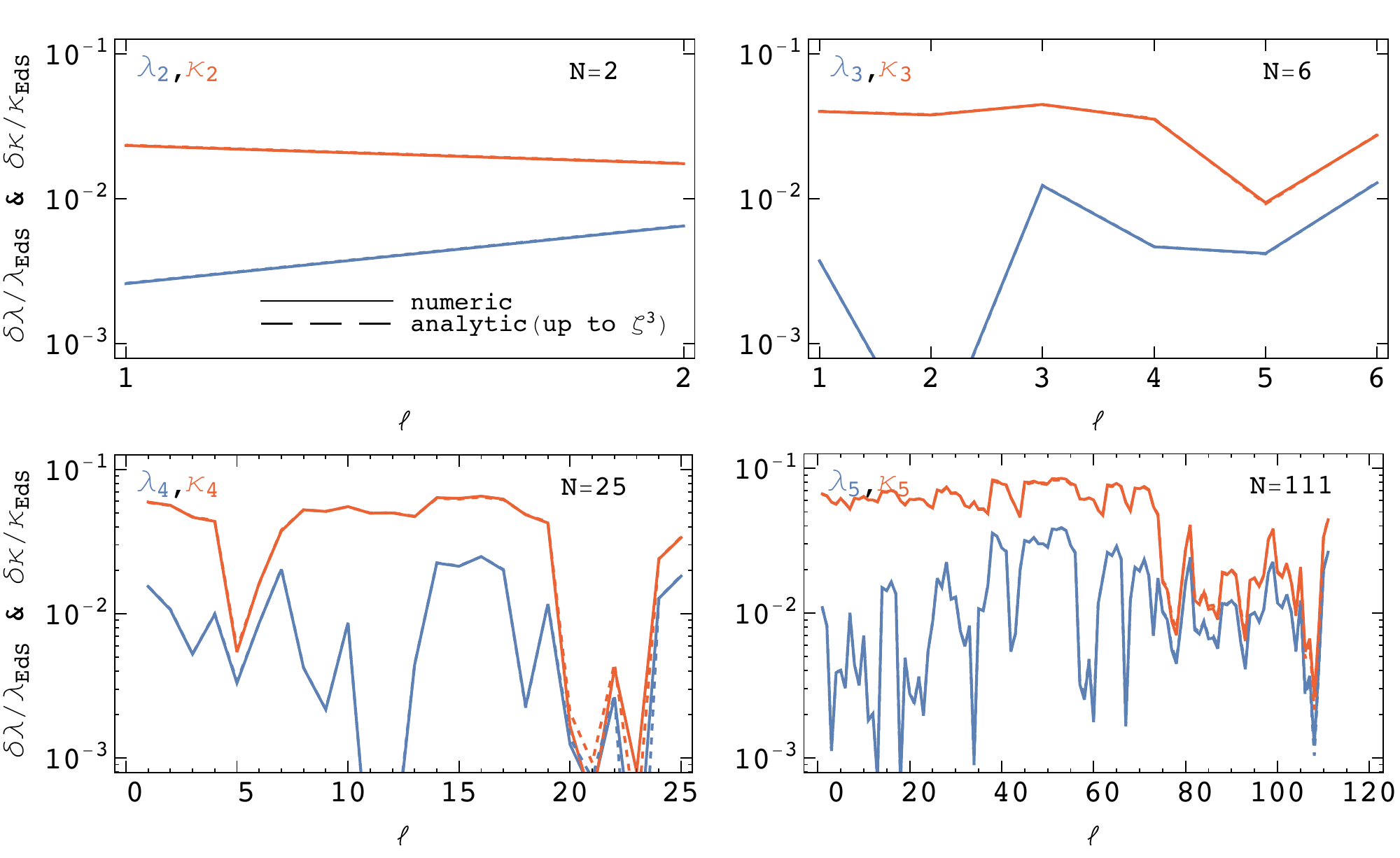}
\caption{
Relative deviation of the $\lambda_n^{(\ell)}$(blue) $\kappa_n^{(\ell)}$(red) coefficients
in $\Lambda$CDM cosmology to the EdS values at the present time, $a=1$.
Shown are both numerical (solid lines) and analytic (dashed lines)
results, calculated up to the third power in $\zeta$.
Different values of $n$ are given in each panel, 
coefficients range is $(n,\ell)=(2,2)$, $(3,6)$, $(4,25)$ and $(5,111)$. 
$\kappa_n^{(\ell)}$'s tend to have larger errors ranging from $0.1\%$
to almost ten percent, while $\lambda_n^{(\ell)}$'s errors are somewhat 
smaller reaching up to a few percent.
This figure also shows that differences between the $\Lambda$CDM and EdS 
coefficients tend to grow with perturbative order $n$.  
}
\label{fig:kappa_lambda_II}
\end{figure}

Given that the number of coefficients at higher orders is large (see Eq.~\eqref{eq:N1-5}), in Figure~\ref{fig:kappa_lambda_I} 
we show the average value of all of the deviations and the typical spread 
(in terms of the one standard deviation). From this, we can observe the trend that, at later times, the deviation of the coefficients from the EdS approximation tends to grow with $n$, (i.e. when considering higher perturbative orders) and the spread of the coefficient values may also grow (i.e. some tend to be close to the EdS values while for others the deviations can be larger). One might wonder how much of a role outliers play  in such analysis. To address this, in Figure~\ref{fig:kappa_lambda_II} we show the relative deviations 
of the $\Lambda$CDM and the EdS results for all the time-dependent coefficients 
at $a=1$, up to $n=5$. Although the $\Delta \lambda_n^{(\ell)}$ are typically 
$\mathcal{O}(1\%)$ in size, $\Delta \kappa_n^{(\ell)}$ can be as large as $\mathcal{O}(10\%)$, 
which is not at all negligible when compared to the precision of upcoming observations.\vspace{0.1cm} 

Motivated by the discussion in the last section, we now embark on a journey to find the 
analytic perturbative solution for the time dependence of the $\lambda_{n}^{(\ell)}$ and $\kappa_{n}^{(\ell)}$ coefficients. 
As shown in Figure~\ref{fig:kappa_lambda_I}, EdS approximation for these coefficients is a good  starting point, and the deviation
are relatively small. It will thus serve us well to use the EdS solution as the result around which to organise the perturbative expansion.
These deviations from the EdS approximation are encoded in the factor $f_-/f_+^2$ in 
the differential Eqs.~\eqref{tk1}, as well as in the higher  order source terms. 
Since this ratio is exactly $-3/2$ in the EdS limit, we introduce the deviation from the EdS value as a small perturbative parameter,
\eq{
\df f (\tau)\equiv \frac{f_{-}(\tau)}{f^2_{+}(\tau)}+\frac{3}{2}.
\label{deltaf def}
}
Figure~\ref{fig:fmfp2}, shows that this dimensionless parameter is $\lesssim0.2$ in absolute terms throughout the evolution of the Universe. 
This is a good indication that a convergent perturbative expansion can be obtained by treating $\delta f$ as a small parameter.
In addition to the expansion in $\delta f$ we are interested in representing $\delta f$ as a power series in $\zeta\equiv\Omega_{\Lambda 0}/\Omega_{m0} e^{3\eta}$
which would allow us to express the final $\lambda_{n}^{(\ell)}$ and $\kappa_{n}^{(\ell)}$ results as a power series in the same variable.
We thus expand  $W^{\,n\,(ij)}_{m1,m2}$ and $U^{\,n\,(ij)}_{m1,m2}$ 
appearing in Eqs.~\eqref{tk1} as
\eq{
W^{n}=W^{n[0]}+W^{n[1]}+W^{n[2]}+\cdots,\qquad
U^{n}=U^{n[0]}+U^{n[1]}+U^{n[2]}+\cdots,
}
where the superscript $[n]$ denotes the perturbative order with respect to $\delta f$, namely
$\mathcal{O}(\delta f^n)$, and the suppressed indices are the same in both sides of the equations. 
Note that order $[0]$ means the solution in the EdS limit. 
This also corresponds to the static limit of the Eqs.~\eqref{tk1}, where we drop the time derivative terms turning these equations into  recursive algebraic equations. In this limit, once the coefficients are combined with the momentum basis $H_n^{(\ell)}$, one just recovers the usual EdS solutions for the $F_n$ and $G_n$ kernels.

For a detailed derivation of the perturbative results, we refer the reader to Appendix~\ref{sec: perturbation of WU}. Here we just note that the  solutions of Eq.~\eqref{tk1} at the order of $\mathcal{O}(\delta f^l)$ can be expressed in the integral form of lower order terms
\eq{
\label{eq:Walpha_l_recursive}
W_\alpha^{n[l]}&=\mathcal{I}_n\left[\left(\partial_\eta +n+\tfrac{1}{2}\right)(\kappa\lambda)^{[l]}
+\delta f\left(\dot{W}^{\,n[l-1]}_{\alpha}+(n-1)W^{\,n[l-1]}_{\alpha}-(\kappa\lambda)^{[l-1]} \right)\right]\, ,
\\
W_\beta^{n[l]}&=\mathcal{I}_n\left[(\kappa\kappa)^{[l]}
+\delta f\left(\dot{W}^{\,n[l-1]}_{\beta}+(n-1)W^{\,n[l-1]}_{\beta} \right)\right]\, ,\non
}
where $\mathcal{I}_n[X]$ is the time functional defined in \eqref{eq:I def}.
 Using the above recursive relations repeatedly, one can obtain the expression for the perturbative solutions~\eqref{eq:Walphan_general} for $W_\alpha^{n}$ and $W_\beta^{n}$. Moreover, to analytically evaluate the integral expressions so obtained, 
we rely on the expansion of $\delta f$ in power  law form. 
In Appendix~\ref{sec:linear_growth} we show how one may obtain the expansion 
of $\delta f$ in powers of $\zeta = \Omega_{\Lambda 0}/\Omega_{m0} e^{3\eta}$. Up to the third order, it reads
\eeq{
\label{df ansatz}
\delta f(\eta) \simeq c_1 \zeta
+ c_2 \zeta^2
+ c_3 \zeta^3 \, , \qquad
c_1=-\frac{3}{22}\,,\  c_2=-\frac{141}{4114}\,,\  c_3=-\frac{9993}{1040842}.
}

Upon performing these steps, we derive the  analytic expressions for $\lambda_n^{(\ell)}$ and $\kappa_n^{(\ell)}$ for a generic choice of cosmological parameters $\Omega_{\Lambda 0}$, $\Omega_{m0}$. 
For instance, at $n=2$, we obtain
\eq{
\label{eq:lamdakappa2}
\lambda_2^{(1)}&=\frac{5}{7}-\frac{c_1}{91}\zeta -\frac{4c_2}{931}\zeta^2 - \frac{2c_3}{875}  \zeta^3 \, , 
&
\lambda_2^{(2)}&=\frac{2}{7}+\frac{c_1}{91}\zeta+\frac{4c_2}{931}\zeta^2 + \frac{2c_3}{875} \zeta^3 \, , 
\\
\kappa_2^{(1)}&=\frac{3}{7}-\frac{5c_1}{91} \zeta -\frac{32c_2}{931}\zeta^2  - \frac{22c_3}{875} \zeta^3 \, , 
&
\kappa_2^{(2)}&=\frac{4}{7}+\frac{5c_1}{91}\zeta +\frac{32c_2}{931}\zeta^2  + \frac{22c_3}{875} \zeta^3 \, , 
\non
}
where we write explicitly only the leading order results in $\delta f$, suppressing the $\mathcal{O}(\delta f^2)$ terms. The results for $n=3$ and $4$, namely  $\lambda_3^{(\ell)}, \kappa_3^{(\ell)}, \lambda_4^{(\ell)}, \kappa_4^{(\ell)}$, are  reported in Eqs.~\eqref{eq:lambda3} and \eqref{eq:lambda4}. Using this perturbative approach, it is straightforward to generate all terms at higher orders.\footnote{\textit{Mathematica} notebook for these coefficients and $H_n$ kernels, up to the fifth order, can be found in the \textit{arXiv} source file of this paper.} It suffices here to derive the ones that will be needed for the two-loop calculation (up to the $\lambda_5^{(\ell)}$ 
and $\kappa_5^{(\ell)}$ coefficients). Given the number of components (recall that for $n=5$ we have 111 terms), we do not report the explicit expression. Nonetheless, in Figs.~\ref{fig:kappa_lambda_I} 
and \ref{fig:kappa_lambda_II} we compare the analytical results at leading order in $\delta f$ to the numerical ones. One observes that our analytic expressions typically achieve 
$\mathcal{O}(10^{-3})$ accuracy at the present time and better accuracy at earlier times. 
Compared to the EdS results, our analytic expressions are about 100 times more accurate.

\section{One- and two-loop power spectra}
\label{sec:one_two_loops}

Equipped with the results of the last section, we are now ready to tackle observables such as 
the matter density and velocity power spectra, as well as the cross power spectrum. 
In the process, we shall develop and illustrate the utility of a systematic method to handle 
infrared and ultraviolet divergences in loop integrals. The equivalence principle, 
fully at work in $\Lambda$CDM, guarantees that specific cancellations will take place 
in the IR configurations of the kernels. Similar cancellations take place due to the 
mass and momentum conservation,  in the absence of which there would be large UV contributions.
Such cancellations between large contributions typically require very high precision, 
thus making numerical integration more difficult and less stable. As we show in the 
rest of this section, the properties required for such cancellations 
are all imprinted in the solutions for the 
$\lambda_n$ and $\kappa_n$ coefficients. The coefficients ``remember'' all the information 
 inherited from their EoM,  and we see the equivalence principle, mass and momentum conservation respected and manifested in the  various limits of the one- and two-loop power spectra that we study below.
One can use these properties in evaluating the loop integrals: we do so by first isolating the leading divergences,  analytically confirming they are cancelled out, and numerically evaluating the remaining ``regularised'' parts of the power spectra.

As observables whose calculation (and target of percent-level precision) 
requires an improvement upon the EdS approximation, we compute 
the one-loop and two-loop order of the following power spectra
\eq{
(2\pi)^3 \df^D_{\vec k + \vec k'} P_{\delta\delta}(k) 
= \la \df(\vec k) \df(\vec k') \ra\, , ~~~
(2\pi)^3 \df^D_{\vec k + \vec k'} P_{\delta\theta}(k)
= \la \df(\vec k) \theta(\vec k') \ra\, , ~~~
(2\pi)^3 \df^D_{\vec k + \vec k'} P_{\theta\theta}(k)
= \la \theta(\vec k) \theta(\vec k') \ra\, .
}

\subsection{One-loop results}
Using the notation introduced in Eq.~\eqref{eq:FGn},
the one-loop results are as follows:
\eq{
\label{eq:PS_one_loop} 
P^\text{1-loop}_{\delta\delta} (k) &
=  \lb \vec{\lambda}_2 \cdot \vec I_{22} \cdot  \vec{\lambda}_2 \rb + 2 \lb \vec{\lambda}_1 \cdot \vec I_{13} \cdot \vec{\lambda}_3 \rb\, ,\\
P^\text{1-loop}_{\delta\theta} (k) &
=  \lb \vec \lambda_2 \cdot \vec I_{22} \cdot \vec \kappa_2 \rb + \lb \vec \lambda_1 \cdot \vec I_{13} \cdot \vec \kappa_3 + 1 \leftrightarrow 3 \rb\, , \non \\
P^\text{1-loop}_{\delta\theta} (k) &
=  \lb \vec \kappa_2 \cdot \vec I_{22} \cdot \vec \kappa_2 \rb + 2 \lb \vec \kappa_1 \cdot \vec I_{13} \cdot \vec \kappa_3 \rb\, , \non
}
where we have defined the scale dependent integrals
\eq{
\vec I_{22} &= 2 \int_{\vec q} \vec H_2(\vec q, \vec k-\vec q) \otimes \vec H_2 (\vec q, \vec k-\vec q) P_{\rm lin}(\vec q) P_{\rm lin}(\vec k - \vec q)\; , \\
\vec I_{13} &= 3 \int_{\vec q} \vec H_1(\vec k) \otimes \vec H_3(\vec k, \vec q, -\vec q) P_{\rm lin}(\vec k) P_{\rm lin}(\vec q) \; . \non
}
Note that, when seen as matrices, these integrals have the properties: $\vec I_{22}^T = \vec I_{22}$ and $\vec I_{13} = \vec I_{31}^T$. 

In this subsection,  we describe an efficient method to compute loop power spectra using the one-loop power spectrum as the simplest example before applying it to two-loop calculations. This method is essentially important to avoid artificial residuals of the physical cancellations and achieve high-precision calculations while saving computational resources. 
Our strategy is simple. We know the integrals $\vec I_{ij}$ in \eqref{eq:PS_one_loop} contain the IR and UV contributions, which eventually cancel. Hence we isolate them as in $\vec I_{ij} = \vec{\tilde I}_{ij} + [\vec I_{ij}]_{\rm IR} + [\vec I_{ij}]_{\rm UV}$, where $\vec{\tilde I}_{ij}$ is the remaining regular part. The cancellations of $[\vec I_{ij}]_{\rm IR}$ and $[\vec I_{ij}]_{\rm UV}$ are analytically confirmed. Then, we focus on the numerical evaluations of the regularised contributions from $\vec{\tilde I}_{ij}$.

 We begin with $\vec I_{22}$. One can see that its IR contributions come from two configurations, namely
$\vec q \to 0$, and $\vec q \to \vec k$.  It is convenient at this point to re-map the second sector 
to the first one (see \cite{Carrasco++:2013a}) as 
\eeq{
\vec I_{22}  = \int_{|\vec q| < |\vec k - \vec q |}+\int_{|\vec q| \geq |\vec k - \vec q |} 
= 4 \int_{\vec q} \vec H_2(\vec q, \vec k-\vec q) \otimes \vec H_2(\vec q, \vec k-\vec q) \Theta( |\vec k - \vec q | - q) P_{\rm lin}(\vec q) P_{\rm lin}(\vec k - \vec q)\, . 
}
 We extract the IR and UV contributions in this integrand.
Using the asymptotic form of the kernels, one can write
\eq{
\vec H_2(\vec q, \vec k-\vec q) \otimes \vec H_2(\vec q, \vec k-\vec q)
\sim 
\begin{cases}
\vec h^{(2)}_{22,{\rm IR}}  \lb \vec k, \hat q \rb \frac{k^2}{q^2} + \vec h^{(1)}_{22,{\rm IR}}  \lb \vec k, \hat q \rb \frac{k}{q} + \mathcal{O}(q^0) \, ,
&{\rm as}~~ q \to 0\, , \\[0.5em]
\vec h^{(4)}_{22,{\rm UV}} \big( \hat k, \vec q \big) \frac{k^4}{q^4} + \mathcal{O}(k^5) \, ,
&{\rm as}~~ k \to 0\, . 
\end{cases}
}
For the explicit form of the $\vec H_n$ operators in the various limits, as well as the asymptotics of  $\vec h^{(n)}_{22} $,
we refer the reader to Appendix \ref{sec: IR and UV limits}.
Having identified both the IR and UV limits of the kernel products, we can introduce the 
regularised version of our integral (we label it $\vec{\tilde I}_{22}$) by subtracting 
these contributions only in the asymptotic regimes. In order to do so, we introduce window functions, $W^{\rm IR}_{22}(k)$ and $W^{\rm UV}_{22}(k)$, which ensure that the appropriate 
asymptotic form is applied only in the IR and UV regimes. 
The regularised integral is thus given by 
\eq{
\vec{\tilde I}_{22} =
\int_{\vec q} 
\bigg[ 4 \vec H_2(\vec q, \vec k-\vec q) & \otimes \vec H_2(\vec q, \vec k-\vec q) \Theta( |\vec k - \vec q | - q) P_{\rm lin}(\vec k - \vec q) \\
& -  4 \lb \vec h^{(2)}_{22,{\rm IR}}  \lb \vec k, \hat q \rb \frac{k^2}{q^2} + \vec h^{(1)}_{22,{\rm IR}} \lb \vec k, \hat q \rb  \frac{k}{q} \rb W^{\rm IR}_{22}(k) P_{\rm lin}(\vec k)
 - 2 \vec h^{(4)}_{22,{\rm UV}} \big( \hat k, \vec q \big) \frac{k^4}{q^4} W^{\rm UV}_{22}(k) P_{\rm lin}(\vec q) \bigg] P_{\rm lin}(\vec q) , \non
}
where one can write $\vec I_{22} = \vec{\tilde I}_{22} + [\vec I_{22}]_{\rm IR} + [\vec I_{22}]_{\rm UV}$, with 
\eq{
[\vec I_{22}]_{\rm IR} &= 4 \int_{\vec q} \lb \vec h^{(2)}_{22,{\rm IR}}  \lb \vec k, \hat q \rb \frac{k^2}{q^2} 
+ \vec h^{(1)}_{22,{\rm IR}} \lb \vec k, \hat q \rb  \frac{k}{q} \rb W^{\rm IR}_{22} P_{\rm lin}(\vec k) P_{\rm lin}(\vec q)
= \lb \vec h^{\rm IR}_{22} W^{\rm IR}_{22} \rb  k^2 \sigma^2_2 P_{\rm lin}(k) \, , \\
[\vec I_{22}]_{\rm UV} &= 2 \int_{\vec q} \vec h^{(4)}_{22,{\rm UV}} \big( \hat k, \vec q \big) \frac{k^4}{q^4} W^{\rm UV}_{22} P_{\rm lin}(\vec q) P_{\rm lin}(\vec q)
=  \lb \vec h^{{\rm UV}}_{22} W^{\rm UV}_{22} \rb k^4 \Sigma^2_2\, . \non
}
Here we have introduced $\Sigma^2_2 = \frac{1}{3} \int_{\vec q} P_{\rm lin}(q)^2/q^2,\ \sigma^2_2 = \frac{1}{3} \int_{\vec q} P_{\rm lin}(q)/q^2$,
and
\eeq{
\vec h^{\rm IR}_{22} = \mat{ 1 & 1 \\ 1 & 1} ~~{\rm and}~~
\vec h^{\rm UV}_{22} = \frac{1}{2}  \mat{ \tfrac{7}{5} & -1 \\ -1 & 3} ,
}
as also shown in Appendix \ref{sec: IR and UV limits}. Note that the UV contribution does not have an additional factor of two since it does not require a re-map in the low $k$ limit.

Let us specify the window functions $W^{\rm IR}_{22}$ and $W^{\rm UV}_{22}$. 
The task we demand of these functions is to effectively restrict the domain of the contribution they are multiplied by  into the appropriate momenta configuration, i.e. high and low $k$ regimes respectively. 
We are free to choose the form of such functions that is best suited for the task at hand. We choose one convenient and simple form 
\eeq{
W^{\rm IR}_{22}(k) = \frac{(k/k_{\rm IR})^4}{1+(k/k_{\rm IR})^4}, ~~{\rm and}~~
W^{\rm UV}_{22}(k) = \frac{1}{1+(k/k_{\rm UV})^6},
} 
with parameters $k_{\rm IR} \approx 0.1{\rm Mpc}/h$ and $k_{\rm UV} \approx 0.2{\rm Mpc}/h$.
It will, of course, be convenient to choose $W^{\rm IR}_{22} = W^{\rm IR}_{13}$, in order 
to quickly arrive at the cancellation of the leading IR contributions in the total one-loop power spectrum.

Let us turn to the $\vec I_{13}$ term, where the asymptotic contributions are
\eeq{
\vec H_1(\vec k) \otimes \vec H_3( \vec k, \vec q,-\vec q)
\sim 
\begin{cases}
\vec h^{(2)}_{13,{\rm IR}} \lb \vec k, \hat q \rb \frac{k^2}{q^2} + \mathcal{O}(q^0) \, , &{\rm as}~~ q \to 0 , \\[0.5em]
\vec h^{(0)}_{13,{\rm UV}}\big( \hat k ,\vec q \big) + \vec h^{(2)}_{13,{\rm UV}}\big( \hat k ,\vec q \big) \frac{k^2}{q^2} + \mathcal{O}(k^4)\, , &{\rm as}~~ k \to 0 .
\end{cases}
}
In an analogous way to 
$\vec {\tilde I}_{22}$, we can introduce the regularised integrals as 
\eq{
\vec{\tilde I}_{13} = 3 \int_{\vec q} 
\bigg[ \vec H_1(\vec k) \otimes &\vec H_3(\vec k, \vec q, -\vec q) \\
& - \vec h^{(2)}_{13,{\rm IR}} \lb \vec k, \hat q \rb \frac{k^2}{q^2} W^{\rm IR}_{13}
- \lb \vec h^{(0)}_{13,{\rm UV}}\big( \hat k ,\vec q \big) + \vec h^{(2)}_{13,{\rm UV}}\big( \hat k ,\vec q \big) \frac{k^2}{q^2} \rb W^{\rm UV}_{13}  \bigg] P_{\rm lin}(\vec k) P_{\rm lin}(\vec q), \non
}
i.e. $\vec I_{13} = \vec{\tilde I}_{13} + [\vec I_{13}]_{\rm IR} + [\vec I_{13}]_{\rm UV}$ with
\eq{
[\vec I_{13}]_{\rm IR} &= 3 \int_{\vec q} \vec h^{(2)}_{13,{\rm IR}} \lb \vec k, \hat q \rb \frac{k^2}{q^2} W^{\rm IR}_{13} P_{\rm lin}(\vec k) P_{\rm lin}(\vec q)
= \lb \vec h^{{\rm IR}}_{13}  W^{\rm IR}_{13}\rb k^2 \sigma^2_2 P_{\rm lin}(k) \, , \\
[\vec I_{13}]_{\rm UV} &= 3 P_{\rm lin}(\vec k) \int_{\vec q} \vec h^{(2)}_{13,{\rm UV}}\big( \hat k ,\vec q \big) \frac{k^2}{q^2} 
W^{\rm UV}_{13} P_{\rm lin}(\vec k) P_{\rm lin}(\vec q)
= \lb \vec h^{\rm UV}_{13} W^{\rm UV}_{13} \rb k^2 \sigma^2_2 P_{\rm lin}(k) , \non
}
where $\vec h^{\rm IR}_{13} =  - \mat{ 1 & 1 & 0 & 0 & 1 & 1}$ and $\vec h^{\rm UV}_{13} = -\mat{ 1, 1, -\frac{12}{5}, 0, 5, 1 }$. Here, since $\vec \lambda_3 \cdot \vec h^{(0)}_{13,{\rm UV}} = \vec \kappa_3 \cdot \vec h^{(0)}_{13,{\rm UV}} =0$,
the term $\vec h^{(0)}_{13,{\rm UV}}$ does not contribute to the power spectrum and is not included in $[\vec I_{13}]_{\rm UV}$.
The remaining UV contribution comes only from the  next-to-leading order term $\vec h^{(2)}_{13,{\rm UV}}$. 

This is of course guaranteed by the mass and momentum conservation of the original EoM  \cite{Bernardeau:2001qr}. We now move to the IR cancellations. As soon as the ``22'' and ``13'' terms are combined, we find
\eeq{
\lb \vec \lambda_2 \cdot [\vec I_{22}]_{\rm IR} \cdot \vec \lambda_2 \rb + 2 \lb \vec \lambda_1 \cdot [\vec I_{13}]_{\rm IR} \cdot \vec \lambda_3 \rb 
= \Big[ \lb \vec \lambda_2 \cdot \vec h^{\rm IR}_{22} \cdot \vec \lambda_2 \rb W^{\rm IR}_{22} 
+ 2 \lb \vec \lambda_1 \cdot \vec h^{\rm IR}_{13} \cdot \vec \lambda_3 \rb  W^{\rm IR}_{13} \Big] k^2 \sigma^2_2 P_{\rm lin}(k)
= 0,
}
where we take $W^{\rm IR}_{22} = W^{\rm IR}_{13}$.  This is of course the same as the usual IR cancellation between $P_{22}$ and $P_{13}$ 
in standard perturbation theory (SPT) \cite{Bernardeau:2001qr}: 
it is guaranteed for equal-time correlators by the equivalence principle, as has been discussed in
\cite{Kehagias+:2013, Peloso+:2013a, Creminelli++:2013a, Peloso+:2013b, Creminelli++:2013b, 
Fujita+:2020, DAmico++:2021}.
The same cancellations take place for the velocity-velocity spectrum and for the velocity-density cross-spectrum. 
The final expression for the one-loop density power spectrum is
\eq{
P^\text{1-loop}_{\delta\delta} (k) 
&=  \lb \vec \lambda_2 \cdot \vec {\mathcal I}_{22} \cdot \vec \lambda_2 \rb + 2 \lb \vec \lambda_1 \cdot \vec {\mathcal I}_{13} \cdot \vec \lambda_3 \rb ,
}
and analogous expressions hold for the other two observables, $P^\text{1-loop}_{\delta\theta}$ and $P^\text{1-loop}_{\theta\theta}$, with the appropriate time-dependent coefficients  in the same way as Eqs.~\eqref{eq:PS_one_loop}. The momentum dependent matrices $\vec {\mathcal I}_{ij}$ that all the three power spectra share at one-loop order are given by
\eeq{
\vec {\mathcal I}_{22} =  \vec {\tilde I}_{22} + \lb \vec h^{\rm UV}_{22} W^{\rm UV}_{22} \rb k^4 \Sigma^2_{4}, 
\qquad{\rm and}\qquad
\vec {\mathcal I}_{13} = \vec {\tilde I}_{13} + \lb \vec h^{\rm UV}_{13} W^{\rm UV}_{13} \rb  k^2 \sigma^2_{2} P_{\rm lin}.
}
We numerically evaluate these regularised expressions, a procedure that circumvents the expensive numerical treatment of the IR and UV cancellations and saves significant computational time.

\begin{figure}[t!]
\centering
\includegraphics[width=0.95\linewidth]{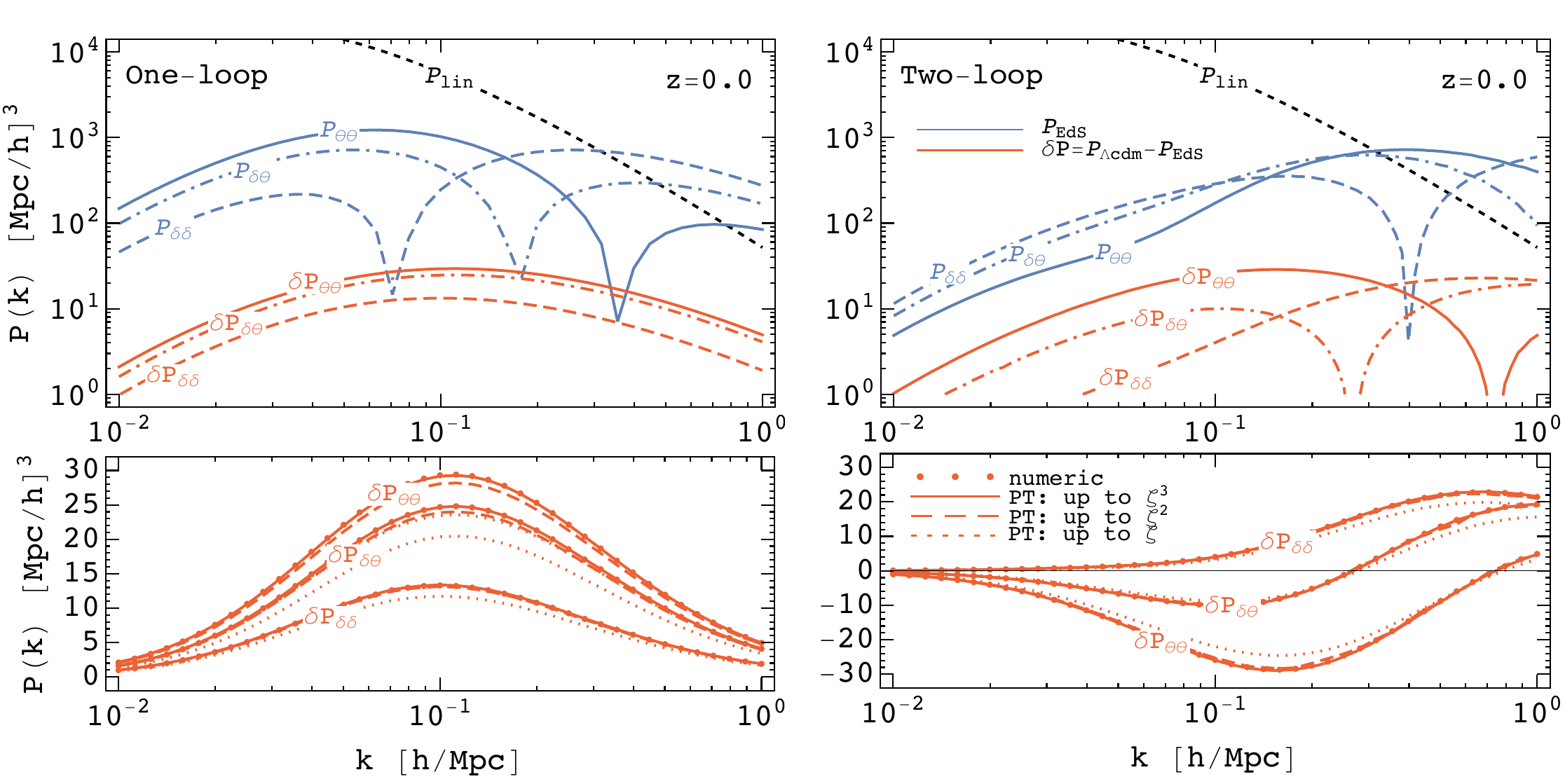}
\caption{
One- (left panel) and two-loop (right panel) contributions to the density-density, density-velocity and velocity-velocity power spectrum. \textit{Upper panels} show the absolute contributions of EdS 
results (blue lines) compared to the $\Lambda$CDM corrections (red lines). We see that the three different spectra $P_{\delta\delta}$ (dashed lines), $P_{\delta\theta}$ (dot-dashed lines) and 
$P_{\theta\theta}$ (solid lines) receive corrections of different sizes, whose 
relative importance is also a function of the scale dependence of the EdS 
terms. \textit{Lower panels} display the $\Lambda$CDM corrections  
$\delta P_{\delta\delta}$, $\delta P_{\delta\theta}$ and $\delta P_{\theta\theta}$
computed using the numerical evaluations of the $\lambda_n$
and $\kappa_n$ coefficients (shown in dots). We also show the 
perturbative time dependence computation as described in Sec.~\ref{sec:PT_coeffs}.
The results including the $\mathcal{O}(\zeta^1)$ (dotted lines),
$\mathcal{O}(\zeta^2)$ (dashed lines)
and $\mathcal{O}(\zeta^3)$ (solid lines) contributions are shown.
Results are shown for redshift $z=0.0$.
}
\label{fig:power_sepctra_I}
\end{figure}

In the left panels of Figure~\ref{fig:power_sepctra_I} we show the one-loop contributions for all these power spectra: $P_{\delta\delta}$, $P_{\delta\theta}$ and  $P_{\theta\theta}$. In particular,
we display the EdS solutions and the corresponding   $\Lambda$CDM correction to the EdS result, i.e. $\delta P_{\delta\delta} = P^{\Lambda{\rm cdm}}_{\delta\delta} -P^{\rm EdS}_{\delta\delta}$
(and equivalently for the other two spectra). As one can see from the figure, 
the one-loop $\Lambda$CDM corrections are from one to two orders of magnitude smaller than the one-loop EdS contributions. However, 
the relevant regime lies in the higher $k$ range ($k\gtrsim 0.1h/{\rm Mpc}$), 
given that is where the one-loop contributions start to be comparable in amplitude to the 
linear result. Moving towards higher redshift the corrections relative to the EdS result decrease further (see Appendix \ref{app:two-loops_higher_z}).

In Figure~\ref{fig:power_sepctra_II}
we show the ratio of the three different total power spectra in $\Lambda$CDM 
relative to the EdS results. The left panels display the one-loop results, where the upper left panel shows the one-loop spectra without adding counterterms to either EdS or $\Lambda$CDM solutions. We see that, at $z=0$, the largest corrections range from a half (for $\delta P_{\delta\delta}$)
to a few percent (for $P_{\theta\theta}$) at scales $k\sim0.4h/{\rm Mpc}$ 
(scales where higher loop results are also relevant).
These results are consistent with the earlier findings shown in \cite{Fasiello+:2016,Donath+:2020,Garny+:2020}). 
In the bottom panel of the same figure 
we plot the effects of the EFT counterterms on the total deviations from $\Lambda$CDM.
First, we note that for the typical values of the counterterms, shown as the 
central lines within the grey bands, the relative difference in the power spectra is lowered. This is expected since the counterterms contributions to the total power spectrum is of the same size as the loop contributions at the relevant scales 
and by construction are equivalent in both the $\Lambda$CDM and EdS case.

Grey bands around each of the three power spectrum lines show the effects of variations (of order 5\%) in the values of the $\Lambda$CDM counterterms. As one can see, assuming the $\sim1\%$ 
accuracy thresholds, the presence of a counterterm  can make up for the deviation between the EdS and the  $\Lambda$CDM result for
the density-density power spectrum. This is not the case for the density-velocity and velocity-velocity power spectra, which
exhibit a noticeably steeper scale dependence.

As the last comment on Figure~\ref{fig:power_sepctra_II}, we note that,  in addition to the results obtained by numerical 
evaluation of the $\lambda_n$ and $\kappa_n$
coefficients (shown as black lines), we also show in the upper left panel the perturbative results given in Eqs.~\eqref{eq:lamdakappa2} 
and \eqref{eq:lambda3}. The profiles corresponding to the $\mathcal{O}(\zeta^1)$ perturbative order are shown explicitly (red lines), and we see that they exhibit up to $0.5\%$ agreement with the full numerical solutions. 
The perturbative solutions accounting up to  $\mathcal{O}(\zeta^3)$ order expansion are not shown as they would be indistinguishable from the full numerical solutions already present in these plots. 

\subsection{Two-loop results}
In the rest of this section we turn our attention to the the two-loop results. 
Writing the perturbative contributions in the separable form, we have
\eq{
P^\text{2-loop}_{\delta\delta}(k) 
&= \lb \vec \lambda_3 \cdot \vec I_{33} \cdot \vec \lambda_3 \rb 
+ 2 \lb \vec \lambda_2 \cdot \vec I_{24} \cdot \vec \lambda_4 \rb
+ 2 \lb \vec \lambda_1 \cdot \vec I_{15} \cdot \vec \lambda_5 \rb \, , \\
P^\text{2-loop}_{\delta\theta} (k) 
&= \lb \vec \lambda_3 \cdot \vec I_{33} \cdot \vec \kappa_3 \rb 
+ \lb \vec \lambda_2 \cdot \vec I_{24} \cdot \vec \kappa_4 + 2 \leftrightarrow 4 \rb
+ \lb \vec \lambda_1 \cdot \vec I_{15} \cdot \vec \kappa_5 + 1 \leftrightarrow 5 \rb \, , \non\\
P^\text{2-loop}_{\theta\theta} (k) 
&= \lb \vec \kappa_3 \cdot \vec I_{33} \cdot \vec \kappa_3 \rb 
+ 2 \lb \vec \kappa_2 \cdot \vec I_{24} \cdot \vec \kappa_4 \rb
+ 2 \lb \vec \kappa_1 \cdot \vec I_{15} \cdot \vec \kappa_5 \rb \, , \non
}
where the two-loop integral functions are
\eq{
\label{eq:Inn_two_loop}
\vec I_{33}
&= 3 \int_{\vec q, \vec p} \bigg( 
2 \vec H_{3}(\vec k - \vec q - \vec p, \vec q, \vec p) \otimes \vec H_3(\vec k - \vec q - \vec p, \vec q, \vec p) P_{\rm lin} (\vec k - \vec q - \vec p) \\
&\hspace{5cm}+ 3 \vec H_3(\vec k, - \vec q, \vec q) \otimes \vec H_3(\vec k, - \vec p, \vec p) P_{\rm lin}(\vec k) \Bigg) P_{\rm lin}(\vec q)P_{\rm lin}(\vec p)\, , \non \\
\vec I_{24}
&=12 \int_{\vec q, \vec p} \vec H_2(\vec k -  \vec q, \vec q ) \otimes \vec H_4( \vec k -  \vec q , \vec q, \vec p, -\vec p) 
P_{\rm lin}( \vec k - \vec q ) P_{\rm lin}( \vec q ) P_{\rm lin}( \vec p )\, , \non\\
\vec I_{15}
&= 15 \int_{\vec q, \vec p} ~ \vec H_1(\vec k) \otimes \vec H_5(\vec k, \vec q, - \vec q, \vec p, -\vec p) P_{\rm lin}(\vec k) P_{\rm lin}( \vec q ) P_{\rm lin}( \vec p )\, . \non
}
and one may verify that $\vec I_{33}^T=\vec I_{33}$, 
$\vec I_{24}= \vec I_{42}^T$ and $\vec I_{15} (k) = \vec I_{51}^T (k)$.

In the two-loop calculation, there are two integration variables $\vec q$ and $\vec p$. UV and IR divergences may result from integrating in both these variables.
After having identified such divergent contributions, our goal is to subtract them from the integrands and implement the cancellation explicitly, in full analogy with 
the one-loop case above. 
The procedure in the two-loop case is somewhat more involved: besides the leading divergencies (when specific limits of both $\vec q$ and $\vec p$ produce a divergent contribution), 
one can have sub-leading terms associated with a specific limit of only one of the two variables, while the contribution from the other stays finite. We provide more details on the UV and IR properties of the two-loop result in Appendix \ref{app:two-loops}, and we briefly summarise some of the key properties below.

Similarly to the one-loop case, the IR contribution extracted from 
the individual two-loop terms ought to cancel as a consequence of the equivalence  
principle and consistency relations (see \cite{Fujita+:2020} 
e.g. for a recent explicit treatment).

Hence one may verify the following cancellation 
\eq{
\vec \lambda_3 \cdot \lb  [\vec I_{33,{\rm I}}]_{{\rm IR}} 
+ [\vec I_{33,{\rm II}}]_{{\rm IR}} \rb \cdot \vec \lambda_3
+ 2  \vec \lambda_2 \cdot [\vec I_{24}]_{\rm IR} \cdot \vec \lambda_4
+ 2  \vec \lambda_1 \cdot [\vec I_{15}]_{\rm IR} \cdot \vec \lambda_5= 0, 
}
and similarly for the cross- and auto- correlations including $\kappa_n$ coefficients. Plugging the expressions derived in Appendix \ref{app:two-loops}, we have
\eq{
\big( \vec \lambda_3 \cdot \vec h^{\rm IR}_{33,{\rm I}} \cdot \vec  \lambda_3 
+ 2  \vec \lambda_2 \cdot \vec h^{\rm IR}_{24,{\rm II}} \cdot \vec \lambda_4 
+ 2  \vec \lambda_1 \cdot \vec h^{\rm IR}_{15} \cdot \vec \lambda_5 \big) W^{\rm IR}_{\rm I} k^2 P_{\rm lin} (\vec k) &= 0\, , \\
\big( \vec \lambda_3 \cdot \vec h^{\rm IR}_{33,{\rm II}} \cdot \vec \lambda_3
+ 2  \vec \lambda_2 \cdot \vec h^{\rm IR}_{24,{\rm I}} \cdot \vec \lambda_4 \big) W^{\rm IR}_{\rm II} k^2 &= 0\, , \non 
}
where we used that $W^{\rm IR}_{33,{\rm I}} = W^{\rm IR}_{24, {\rm II}} = W^{\rm IR}_{15} = W^{\rm IR}_{\rm I}$
and $W^{\rm IR}_{33,{\rm II}} = W^{\rm IR}_{24, {\rm I}} = W^{\rm IR}_{\rm II}$. 
The fact that the cancellation occurs independently in two different terms  was also pointed 
out in \cite{Carrasco++:2013a}. More explicitly, we can write 
\eq{
\vec \lambda_3 \cdot \vec{\hat h}^{(2)}_{33,{\rm I, IR}} \cdot \vec  \lambda_3 
+ 2  \vec \lambda_2 \cdot \vec{\hat h}^{(2)}_{24,{\rm II, IR}} \cdot \vec \lambda_4 
+ 2  \vec \lambda_1 \cdot \vec{\hat h}^{(2)}_{15,{\rm IR}} \cdot \vec \lambda_5 &= 0\, , \\
\vec \lambda_3 \cdot \vec{\hat h}^{(2)}_{33,{\rm II, IR}} \cdot \vec \lambda_3
+ 2  \vec \lambda_2 \cdot \vec{\hat h}^{(2)}_{24,{\rm I, IR}}  \cdot \vec \lambda_4 &= 0\, . \non 
}
In addition, IR and UV cancellations take place as a consequence of 
the mass and momentum conservation. Thus, in addition to the condition
$\vec \lambda_3 \cdot [\vec H_{3,{\rm I}}]^{(0)}_{\rm UV} = \vec \kappa_3 \cdot [\vec H_{3,{\rm I}}]^{(0)}_{\rm UV} =0$,
one may also verify how the contributions from $\vec h^{(0)}_{15}$ vanish after contractions 
with the $\vec \lambda_5$ and $\vec \kappa_5$ coefficients. 

Using such properties in the IR and UV regimes as well as the appropriately defined window functions $W^{\rm IR}$ and $W^{\rm UV}$, one may introduce the regularised integrals $\vec {\tilde I}_{33,{\rm I}}$,
$\vec {\tilde I}_{33,{\rm II}}$, $\vec {\tilde I}_{24}$ and  $\vec {\tilde I}_{15}$, which are shown explicitly in Appendix \ref{app:two-loops}. 
These steps mirror the procedure we employed in the one-loop case and motivate our
introducing the regularised two-loop expressions,
\eq{
P^\text{2-loop}_{\delta\delta}(k) 
&= \lb \vec \lambda_3 \cdot \vec {\mathcal I}_{33} \cdot \vec \lambda_3 \rb 
+ 2 \lb \vec \lambda_2 \cdot \vec {\mathcal I}_{24} \cdot \vec \lambda_4 \rb
+ 2 \lb \vec \lambda_1 \cdot \vec {\mathcal I}_{15} \cdot \vec \lambda_5 \rb \, ,
}
and similarly for $P^\text{2-loop}_{\delta\theta}$ and $P^\text{2-loop}_{\theta\theta}$.
The regularised integral functions $\vec {\mathcal I}_{ij}$, including the  sub-leading UV contributions that are isolated and subsequently computed, read
\eq{
\vec {\mathcal I}_{33} &=  \vec {\tilde I}_{33,{\rm I}} + \vec {\tilde I}_{33,{\rm II}} 
+ \lb \vec h^{\rm UV}_{33,{\rm I}} W^{\rm UV}_{33,{\rm I}} \rb k^4 P_{\rm lin}(k)
+ \lb  \vec h^{\rm UV}_{33,{\rm II}} W^{\rm UV}_{33,{\rm II}} \rb k^4 \, , \\
\vec {\mathcal I}_{24} &= \vec {\tilde I}_{24} + \lb \vec h^{\rm UV}_{24} W^{\rm UV}_{24} \rb  k^4 \, ,\non\\
\vec {\mathcal I}_{15} &= \vec {\tilde I}_{15} + \lb \vec h^{\rm UV}_{15} W^{\rm UV}_{15} \rb  k^2 P_{\rm lin} \, , \non
}
where all the terms are written explicitly in Appendix \ref{app:two-loops}.

\begin{figure}[t!]
\centering
\includegraphics[width=0.9\linewidth]{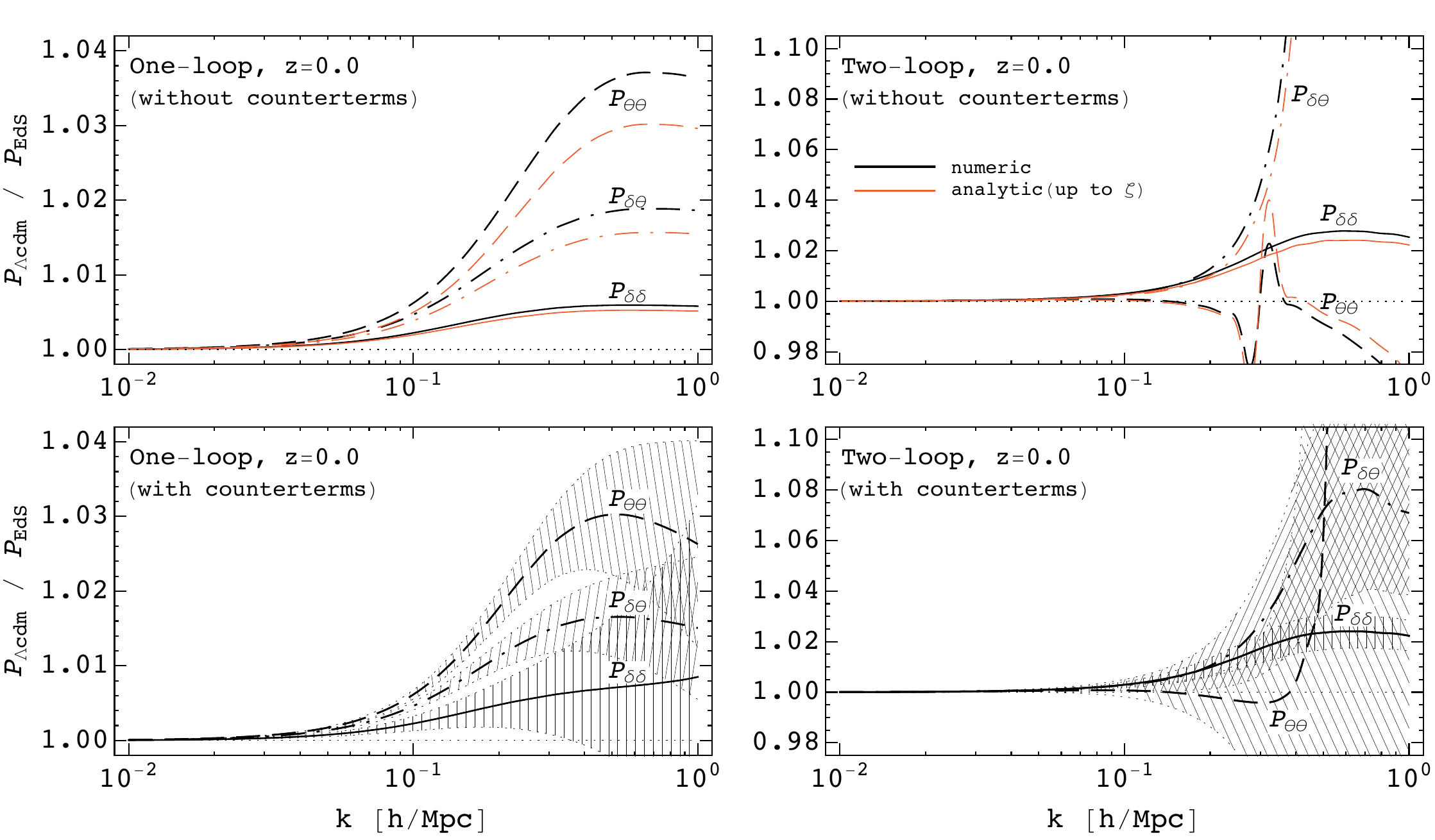}
\caption{
The ratio of the $\Lambda$CDM and the EdS density-density (solid lines), 
density-velocity (dot-dashed lines), velocity-velocity (dashed lines) 
power spectrum at the redshift $z=0.0$. \textit{Upper panels} show the ratios of these spectra without any counterterms.  Black lines denote spectra computed by numerically evaluating the $\lambda_n$ and $\kappa_n$ coefficients, while the orange lines show the perturbatively evaluated coefficients up to the linear corrections in $\zeta$. Adding the corrections up to order $\zeta^3$ would superimpose the perturbative results onto the numerical ones (black lines).  \textit{Lower panels}
show the ratio of the power spectra when including the leading EFT counterterms ($\sim k^2P_{\rm lin}$), the latter having been chosen to roughly match the realistic values ($c_{\delta\delta}^2\approx 3.0$, $c_{\delta\theta}^2\approx -1.0$,
 $c_{\theta\theta}^2\approx -1.5$). Grey hashed bands centred on each of the lines indicate the $5\%$ variation in the values for the counterterms.
}

\label{fig:power_sepctra_II}
\end{figure}

In the right panels of Figure~\ref{fig:power_sepctra_I}, we show the two-loop contributions for
the three power spectra $P_{\delta\delta}$, $P_{\delta\theta}$ and  $P_{\theta\theta}$. The two-loop $\Lambda$CDM corrections 
are typically one to two orders of magnitude smaller than the EdS contributions. However, they can in fact dominate in the regimes where EdS contributions have zero crossing. 

Figure~\ref{fig:power_sepctra_II} shows the ratio of the three total power spectra in $\Lambda$CDM relative to the EdS results. In the right panels, we plot our two-loop results.  The upper right panel displays the two-loop spectra without the effect of counterterms either in the EdS or $\Lambda$CDM case. The deviations range from a few percent  (for $\delta P_{\delta\delta}$) to a dozen percents (for e.g. $P_{\delta\theta}$) 
at scales $k\gtrsim0.4h/{\rm Mpc}$.\footnote{These results also agree with the numerical results obtained in the reference \cite{Garny:2022}.} In the bottom panel of the same figure we include the EFT counterterms for both $P_{\rm EdS}$ and $P_{\Lambda {\rm cdm}}$.  For two-loops, this is done so that the additional counterterms only cancel 
the $k^2P_{\rm lin}$ contributions, and so we effectively only have counterterms that are already present at one-loop order. \newline

These are shown as the central lines within the grey bands, while the bands themselves represent the effects of variations of the $\Lambda$CDM counterterms by 5\%. One can see how the addition of  counterterms can significantly change the relative differences between the  $\Lambda$CDM and EdS results, reducing it to below ten percent on most of the scales of interest. We stress the high sensitivity of these lines on the values of the counterterms. This is especially so for  $P_{\theta\theta}$, in which case the values of the counterterms affect the zero crossing of the total two-loop power spectrum prediction. 

As in the case of one-loop results, we also compare our two-loop numerical results (i.e. those obtained using the numerical values for $\lambda_n$ and $\kappa_n$) to their perturbative counterpart. These are shown both in Figure~\ref{fig:power_sepctra_I}
and in Figure~\ref{fig:power_sepctra_II} as orange lines. We see that adding the linear corrections in the $\zeta$ parameter reaches roughly 
a few percent agreements with the full numerical results, while adding corrections up to $\zeta^3$ renders the results essentially indistinguishable from the numerical findings. 

 Before closing this section, a few comments on computational methods are in order.
Our proposed method for dealing with the cancellation of these divergences differs from the one suggested in 
\cite{Carrasco++:2013a} in that it explicitly subtracts the contributions that are cancelled 
at the level of the integrand(s). In the case of EdS, the difference between the two recipes is not particularly noteworthy: the cancellations, encoded in the  analytic coefficients in the $F_n$ and $G_n$ kernels, can be implemented with machine-level accuracy since they amount to subtractions of simple fractions. 
In $\Lambda$CDM the time coefficients, when obtained numerically, are computed with finite 
accuracy which can generate some spurious remainders in the cancellations. These can spoil the accuracy when evaluating the various integrals. For this reason, it is quite useful to implement subtraction and cancellation of the key contributions analytically. 
We also note that in our perturbative treatment of the coefficients $\lambda_n$ and $\kappa_n$ given in Eqs.~\eqref{eq:lamdakappa2} and in Appendix~\ref{sec: perturbation of WU} 
these cancellations are enforced at each $\zeta$ order. Since the corresponding pre-factors of $\zeta$ powers are also given as fractions, the implicit method proposed in \cite{Carrasco++:2013a} 
may also be applied at each $\zeta$ order.
 We are able to confirm that the power spectra with the perturbative coefficients computed in this manner reproduce the same results as 
 our proposed method, thus providing yet another consistency check for our treatment of the regularised integral functions.

\section{Discussion and Conclusions} 
\label{sec:conclusion}

\noi Perturbative approaches to structure formation allow us to develop controlled 
analytical predictions on the physics of mildly non-linear scales. Although limited in its reach to large scales, the perturbative scheme provides a clean and systematic treatment of LSS dynamics. 
In particular, it is the ideal framework to highlight the role of symmetries and related properties, such as the equivalence principle, mass and momentum conservation (see \cite{Fujita+:2020, DAmico++:2021}), 
in the construction of cosmological correlations, the key observables in large scale structure. 

In this work, we develop exact, separable solutions for PT kernels of density 
and velocity fields in $\Lambda$CDM cosmology. 
So far, such explicit solutions have been obtained only within the EdS approximation, with the extensions to $\Lambda$CDM  worked out analytically only to lower orders.
In this work, we presented a recursive solution valid up to arbitrary order in perturbation theory, 
providing in particular an algorithm to obtain separable solutions for the $F_n$ and $G_n$ kernels at each perturbative order $n$. 

The solutions building blocks are elements of the basis of the momentum dependent operators $\vec H_n$. The (upper limit on the) dimension of the basis depends on the perturbative order $n$ and is given by the number $N(n)$, for which we also provide the explicit recursive expression. To obtain the full kernels $F_n$ and $G_n$, this momentum operator basis has to be appropriately ``contracted'' with the time-dependent coefficients for matter and velocity fields, respectively dubbed $\vec \lambda_n$ and $\vec \kappa_n$. 

We arrive at the solutions for the time coefficients following two different paths. First, starting from the implicit integral solutions obtained in \cite{Fasiello+:2016}, we obtain the recursive integral solution  for $\vec \lambda_n$ and $\vec \kappa_n$.
We show that these can be recast as the solutions to a set of coupled differential equations, which we find quite suitable for numerical treatment.  We are then able to compute our numerical benchmark solutions, which we use in the remainder of our analysis. The analysis of the differential equations
makes it clear that the ``clock'', i.e. the time evolution, is set by the combination of growth rates $f_-/f_+^2$, whose time dependence in $\Lambda$CDM cosmology is completely determined by the new variable $\zeta = \Omega_{\Lambda,0}/\Omega_{m,0} D_+$. This implies that using $\zeta$ as the time variable 
in solving for the $\vec \lambda_n$ and $\vec \kappa_n$ coefficients one obtains solutions valid universally in $\Lambda$CDM, that is irrespective of the choice of cosmological parameters. This significantly simplifies the computational task involved in the cosmological parameter search.

As an alternative path to the solution, equipped with the differential equations we develop an analytic  perturbative solution for the $\vec \lambda_n$ and $\vec \kappa_n$ coefficients.
The starting point of this perturbative solution lies in the observation that the EdS approximation, a static solution to the set of our differential equations, is an excellent (yet insufficient) approximation to the full set of the $\vec \lambda_n$ and $\vec \kappa_n$ 
coefficients, especially at lower orders. This suggests that we organise the perturbative treatment around the parameter $\delta f = 3/2 + f_-/f_+^2$. We have obtained the general perturbative solution, laying the basis for an iterative path to the time coefficients. We then applied our algorithm to derive  solutions up to the leading correction in $\delta f$, and implemented our procedure all the way to the $\vec \lambda_5$ and $\vec \kappa_5$  coefficients needed for the two-loop power spectrum calculations.

The final form of these perturbative solutions is given  in terms of the variable $\zeta$ so as to fully capture the dependence on cosmological parameters. We investigated the agreement of our perturbative solutions with the full numerical evaluation and found a perfect agreement 
for all coefficients up to $n=5$ if terms up to third order in $\zeta$ are included. We also note that the perturbative solutions exhibit, at each order in $\zeta$, similar behaviour as the EdS solutions when it comes to IR cancellations and properties that stem from mass and momentum conservation. This makes them particularly well suited  for use in the numerical evaluation of higher loop power spectra that rely on accurate  cancellations in their integrands. Let us also mention that these findings may be generalised to beyond-$\Lambda$CDM scenarios, something we will address in future work.

\bigskip

As an application of our results, we compute one- and two-loop 
matter and velocity auto- and cross-power spectra.  
We compare our solutions and explore the differences 
between the EdS and $\Lambda$CDM solutions. The results 
are summarized in Figures~\ref{fig:power_sepctra_I} and
\ref{fig:power_sepctra_II}. We find that some care has to be exerted in quantifying these differences since a fraction of the effect  in the one- and two-loop contributions can be re-absorbed in 
the EFT counterterms, which can be treated as free coefficients of the perturbative loop expansion. Specifically, in the one-loop case, the difference in the density-density power spectrum 
between $\Lambda$CDM and EdS  
can be fully reabsorbed by the counterterms 
at the scales and accuracy of interest. 
For the velocity statistics, the deviation is instead more pronounced, at the level of a few percent, even when fully engaging counterterms.
At two-loop the $\Lambda$CDM deviation from the EdS result increases to a couple of percent, which, to a large extent, can again be covered by counterterms. 
In terms of the velocity related statistics, the results depend rather heavily on the 
numerical values of the required counterterms, with the latter requiring calibration against N-body simulations, something that goes beyond the scope of our analysis. Nevertheless, our work clearly shows how the deviation could reach ten percent at the scales of interest, as demonstrated in Figure~\ref{fig:power_sepctra_II}. 

\bigskip

Beyond their use in higher loop calculations, our results are also readily applicable in tackling higher $n$-point functions, such as the bispectrum, the trispectrum, etc. Remarkably, these observables  are sensitive to the $\Lambda$CDM deviations from the EdS approximation already at tree-level (see \cite{Sefusatti+:2011} and also \cite{Steele+:2020, Steele+:2021} for the recent investigations).

Lastly, we ought to comment on the fact that the dark matter density and velocity are not directly observable but act as the building components within the more general framework of the biased  tracers of large scale structure (see, e.g. \cite{Desjacques++:2016}).  
Given the additional dynamics (and degeneracies) associated with the presence of the bias coefficients, one ought to take into account what survives of the discrepancies such as the one between the EdS and $\Lambda$CDM solutions at the level of the observables. This has been recently explored 
in \cite{Donath+:2020} at one loop order. It would be quite interesting to do the corresponding analysis at two loops. This will be possible upon deriving the two-loop results for biased tracers in redshift space. This is yet another line of investigation we plan to pursue in the near future.
It is possible that deviations like the ones studied here might well bias our
parameter inference and would thus need to be included in the budget of possible theoretical systematic errors. It is
important to keep this budget to a minimum given that parameter tensions of several sigmas are nowadays a familiar occurrence in data-driven cosmology.

\begin{acknowledgments}
M.F. would like to acknowledge support from the “Atracción de Talento” grant 2019-T1/TIC15784, his work is partially supported by the Spanish Research Agency (Agencia
Estatal de Investigación) through the Grant IFT Centro de Excelencia Severo Ochoa
No CEX2020-001007-S, funded by MCIN/AEI/10.13039/501100011033.
T.F. acknowledges the support by the Grant-in-Aid for Scientific Research Fund of the JSPS~18K13537 and 20H05854.
Z.V. acknowledges the support of the Kavli Foundation.
\end{acknowledgments}

\appendix

\section{Linear growth and decay factors and rates}
\label{sec:linear_growth}

In this appendix, we summarise the linear growth solutions and derive some of the results in a form useful for our subsequent computation. In particular, we derive a specific form in which we can expand the linear growth rate combination $f_-/f_+^2$ around the EdS value. 
 
We start from the well-known solutions for $D_\pm$, these are
\eeq{
\label{eq:D+D- solutions}
D_+=\frac{5}{2}H_0^2 \Omega_{\rm m0} H(a)\int^a_0
\frac{{\rm d} \tilde{a}}{\tilde{a}^3 H^3(\tilde{a})} \, ,
\qquad
D_- = \frac{H(a)}{H_0} \, ,
}
where $\Omega_{\rm m0}$ and $\Omega_{\Lambda 0}$ are the current energy fractions 
of dark matter and the cosmological constant, respectively. The Hubble parameter is given by
$H(a) = H_0 \sqrt{\Omega_{m0} a^{-3} + \Omega_{\Lambda 0}}$, where the radiation component can be ignored.
Changing the time variable into $q\equiv a^{3}\Omega_{\Lambda 0}/\Omega_{m0}$
and performing the integral in $D_+$, we find
\begin{align}
\label{eq:D+ solution}
    \hat{D}_+&\equiv \left(\frac{\Omega_{\Lambda0}}{\Omega_{m0}}\right)^{\frac{1}{3}}D_+
    =q^{\frac{1}{3}}\, {}_2F_1\left(1,\frac{1}{3},\frac{11}{6},-q\right)\, ,
\end{align}
where we introduced $\hat{D}_+$, which depends only on $q$, and ${}_2F_1(a,b,c,z)$ is the hypergeometric function. Using the above equations, we can compute 
$\delta f = f_{-}/f^2_{+}+3/2 = (\dd\ln H/\dd\ln a)(D_+/a)^2 (\dd D_+/\dd a)^{-2}+3/2$ 
obtaining 
\eeq{
\delta f =\frac{3}{2}\left[1-4(1+q)\left(\frac{5}{{}_2F_1(1,1/3,11/6,-q)}-3\right)^{-2} \right] \, .
}
This expression is useful for the purpose of numerical treatments. However, we choose to further reduce $\hat{D}_+$ and $\delta f$ to obtain simplified expressions.
Expanding $\hat{D}_+$ around $q=0$, one finds
\eeq{
    \hat{D}_+\simeq q^{\frac{1}{3}}\left[
    1-\epsilon \frac{2}{11}q+\epsilon^2 \frac{16}{187}q^2-\epsilon^3 \frac{224}{4301}q^3+\ldots
    \right] \, ,
}
where we inserted a bookkeeping parameter $\epsilon$.
Inverting the above expression by plugging an ansatz, 
$q=Q_0+\epsilon Q_1+\epsilon^2 Q_2+\epsilon^3 Q_3+\cdots$,
and solving for $Q_n$ at each order in $\epsilon$, we obtain
\eeq{
    q\simeq \,\hat{D}_+^{3}+\epsilon\frac{6}{11}\hat{D}_+^{6}
    +\epsilon^2 \frac{492}{2057}\hat{D}_+^{9}+\epsilon^3\frac{50216}{520421}\hat{D}_+^{12}+\ldots \, .
    \label{perturbative q}
}
We can also expand $\delta f$ around $q=0$ to obtain
\eeq{
    \delta f \simeq -\frac{3}{22}q
    +\epsilon\frac{15}{374}q^2
    -\epsilon^2\frac{21585}{1040842}q^3
    +\epsilon^3 \frac{74212575}{5644486166}q^4+\cdots \, .
}
By substituting Eq.~\eqref{perturbative q} into this equation, one finds an analytic expression for $\delta f$ in powers of $D_+$,
\eeq{
\label{eq:df_taylor}
    \delta f \simeq c_1 \lb \frac{\Omega_{\Lambda 0}}{\Omega_{m0}} \rb D_+^3 
    +c_2 \lb \frac{\Omega_{\Lambda 0}}{\Omega_{m0}} \rb^2 D_+^6 
    + c_3 \lb \frac{\Omega_{\Lambda 0}}{\Omega_{m0}} \rb^3 D_+^9 
    + c_4 \lb \frac{\Omega_{\Lambda 0}}{\Omega_{m0}} \rb^4 D_+^{12} +\ldots,
}
with $c_1, c_2, c_3$ given below Eq.~\eqref{df ansatz} and $c_4=-\frac{15954399}{5644486166}$.
Here we omit the higher order correction $\mathcal{O}(D_+^{15})$ and set $\epsilon=1$. 
We find that $c_4$ and the higher order terms obtained in this manner do not significantly improve the fit to $\delta f$, as can also be deduced from Figure~\ref{fig:fmfp2}. 
In this work it will therefore suffice to truncate the expansion so as to include  the $c_1, c_2$ and $c_3$ coefficients. Nevertheless, we note that the extension to higher orders is straightforward within the formalism developed in this work.

\section{Direct integral solutions of the dark matter kernels}
\label{sec:direct integral solution}

In Section \ref{sec:eom} we have presented the integral solutions for the dark matter kernels.
Here we return to these findings providing further  details for each component of the solution. 
We start from the EoMs and briefly review the solutions obtained in \cite{Fasiello+:2016}.
With the symmetrized kernels $F_n(\vq_1,..,\vq_n, \eta)$ and $G_n(\vq_1,..,\vq_n, \eta)$ 
in Eq.~\eqref{eq: sym kernels}, the EoMs given in Eq.~\eqref{eq:eom_v1} read
\eq{
\label{eq:FsGs}
\dot{F}_{n} + n\,{F}_{n} - {G}_{n} 
&= h^{(n)}_\alpha (\vq_1,..,\vq_n, \eta)\, , \\ 
\dot{G}_{n} + (n-1)\,{G}_{n} 
-\frac{f_{-}}{f^2_{+}}\lb{G}_{n} - {F}_{n} \rb
&= h^{(n)}_\beta (\vq_1,..,\vq_n, \eta)\, , \non
}
where we used the shorthand notation $\dot{ } = \partial/\partial \eta$ (remember that $\eta \equiv \ln D_+$). 
The functions $f_{+,-}$ are defined as $f_{+,-}\equiv{\rm d}\ln D_{+,-}/{\rm d}\ln a$.
The source terms are given by:
\eq{
\label{eq:hahb}
h^{(n)}_\alpha(\vq_1,..,\vq_n, \eta) 
& =  \sum_{\pi- {\rm all}} \sum^{n-1}_{m=1}\alpha({\vp_m,\vp_{n-m}})
G_m(\vq_1,..,\vq_m,\eta) F_{n-m}(\vq_{m+1},..,\vq_n,\eta) \\
&= \sum^{n-1}_{m=1}\frac{m!(n-m)!}{n!} \sum_{\pi-{\rm cross}}\,\alpha({\vp_m},{\vp_{n-m}}) 
G_m F_{n-m} \non \\
&=  \df^K_{\frac{n}{2} , \lfloor \frac{n}{2} \rfloor} \frac{(n/2!)^2}{ n!}
\sum_{\pi-{\rm cross}}\, \alpha({\vp_{n/2}},{\vp_{n/2}}) 
G_{n/2} F_{n/2} \non\\
&~~~~ +
\sum^{\left \lfloor (n-1)/2 \right \rfloor }_{m=1}\frac{m! (n-m)!}{n!} 
 \sum_{\pi-{\rm cross}}\,\Big[ \alpha({\vp_m},{\vp_{n-m}}) G_m F_{n-m}
 +\alpha({\vp_{n-m}},{\vp_m}) G_{n-m} F_m \Big]\; ,  \non\\
h^{(n)}_\beta(\vq_1,..,\vq_n, \eta) 
&= \sum_{\pi- {\rm all}} \sum^{n-1}_{m=1}\beta({\vp_m},{\vp_{m-n}})
G_m(\vq_1,..,\vq_m,\eta)  G_{n-m}(\vq_{m+1},..,\vq_n,\eta)\, , \non\\
&=
\df^K_{\frac{n}{2} , \lfloor \frac{n}{2} \rfloor} \frac{(n/2!)^2}{ n!}
\sum_{\pi-{\rm cross}}\,\beta({\vp_{n/2},\vp_{n/2}}) \, G_{n/2} G_{n/2} \non\\
&~~~~ +
2 \sum^{\left \lfloor (n-1)/2 \right \rfloor }_{m=1}\frac{m! (n-m)!}{n!}
\sum_{\pi-{\rm cross}}\,\beta({\vp_m,\vp_{n-m}}) \, G_m G_{n-m}\, . \non
}
where the subscript ``${\pi{\rm-all}}$'' stands for symmetrization over 
all momenta $\{{\bf q}_1\dots {\bf q}_n\}$ while ${\pi-{\rm cross}} $
indicates  permutations that exchange the momenta in the $\{1 \dots m \}$ 
set with those in the $\{m+1 \dots n\}$ set. In the last line of Eq.~\eqref{eq:hahb} 
the double counting for the case when $n$ is even has been removed. 
The following definitions have also been employed: 
$\vp_m=\vq_1+..+\vq_m$; $\vp_{n-m}=\vq_{m+1}+..+\vq_n$. 
Combining the two expressions in Eq.~\eqref{eq:FsGs} one readily obtains 
the EoM for the first kernel $F_n$:
\eeq{
\ddot{F}_{n}+ \dot{F}_{n}\lb2n-1 -\frac{f_{-}}{f_{+}^2} \rb
+(n-1)F_{n}\lb n-\frac{f_{-}}{f_{+}^2}\rb = 
h^{(n)}_{\beta}+
\lb n-1 -\frac{f_{-}}{f_{+}^2}\right)h^{(n)}_{\alpha}  +\dot{h}^{(n)}_{\alpha}\, , 
}
whose solution reads:
\eeq{
\label{Fnin}
F_{n}(\eta)=\int_{-\infty}^{\eta} \dd\tilde{\eta} \, 
 e^{(n-1)(\tilde{\eta}-\eta)}\frac{\tilde{f}_{+}}{\tilde{f}_{+}-\tilde{f}_{-}} 
\Bigg[\left(\tilde{h}^{(n)}_{\beta}-\frac{\tilde{f}_{-}}{\tilde{f}_{+}}\tilde{h}^{(n)}_{\alpha}  \right)
+e^{\tilde{\eta}-\eta}\,\frac{D_{-}(\eta)}{D_{-}(\tilde\eta)} 
\left(\tilde{h}^{(n)}_{\alpha} -\tilde{h}^{(n)}_{\beta}\right)\Bigg]\, ,
}
where in deriving the above we have used Eq.~\eqref{eq:Dpm EoM}
for the growing and decaying solutions for the linear growth factor $D_\pm(\eta)$. 
Using again the first expression in Eq.~\eqref{eq:FsGs} one  
arrives at the solution for the $G$ kernels:
\eeq{
\label{Gnin}
G_{n}(\eta)=\int_{-\infty}^{\eta} \dd\tilde{\eta} \, 
e^{(n-1)(\tilde{\eta}-\eta)}\frac{\tilde{f}_{+}}{\tilde{f}_{+}-\tilde{f}_{-}}
\Bigg[\left(\tilde{h}^{(n)}_{\beta}-\frac{\tilde{f}_{-}}{\tilde{f}_{+}}\tilde{h}^{(n)}_{\alpha}  \right)
+e^{\tilde{\eta}-\eta}\frac{f_{-}}{f_{+}}\frac{D_{-}(\eta)}{D_{-}(\tilde{\eta})} 
\left(\tilde{h}^{(n)}_{\alpha} -\tilde{h}^{(n)}_{\beta}\right)\Bigg]\, ,
}
where again a function with tilde depends not on the variable 
$a$ but the variable $\tilde a$ (e.g. $\tilde{D}_+\equiv D_+(\tilde a)$). Note that the time-dependent coefficients of $\tilde{h}_\alpha^{(n)}$ and $\tilde{h}_\beta^{(n)}$ in the integrands of Eqs. (\ref{Fnin}) and (\ref{Gnin}), 
also require integration, thus implying recursive 
 time integrals, something that is far from ideal for a fast evaluation.
Eqs. (\ref{Fnin}) and (\ref{Gnin}) are the integral solutions, 
to all orders, as first derived in \cite{Fasiello+:2016}.
Changing the time variable in favour of the scaling factor $a$, we can express these solutions in the form 
\eq{
F_{n}(\vq_1,..,\vq_n,a)
&=\int_{0}^{a}   \frac{\dd\tilde{a}}{\tilde a} 
 \Big( w_\alpha^{(n)}(a, \tilde a) h^{(n)}_{\alpha} (\vq_1,..,\vq_n,\tilde a) 
 + w_\beta^{(n)}(a, \tilde a) h^{(n)}_{\beta}  (\vq_1,..,\vq_n,\tilde a) \Big)\, ,
\\
G_{n}(\vq_1,..,\vq_n,a)
&=\int_{0}^{a}   \frac{\dd\tilde{a}}{\tilde a} 
 \Big( u_\alpha^{(n)}(a, \tilde a) h^{(n)}_{\alpha} (\vq_1,..,\vq_n,\tilde a) 
 + u_\beta^{(n)}(a, \tilde a) h^{(n)}_{\beta}  (\vq_1,..,\vq_n,\tilde a) \Big)\, ,\non
}
as was presented in the Eq.~\eqref{Fn-tdep}. The Green's function
components are given by 
\eq{
\label{omegaabn}
w^{(n)}_\alpha(a, \tilde a)  &= w^{(n)}(a, \tilde a)   \Big( 1-  \delta(\tilde a)  d (a, \tilde a)  \Big)\, , \qquad
w^{(n)}_\beta(a, \tilde a)  = - w^{(n)}(a, \tilde a)   \Big( 1 -  d (a, \tilde a)  \Big)\, ,  \\ 
u^{(n)}_\alpha(a, \tilde a)  &= w^{(n)}(a, \tilde a)  \Big( \delta(a) - \delta(\tilde a) d (a, \tilde a)  \Big)\, , \qquad
u^{(n)}_\beta(a, \tilde a)  = - w^{(n)}(a, \tilde a)   \Big( \delta(a) - d(a, \tilde a)   \Big)\, , \non
 }
where we introduced the quantities
\eeq{
w^{(n)} (a, \tilde a) = \lb \frac{\tilde D_+}{D_+} \rb^{n} \frac{\tilde{f}^2_{+}}{\tilde{f}_{+}-\tilde{f}_{-}} \frac{D_{-}}{\tilde D_{-}}\, ,
~~~ 
d (a, \tilde a) =  \frac{ \tilde D_{-}}{D_{-}}  \frac{D_+}{\tilde D_+}\, , 
~~~
\delta (a) =   \frac{f_{-}}{f_{+}}\, . 
}
Moreover, we find that $w^{(n)}_{\alpha,\beta}$ and $u^{(n)}_{\alpha,\beta}$ satisfy a simple relation,
\eq{\label{eq:wu relation}
w_\alpha^{(n)}(a, \tilde a)+w_\beta^{(n)}(a, \tilde a)
=
u_\alpha^{(n)}(a, \tilde a)+u_\beta^{(n)}(a, \tilde a)
=  w^{(n)} d (1- \tilde \delta)
= \left(\frac{\tilde{D}_+}{D_+}\right)^{n-1}\tilde{f}_+\; .
}
Eq.~\eqref{eq:wu relation} indicates that not all of the Green's functions are independent: there are relations between them that can be obtained at each order $n$.  
We see how these come to play when computing the one- and two- 
loop power spectra in Section \ref{sec:PT_coeffs}.

\section{Derivation of the separable kernels}
\label{sec:derivtion of kernels}

In Appendix \ref{sec:direct integral solution} 
we have shown the explicit integral form of the solutions for the $F_n$ 
and $G_n$ kernels. In this appendix, we recast such solutions 
into separable form $F_n = \boldsymbol{\lambda}_n \cdot \vec H_n$ and 
$G_n = \boldsymbol{\kappa}_n \cdot \vec H_n$, as indicated in Section \ref{sec:eom}.
Plugging Eqs.~\eqref{eq:hahb} and \eqref{eq:FGn} into the right hand side 
of Eq.~\eqref{Fn-tdep}, and separately re-organizing the momentum-dependent 
and time-dependent terms, one finds
\begingroup
\allowdisplaybreaks
\eq{
F_{n}(a)
&=  \df^K_{\frac{n}{2} , \lfloor \frac{n}{2} \rfloor } 
\sum_{i=1}^{N(n/2)} \sum_{j=1}^{N(n/2)} W^{(ij)}_{\alpha; n/2,n/2} [h_{\alpha}]_{n/2, n/2}^{(ij)} 
+ \df^K_{\frac{n}{2} , \lfloor \frac{n}{2} \rfloor } 
\sum_{i=1}^{N(n/2)} \sum_{j=i}^{N(n/2)} \Big[ 2  -  \df^K_{ij} \Big] W^{(ij)}_{\beta; n/2,n/2} [h_{\beta}]_{n/2,n/2}^{(ij)} 
\label{eq:FG_sol_3sums}\\
&~~ + \sum^{\left \lfloor (n-1)/2 \right \rfloor }_{m=1}
 \sum_{i=1}^{N(m)} \sum_{j=1}^{N(n-m)}\Big( 
W^{(ij)}_{\alpha; m,n-m} [h_{\alpha}]_{m,n-m}^{(ij)} 
+ W^{(ji)}_{\alpha; n-m,m} [h_{\alpha}]_{n-m, m}^{(ji)} 
+ 2 W^{(ij)}_{\beta; m,n-m} [h_{\beta}]_{m,n-m}^{(ij)}  \Big)\, ,  \non \\
G_{n}(a)
&=  \df^K_{\frac{n}{2} , \lfloor \frac{n}{2} \rfloor } 
\sum_{i=1}^{N(n/2)} \sum_{j=1}^{N(n/2)} U^{(ij)}_{\alpha; n/2,n/2} [h_{\alpha}]_{n/2, n/2}^{(ij)} 
+ \df^K_{\frac{n}{2} , \lfloor \frac{n}{2} \rfloor } 
\sum_{i=1}^{N(n/2)} \sum_{j=i}^{N(n/2)} \Big[ 2  -  \df^K_{ij} \Big] U^{(ij)}_{\beta; n/2,n/2} [h_{\beta}]_{n/2,n/2}^{(ij)} 
 \non \\
&~~ + \sum^{\left \lfloor (n-1)/2 \right \rfloor }_{m=1}
 \sum_{i=1}^{N(m)} \sum_{j=1}^{N(n-m)}\Big( 
U^{(ij)}_{\alpha; m,n-m} [h_{\alpha}]_{m,n-m}^{(ij)} 
+ U^{(ji)}_{\alpha; n-m,m} [h_{\alpha}]_{n-m, m}^{(ji)} 
+ 2 U^{(ij)}_{\beta; m,n-m} [h_{\beta}]_{m,n-m}^{(ij)}  \Big)\, , \non
}
\endgroup
with the momentum basis source terms $[h_\alpha]$ and $[h_\beta]$ defined as
\eq{
\label{eq:[h] def_II}
[h_{\alpha}]_{m,n-m}^{(ij)} (\vq_1,..,\vq_n) &=  \frac{m!(n-m)!}{n!}  \sum_{\pi- {\rm cross}}  \alpha({\vq_m,\vq_{n-m}})
 H_m^{(i)} (\vec q_1, .., \vec q_m) H_{n-m}^{(j)} (\vec q_{m+1}, .., \vec q_n)\, ,  \\
[h_{\beta}]_{m,n-m}^{(ij)} (\vq_1,..,\vq_n) &= \frac{m!(n-m)!}{n!}  \sum_{\pi- {\rm cross}}  \beta  (\vq_{m},\vq_{n-m}) 
H_m^{(i)} (\vec q_1, .., \vec q_m) H_{n-m}^{(j)} (\vec q_{m+1}, .., \vec q_n)\, .  \non
}
We have also introduced the time-dependent coefficients
\eq{
\label{eq:WU_defs}
W^{n,(ij)}_{\alpha; m_1,m_2}(a)&= \int_{0}^{a} \frac{\dd\tilde{a}}{\tilde a} 
w^{(n)}_\alpha(a, \tilde a) \kappa_{m_1}^{(i)}(\tilde a) \lambda_{m_2}^{(j)}(\tilde a) \, , \qquad
W^{n,(ij)}_{\beta; m_1,m_2}(a) = \int_{0}^{a}  \frac{\dd\tilde{a}}{\tilde a} 
w^{(n)}_\beta(a, \tilde a) \kappa_{m_1}^{(i)}(\tilde a) \kappa_{m_2}^{(j)}(\tilde a) \, , \\
U^{n,(ij)}_{\alpha; m_1,m_2}(a)&= \int_{0}^{a}  \frac{\dd\tilde{a}}{\tilde a} 
u^{(n)}_\alpha(a, \tilde a) \kappa_{m_1}^{(i)}(\tilde a) \lambda_{m_2}^{(j)}(\tilde a) \, , \qquad
U^{n,(ij)}_{\beta; m_1,m_2}(a) = \int_{0}^{a}  \frac{\dd\tilde{a}}{\tilde a} 
u^{(n)}_\beta(a, \tilde a) \kappa_{m_1}^{(i)}(\tilde a) \kappa_{m_2}^{(j)}(\tilde a)\, , \non
}
which are constructed from to the Green's functions introduced in Eq.~\eqref{omegaabn}, 
as well as from the lower order coefficients $\lambda_n$ and $\kappa_n$.

Note that Eqs.~\eqref{eq:FG_sol_3sums} are already in the form required in Eq.~\eqref{eq:FGn}. 
However, there remains to be chosen a counting algorithm that systematically maps (bijectively)
the $[h_{\alpha}]^{(ij)}_{m,n-m}$ and $[h_{\beta}]^{(ij)}_{m,n-m}$ operators to the $H_n^{(\ell)}$ operators. 
The running of the various indices in the $h$ operators will correspond to the running of the index 
$\ell$ in $H_n^{(\ell)}$ according to $\ell=1,2,\ldots,N(n)$ at any given order $n$. 
We shall employ the algorithm based on the following five ``counting'' functions:
\eq{
\label{eq:phiis}
\phi_1(n,i,j) &= N\lb \tfrac{n}{2}\rb (i-1) + j ,\qquad
\phi_2(n,i,j) = \lb N\lb \tfrac{n}{2}\rb \rb^2 - \frac{1}{2} i \lb i - 1 \rb + \phi_1(n,i,j)  ,\\
\phi_3(n,m,i,j) &=  \df^K_{\frac{n}{2} , \lfloor \frac{n}{2} \rfloor } \frac{1}{2} N \lb \tfrac{n}{2}\rb \big( 3 N \lb \tfrac{n}{2}\rb +  1 \big)
 + \sum_{k=1}^{m-1} N(k)N(n-k) + (i-1) N(n-m) + j ,\non\\
\phi_4(n,m,i,j) &= \sum_{k=1}^{\left \lfloor (n-1)/2 \right \rfloor }  N(k)N(n-k) + \phi_3(n,m,i,j) ,\qquad
\phi_5(n,m,i,j) = 2 \sum_{k=1}^{\left \lfloor (n-1)/2 \right \rfloor }  N(k)N(n-k) + \phi_3(n,m,i,j). \non
}
For a given $n$, the $\phi_{1,\dots, 5}$ counters run through all the relevant values 
of the indices $\{m,i,j\}$, eventually covering\footnote{One may choose, for example, 
to first fix a value for $m$ starting with the lowest possible, $m=1$, then do the same 
for the index $i$, and run through the index $j$, again running from the lowest to the 
highest value allowed etc. So long as this is done consistently, no ambiguity arises in 
the process.} all the $N(n)$ entries and mapping all of the $[h_{\alpha}]^{(ij)}_{m,n-m}$ 
and $[h_{\beta}]^{(ij)}_{m,n-m}$ operators to $H_n^{(\ell)}$ operators.
As shown in Eq.~\eqref{eq: H exp}, using this mapping, we obtain a closed 
 expression 
\eq{
H_n^{(\ell)}(\vq_1,..,\vq_n) &= \df^K_{\frac{n}{2} , \lfloor \frac{n}{2} \rfloor } 
\sum_{i=1}^{N(n/2)} \Bigg[ \sum_{j=1}^{N(n/2)} [h_{\alpha}]_{\frac{n}{2},\frac{n}{2}}^{(ij)} \delta^K_{\ell, \phi_1}
+ \sum_{j=i}^{N(n/2)} \Big[ 2  -  \df^K_{ij} \Big] [h_{\beta}]_{\frac{n}{2},\frac{n}{2}}^{(ij)} \delta^K_{\ell, \phi_2} \Bigg]  \\
&\hspace{0.9cm} + \sum^{\left \lfloor (n-1)/2 \right \rfloor }_{m=1} \sum_{i=1}^{N(m)} \sum_{j=1}^{N(n-m)}
\Big( [h_{\alpha}]_{m,n-m}^{(ij)}\delta^K_{\ell, \phi_3}+ [h_{\alpha}]_{n-m, m}^{(ji)} \delta^K_{\ell, \phi_4} 
+ 2 [h_{\beta}]_{m,n-m}^{(ij)} \delta^K_{\ell, \phi_5} \Big) \, , \non
}
with the initial terms $H_2^{(1)}=\alpha$ and $H_2^{(2)}=\beta$, 
and one can systematically compute the higher momentum operators 
using definition of the sourcing terms $[h_{\alpha}]$
and $[h_{\beta}]$
given in Eq.~\eqref{eq:[h] def_II}.

Once this mapping is chosen for the $H_n^{(\ell)}$ operators, it immediately 
fixes the mapping between the $\lambda_n$($\kappa_n$) and 
$W$ ($U$) time coefficients. 
Explicitly we have
\eq{
\label{eq:lambdakappa_II}
\lambda_n^{(\ell)}(a) &= \df^K_{\frac{n}{2} , \lfloor \frac{n}{2} \rfloor }  \sum_{i=1}^{N(n/2)}
\Bigg[ \sum_{j=1}^{N(n/2)}  
W^{(ij)}_{\alpha; n/2,n/2}\delta^K_{\ell, \phi_1}
+ \sum_{j=i}^{N(n/2)}  W^{(ij)}_{\beta; \frac{n}{2},\frac{n}{2}} \delta^K_{\ell, \phi_2} \Bigg]  \\
&\hspace{0.9cm} + \sum^{\left \lfloor (n-1)/2 \right \rfloor }_{m=1}
 \sum_{i=1}^{N(m)} \sum_{j=1}^{N(n-m)}\Big( W^{(ij)}_{\alpha; m,n-m} \delta^K_{\ell, \phi_3}
+ W^{(ji)}_{\alpha; n-m,m} \delta^K_{\ell, \phi_4}
+ W^{(ij)}_{\beta; m,n-m} \delta^K_{\ell, \phi_5} \Big) \, , \non\\
\kappa_n^{(\ell)}(a) &= \df^K_{\frac{n}{2} , \lfloor \frac{n}{2} \rfloor }  \sum_{i=1}^{N(n/2)} 
\Bigg[ \sum_{j=1}^{N(n/2)}  U^{(ij)}_{\alpha; n/2,n/2} \delta^K_{\ell, \phi_1}
+ \sum_{j=i}^{N(n/2)}  U^{(ij)}_{\beta; \frac{n}{2},\frac{n}{2}} \delta^K_{\ell, \phi_2} \Bigg]  \non \\
&\hspace{0.9cm} + \sum^{\left \lfloor (n-1)/2 \right \rfloor }_{m=1}
 \sum_{i=1}^{N(m)} \sum_{j=1}^{N(n-m)}\Big( U^{(ij)}_{\alpha; m,n-m} \delta^K_{\ell, \phi_3}
+ U^{(ji)}_{\alpha; n-m,m} \delta^K_{\ell, \phi_4}
+ U^{(ij)}_{\beta; m,n-m} \delta^K_{\ell, \phi_5} \Big) \, , \non
}
with $W$ and $U$ coefficients given in Eq.~\eqref{eq:WU_defs}.
We have thus achieved the separation of the $F_n$ and $G_n$
kernels in the purely momentum dependent operators $H_n$
and time dependent coefficients $\lambda_n$ and $\kappa_n$. 
However, the time coefficients are still determined by the 
recursive time integrals given in Eq.~\eqref{eq:WU_defs}.
In order to facilitate the evaluation of these coefficients  
we can recast these integral expressions into differential equations.
After some manipulation of our integral solutions, we obtain 
\eq{
\label{eq:WU_diffs_app}
\dot{W}^{\,n\,(ij)}_{\alpha; m_1,m_2} + n {W}^{\,n\,(ij)}_{\alpha; m_1,m_2} - {U}^{\,n\,(ij)}_{\alpha; m_1,m_2} 
& =\kappa^{(i)}_{m_1} \lambda^{(j)}_{m_2}\, , \\
\dot{W}^{\,n\,(ij)}_{\beta; m_1,m_2} + n {W}^{\,n\,(ij)}_{\beta; m_1,m_2} - {U}^{\,n\,(ij)}_{\beta; m_1,m_2} &= 0 \, , \non\\
\dot{U}^{\,n\,(ij)}_{\alpha; m_1,m_2} + (n-1) {U}^{\,n\,(ij)}_{\alpha; m_1,m_2} -\frac{f_{-}}{f_{+}^{2}} 
\Big[  {U}^{\,n\,(ij)}_{\alpha; m_1,m_2} - {W}^{\,n\,(ij)}_{\alpha; m_1,m_2}\Big] &= 0\, , \non\\
\dot{U}^{\,n\,(ij)}_{\beta; m_1,m_2} + (n-1) {U}^{\,n\,(ij)}_{\beta; m_1,m_2} -\frac{f_{-}}{f_{+}^{2}} 
\Big[  {U}^{\,n\,(ij)}_{\beta; m_1,m_2} - {W}^{\,n\,(ij)}_{\beta; m_1,m_2}\Big] &= \kappa^{(i)}_{m_1} \kappa^{(j)}_{m_2}\, , \non
}
which we have also presented in Eq.~\eqref{tk1}. 
This form is convenient for the recursive numerical treatment using the 
$\lambda_1^{(1)}=\kappa_1^{(1)}=1$ initial conditions. 
These expressions are used in Section~\ref{sec:PT_coeffs}
to compare the results with the analytic perturbative treatment 
derived in Appendix~\ref{sec: perturbation of WU}.

Before we close this appendix and turn our attention towards 
obtaining the analytic solutions for these time coefficients,
we note that it is useful to count the number of coefficients 
at each perturbative order $n$. 
As we have indicated when postulating the ansatz in Eq.~\eqref{eq:FGn},
we expect the the number of the basis elements $N(n)$ to be a function of 
the perturbative order. Thus, counting the terms given either in the Eq.~\eqref{eq: H exp}, 
or equivalently in Eq.~\eqref{eq:lambdakappa} gives  
\eeq{
\label{eq:N(n) exp}
N(n) = \df^K_{\frac{n}{2},\lfloor \frac{n}{2} \rfloor} 
\frac{1}{2} N \lb \tfrac{n}{2}\rb \big( 3 N \lb \tfrac{n}{2}\rb +  1 \big)
+ 3 \sum_{m=1}^{\left \lfloor (n-1)/2 \right \rfloor } N(m)N(n-m)\, , 
}
which, up to the fifth order, yields the numbers given in Eq.~\eqref{eq:N1-5}.
As noted in Section~\ref{sec:eom},
not all of these time coefficients are independent, there are indeed several constraints that effectively reduce the dimension of the operators basis down from the upper bound $N(n)$.

\section{Perturbative solution of the time-dependence}
\label{sec: perturbation of WU}

In Section~\ref{sec:PT_coeffs} we have summarised the perturbative solution 
for the time-dependent coefficients $\lambda_n^{(\ell)}$ and $\kappa_n^{(\ell)}$.
Here we present the derivation. In this appendix, we show how we can 
analytically invert the differential equations given in Eqs.~\eqref{tk1} (and Eqs.~\eqref{eq:WU_diffs_app})
and thus represent the solution as a power expansion in $\delta f$ parameter. 

We start by eliminating $U_{\alpha,\beta}^n$ from the equations given in Eqs.~\eqref{tk1}, 
to find the EoMs for $W_{\alpha,\beta}^n$. As expected, we end up with the second 
order differential equations that read
\eq{
\ddot{W}^{\,n\,(ij)}_{\alpha; m_1,m_2}+\left(2n+\frac{1}{2}-\delta f\right)\dot{W}^{\,n\,(ij)}_{\alpha; m_1,m_2}
+\left(n^2+\frac{n-3}{2}-(n-1)\delta f\right)W^{\,n\,(ij)}_{\alpha; m_1,m_2}
&=\left(\partial_\eta +n+\frac{1}{2}-\delta f\right)\kappa_{m_1}^{(i)}\lambda_{m_2}^{(j)}\, ,
\\
\ddot{W}^{\,n\,(ij)}_{\beta; m_1,m_2}+\left(2n+\frac{1}{2}-\delta f\right)\dot{W}^{\,n\,(ij)}_{\beta; m_1,m_2}
+\left(n^2+\frac{n-3}{2}-(n-1)\delta f\right)W^{\,n\,(ij)}_{\beta; m_1,m_2}
&=\kappa_{m_1}^{(i)}\kappa_{m_2}^{(j)} \, . \non
}
Further on we suppress the indices, $n, m_1, m_2$ and $(ij)$, that are not relevant for the 
following calculation, given that they stay the same on the right and the left side of the equations.
We can reorganise the above equations perturbatively w.r.t. $\delta f$, recasting them as
\eq{
\label{eqc2}
\ddot{W}^{\,n}_{\alpha}+\left(2n+\frac{1}{2}\right)\dot{W}^{\,n}_{\alpha}+\left(n^2+\frac{n}{2}-\frac{3}{2}\right)W^{\,n}_{\alpha}&=\left(\partial_\eta +n+\frac{1}{2}\right)\kappa\lambda
+\delta f\left(\dot{W}^{\,n}_{\alpha}+(n-1)W^{\,n}_{\alpha}-\kappa\lambda \right) \, ,
\\
\ddot{W}^{\,n}_{\beta}+\left(2n+\frac{1}{2}\right)\dot{W}^{\,n}_{\beta}+\left(n^2+\frac{n}{2}-\frac{3}{2}\right)W^{\,n}_{\beta}&=\kappa\kappa
+\delta f\left(\dot{W}^{\,n}_{\beta}+(n-1)W^{\,n}_{\beta} \right) \, . \non
}
We note that the structure of the left hand side in both these equations is exactly the same. 
It is straightforward to solve this type of equation,
\eq{
\label{eq:I def}
&\ddot{W}^{\,n}_{\gamma}(\eta)+\left(2n+\frac{1}{2}\right)\dot{W}^{\,n}_{\gamma}(\eta)
+\left(n^2+\frac{n}{2}-\frac{3}{2}\right)W^{\,n}_{\gamma}(\eta)=S^{\,n}_\gamma(\eta)\, , \\
&\hspace{3cm} \Longrightarrow \quad
W^{\,n}_\gamma(\eta) = \mathcal{I}_n[S_\gamma^{\,n}]\equiv 
\frac{2}{5}\int^\eta_{-\infty}{\rm d}\eta' 
\left[e^{(n-1)(\eta'-\eta)}-e^{(n+\tfrac{3}{2})(\eta'-\eta)}\right]S^{\,n}_\gamma(\eta') \, ,
\non
}
where $\gamma=\alpha,\beta$ and $S^n_\gamma$ denotes the source term, including
the one proportional to $\delta f$. Note that, in the process above, we have also defined the 
functional integral $\mathcal{I}_n$. Although one initially includes other terms with  integration 
constants $c_1 e^{-(n+3/2)\eta}+c_2 e^{(1-n)\eta}$ in the general solution, 
these ought to be set to zero in our case in order to reproduce the known EdS results.
This is, of course, equivalent to setting the initial conditions to the EdS values.
In particular, for a constant source term, the time integration is performed as follows
\eq{\mathcal{I}_n[\mathcal{C}]=
\frac{2}{5}\mathcal{C}\int^\eta_{-\infty}{\rm d}\eta' 
\left[e^{(n-1)(\eta'-\eta)}-e^{(n+\tfrac{3}{2})(\eta'-\eta)}\right]
=\frac{2\mathcal{C}}{(2n+3)(n-1)}\, . 
}
We seek to obtain the perturbative solution using this integral representation.
As noted in Section~\ref{sec:PT_coeffs} we expand our solutions in $\delta f$ as 
\eq{
W^{n}=W^{n[0]}+W^{n[1]}+W^{n[2]}+\cdots,\qquad
}
where the superscript $[n]$ denotes the order in powers of $\delta f$, 
with $[0]$ representing the solution in the EdS limit. 
We obtain the recursive relations 
\eq{
\label{eq:Walpha_l_recursive}
W_\alpha^{n[l]}&=\mathcal{I}_n\left[\left(\partial_\eta +n+\tfrac{1}{2}\right)(\kappa\lambda)^{[l]}
+\delta f\left(\dot{W}^{\,n[l-1]}_{\alpha}+(n-1)W^{\,n[l-1]}_{\alpha}-(\kappa\lambda)^{[l-1]} \right)\right]\, ,
\\
W_\beta^{n[l]}&=\mathcal{I}_n\left[(\kappa\kappa)^{[l]}
+\delta f\left(\dot{W}^{\,n[l-1]}_{\beta}+(n-1)W^{\,n[l-1]}_{\beta} \right)\right]\, ,\non
}
where $(xy)^{[l]}=x^{[0]}y^{[l]}+x^{[1]}y^{[l-1]}+x^{[2]}y^{[l-2]}+\cdots+x^{[l]}y^{[0]}$ and $(\kappa\lambda)^{[-1]}=0$.
To further reduce the expressions, it is useful to define the second functional expression 
\eq{
\label{eq:Itil def}
\tilde{\mathcal{I}}_n[X]\equiv \left(\partial_\eta +n-1\right) \mathcal{I}_n[\delta f\, X]
=\frac{2n+1}{5}\int^\eta_{-\infty}{\rm d}\eta' e^{(n+\tfrac{3}{2})(\eta'-\eta)}
\delta f(\eta')X(\eta')\, .
}
Repeatedly using the above recursive relation, one finds
\eq{
\label{Wbetal solution}
\dot{W}^{\,n[l]}_{\beta}+(n-1)W^{\,n[l]}_{\beta}
&=\tilde{\mathcal{I}}_n\left[\frac{(\kappa\kappa)^{[l]}}{\delta f}+\dot{W}^{\,n[l-1]}_{\beta}+(n-1)W^{\,n[l-1]}_{\beta}\right] \\
&=\tilde{\mathcal{I}}_n\left[\frac{(\kappa\kappa)^{[l]}}{\delta f}+
\tilde{\mathcal{I}}_n\left[\frac{(\kappa\kappa)^{[l-1]}}{\delta f}+
\dot{W}^{\,n[l-2]}_{\beta}+(n-1)W^{\,n[l-2]}_{\beta}\right]\right] \non\\
&=\sum_{k=1}^{l}\tilde{\mathcal{I}}_n^k\left[\frac{(\kappa\kappa)^{[l-k+1]}}{\delta f}\right]+(n-1)W^{\,n[0]}_{\beta}\tilde{\mathcal{I}}_n^l[1]\, ,\non
}
where $\dot{W}^{\,n[0]}_{\beta}=0$ and $\tilde{\mathcal{I}}_n^k[X]$ means the recursive operations of $\tilde{\mathcal{I}}_n$ by $k$ times, for instance, 
$\tilde{\mathcal{I}}_n^3[X]=\tilde{\mathcal{I}}_n[\tilde{\mathcal{I}}_n[\tilde{\mathcal{I}}_n[X]]]$.
In the case of $W_\alpha^{n[l]}$, $(\kappa\kappa)^{[l]}$ in Eq.~\eqref{Wbetal solution} is replaced by 
$\left(\partial_\eta +n+\tfrac{1}{2}\right)(\kappa\lambda)^{[l]}-\delta f (\kappa\lambda)^{[l-1]}$.
Plugging these expressions back into Eq.~\eqref{eq:Walpha_l_recursive},
and restoring the suppressed indices, we obtain the general solutions
\eq{
\label{eq:Walphan_general}
W^{\,n\,(ij)}_{\alpha; m_1,m_2}&=
\left(1-\frac{2n-2}{2n+1}\sum_{l=1}^{\infty}\mathcal{I}_n\left[\delta f\, \tilde{\mathcal{I}}_n^{l-1}[1]\right]\right)W^{\,n\,(ij)\,[0]}_{\alpha; m_1,m_2}  \\
&\hspace{0.9cm} +\sum_{l=1}^{\infty} \mathcal{I}_n\left[
\left(\partial_\eta +n+\tfrac{1}{2}\right)\left(\kappa_{m_2}^{(i)}\lambda_{m_2}^{(j)}\right)^{[l]}\right]
-\sum_{l=2}^{\infty} \mathcal{I}_n\left[\delta f \left(\kappa_{m_2}^{(i)}\lambda_{m_2}^{(j)}\right)^{[l-1]}\right]  \non\\
&\hspace{0.9cm} +
\sum_{l=2}^{\infty} \mathcal{I}_n\left[
\delta f\sum_{k=1}^{l-1}\tilde{\mathcal{I}}_n^k\left[\delta f^{-1} \left(\partial_\eta +n+\tfrac{1}{2}\right)\left(\kappa_{m_2}^{(i)}\lambda_{m_2}^{(j)}\right)^{[l-k]}\right]
-\delta f\sum_{k=1}^{l-2}\tilde{\mathcal{I}}_n^k\left[ \left(\kappa_{m_2}^{(i)}\lambda_{m_2}^{(j)}\right)^{[l-k-1]}\right]\right]\, , \non\\
W^{\,n\,(ij)}_{\beta; m_1,m_2}&=
\left(1+(n-1)\sum_{l=1}^{\infty}\mathcal{I}_n\left[\delta f\, \tilde{\mathcal{I}}_n^{l-1}[1]\right]\right)W^{\,n\,(ij)\,[0]}_{\beta; m_1,m_2}  \non\\
&\hspace{0.9cm} +\sum_{l=1}^{\infty} \mathcal{I}_n\left[
\left(\kappa_{m_1}^{(i)}\kappa_{m_2}^{(j)}\right)^{[l]}\right]
+\sum_{l=2}^{\infty} \mathcal{I}_n\left[\delta f 
\sum_{k=1}^{l-1}\tilde{\mathcal{I}}_n^k\left[\delta f^{-1} \left(\kappa_{m_1}^{(i)}\kappa_{m_2}^{(j)}\right)^{[l-k]}\right]
\right] \, ,\non
}
where we manipulated the terms with $(\kappa\lambda)^{[0]}$.
The results in the EdS limit can be found from \eqref{eq:Walpha_l_recursive} for $l=0$,
\eq{
W_{\alpha}^{n[0]}=\mathcal{I}_n\left[\left(n+\tfrac{1}{2}\right)(\kappa\lambda)^{[0]}\right]=\frac{(2n+1)\kappa^{[0]}\lambda^{[0]}}{(2n+3)(n-1)},
\qquad
W_{\beta}^{n[0]}=\mathcal{I}_n\left[(\kappa\kappa)^{[0]}\right]=\frac{2\kappa^{[0]}\kappa^{[0]}}{(2n+3)(n-1)}\, ,
}
where we used the fact that the zero-th order quantities are constant, $\kappa^{[0]},\lambda^{[0]}=const$.

In order to utilise these general solutions and employ the explicit form of the $\mathcal{I}_n$ and $\tilde{\mathcal{I}}_n$
integrals we use the expanded form of $\delta f$ that we put forward in Eq.~\eqref{df ansatz} and derived in 
Appendix~\ref{sec:linear_growth}.  Plugging this approximated expression into \eqref{eq:Walphan_general} 
and then using \eqref{tk1}, at leading order in $\delta f$ 
we obtain for $n=2$
\eq{
\lambda_2^{(1)}&=\frac{5}{7}-\frac{c_1}{91}\zeta -\frac{4c_2}{931}\zeta^2 - \frac{2c_3}{875}  \zeta^3 \, , 
&
\lambda_2^{(2)}&=\frac{2}{7}+\frac{c_1}{91}\zeta+\frac{4c_2}{931}\zeta^2 + \frac{2c_3}{875} \zeta^3 \, , 
\\
\kappa_2^{(1)}&=\frac{3}{7}-\frac{5c_1}{91} \zeta -\frac{32c_2}{931}\zeta^2  - \frac{22c_3}{875} \zeta^3 \, , 
&
\kappa_2^{(2)}&=\frac{4}{7}+\frac{5c_1}{91}\zeta +\frac{32c_2}{931}\zeta^2  + \frac{22c_3}{875} \zeta^3 \, , 
\non
}
shown also in Eq.~\eqref{eq:lamdakappa2}. 
In the same way, we compute the $n=3$ time-dependent coefficients 
\begin{align}
\label{eq:lambda3}
    \lambda_3^{(1)}&=\frac{5}{18}-\frac{29 c_1}{4725}\zeta-\frac{22 c_2}{9261}\zeta^2-\frac{118 c_3}{93555}\zeta^3\, ,
    &
    \lambda_3^{(2)}&=\frac{1}{9}+\frac{c_1}{4725}\zeta-\frac{5c_2}{18522}\zeta^2-\frac{2 c_3}{8505}\zeta^3\, ,
    \\
    \lambda_3^{(3)}&=\frac{1}{6}-\frac{19c_1}{1575}\zeta-\frac{31c_2}{6174}\zeta^2-\frac{86c_3}{31185}\zeta^3\, ,
    &
    \lambda_3^{(4)}&=\frac{2}{9}+\frac{29 c_1}{4725}\zeta+\frac{22c_2}{9261}\zeta^2+\frac{118c_3}{93555}\zeta^3\, ,
    \notag\\
    \lambda_3^{(5)}&=\frac{1}{21}+\frac{22 c_1}{20475}\zeta+\frac{85 c_2}{117306}\zeta^2+\frac{368 c_3}{779625}\zeta^3 \, ,
    &
    \lambda_3^{(6)}&=\frac{4}{63}+\frac{298 c_1}{61425}\zeta+\frac{338 c_2}{175959}\zeta^2+\frac{2396 c_3}{2338875}\zeta^3\, ,
    \notag\\
    \kappa_3^{(1)}&=\frac{5}{42}-\frac{529 c_1}{20475}\zeta-\frac{334 c_2}{19551}\zeta^2-\frac{10018 c_3}{779625}\zeta^3\, ,
    &
    \kappa_3^{(2)}&=\frac{1}{21}-\frac{199 c_1}{20475}\zeta-\frac{263 c_2}{39102}\zeta^2-\frac{362 c_3}{70875}\zeta^3\, ,
    \notag\\
    \kappa_3^{(3)}&=\frac{1}{14}-\frac{17 c_1}{975}\zeta-\frac{141 c_2}{13034}\zeta^2-\frac{2066 c_3}{259875}\zeta^3\, ,
    &
    \kappa_3^{(4)}&=\frac{2}{21}-\frac{53 c_1}{2925}\zeta-\frac{254 c_2}{19551}\zeta^2-\frac{7802 c_3}{779625}\zeta^3\, ,
    \notag\\
    \kappa_3^{(5)}&=\frac{1}{7}+\frac{44 c_1}{6825}\zeta+\frac{85 c_2}{13034}\zeta^2+\frac{1472 c_3}{259875}\zeta^3\, ,
    &
    \kappa_3^{(6)}&=\frac{4}{21}+\frac{596 c_1}{20475}\zeta+\frac{338 c_2}{19551}\zeta^2+\frac{9584 c_3}{779625}\zeta^3\, .
    \notag
\end{align}
and the result for $n=4$
\begingroup
\allowdisplaybreaks
\begin{align}
\label{eq:lambda4}
\lambda_4^{(1)}&=\frac{45}{539}-\frac{900 c_1 \zeta }{119119}-\frac{2272 c_2 \zeta ^2}{706629}-\frac{3496 c_3 \zeta ^3}{1953875}
\,,&
\lambda_4^{(2)}&=\frac{18}{539}-\frac{489 c_1 \zeta }{238238}-\frac{724 c_2 \zeta ^2}{706629}-\frac{2381 c_3 \zeta ^3}{3907750}
\,,\\
\lambda_4^{(3)}&=\frac{60}{539}+\frac{2425 c_1 \zeta }{714714}+\frac{3232 c_2 \zeta ^2}{2119887}+\frac{10147 c_3 \zeta ^3}{11723250}
\,,&
\lambda_4^{(4)}&=\frac{24}{539}+\frac{947 c_1 \zeta }{357357}+\frac{2032 c_2 \zeta ^2}{2119887}+\frac{2861 c_3 \zeta ^3}{5861625}
\,,\notag\\
\lambda_4^{(5)}&=\frac{6}{539}-\frac{32 c_1 \zeta }{119119}+\frac{188 c_2 \zeta ^2}{4946403}+\frac{19 c_3 \zeta ^3}{279125}
\,,&
\lambda_4^{(6)}&=\frac{8}{539}+\frac{257 c_1 \zeta }{357357}+\frac{5680 c_2 \zeta ^2}{14839209}+\frac{197 c_3 \zeta ^3}{837375}
\,,\notag\\
\lambda_4^{(7)}&=\frac{32}{1617}+\frac{856 c_1 \zeta }{357357}+\frac{14144 c_2 \zeta ^2}{14839209}+\frac{424 c_3 \zeta ^3}{837375}
\,,&
\lambda_4^{(8)}&=\frac{5}{66}-\frac{223 c_1 \zeta }{117810}-\frac{244 c_2 \zeta ^2}{334719}-\frac{13 c_3 \zeta ^3}{33495}
\,,\notag\\
\lambda_4^{(9)}&=\frac{1}{33}-\frac{43 c_1 \zeta }{117810}-\frac{149 c_2 \zeta ^2}{669438}-\frac{3 c_3 \zeta ^3}{22330}
\,,&
\lambda_4^{(10)}&=\frac{1}{22}-\frac{31 c_1 \zeta }{13090}-\frac{179 c_2 \zeta ^2}{223146}-\frac{13 c_3 \zeta ^3}{33495}
\,,\notag\\
\lambda_4^{(11)}&=\frac{2}{33}+\frac{13 c_1 \zeta }{117810}-\frac{50 c_2 \zeta ^2}{334719}-\frac{3 c_3 \zeta ^3}{22330}
\,,&
\lambda_4^{(12)}&=\frac{1}{77}-\frac{c_1 \zeta }{85085}-\frac{43 c_2 \zeta ^2}{4239774}-\frac{c_3 \zeta ^3}{76125}
\,,\notag\\
\lambda_4^{(13)}&=\frac{4}{231}+\frac{373 c_1 \zeta }{765765}+\frac{530 c_2 \zeta ^2}{6359661}+\frac{c_3 \zeta ^3}{76125}
\,,&
\lambda_4^{(14)}&=\frac{5}{154}-\frac{6469 c_1 \zeta }{1531530}-\frac{12352 c_2 \zeta ^2}{6359661}-\frac{313 c_3 \zeta ^3}{279125}
\,,\notag\\
\lambda_4^{(15)}&=\frac{1}{77}-\frac{2449 c_1 \zeta }{1531530}-\frac{9743 c_2 \zeta ^2}{12719322}-\frac{249 c_3 \zeta ^3}{558250}
\,,&
\lambda_4^{(16)}&=\frac{3}{154}-\frac{1439 c_1 \zeta }{510510}-\frac{5185 c_2 \zeta ^2}{4239774}-\frac{193 c_3 \zeta ^3}{279125}
\,,\notag\\
\lambda_4^{(17)}&=\frac{2}{77}-\frac{4601 c_1 \zeta }{1531530}-\frac{9446 c_2 \zeta ^2}{6359661}-\frac{489 c_3 \zeta ^3}{558250}
\,,&
\lambda_4^{(18)}&=\frac{3}{77}+\frac{16 c_1 \zeta }{36465}+\frac{1741 c_2 \zeta ^2}{4239774}+\frac{81 c_3 \zeta ^3}{279125}
\,,\notag\\
\lambda_4^{(19)}&=\frac{4}{77}+\frac{394 c_1 \zeta }{109395}+\frac{9026 c_2 \zeta ^2}{6359661}+\frac{632 c_3 \zeta ^3}{837375}
\,,&
\lambda_4^{(20)}&=\frac{5}{693}-\frac{944 c_1 \zeta }{11486475}+\frac{218 c_2 \zeta ^2}{4946403}+\frac{6857 c_3 \zeta ^3}{135654750}
\,,\notag\\
\lambda_4^{(21)}&=\frac{2}{693}-\frac{239 c_1 \zeta }{11486475}+\frac{277 c_2 \zeta ^2}{14839209}+\frac{1384 c_3 \zeta ^3}{67827375}
\,,&
\lambda_4^{(22)}&=\frac{1}{231}-\frac{334 c_1 \zeta }{3828825}+\frac{311 c_2 \zeta ^2}{14839209}+\frac{17 c_3 \zeta ^3}{587250}
\,,\notag\\
\lambda_4^{(23)}&=\frac{4}{693}-\frac{181 c_1 \zeta }{11486475}+\frac{620 c_2 \zeta ^2}{14839209}+\frac{37 c_3 \zeta ^3}{880875}
\,,&
\lambda_4^{(24)}&=\frac{2}{231}+\frac{2434 c_1 \zeta }{3828825}+\frac{1553 c_2 \zeta ^2}{4946403}+\frac{4111 c_3 \zeta ^3}{22609125}
\,,\notag\\
\lambda_4^{(25)}&=\frac{8}{693}+\frac{14356 c_1 \zeta }{11486475}+\frac{7444 c_2 \zeta ^2}{14839209}+\frac{18292 c_3 \zeta ^3}{67827375}
\,,&
\kappa_4^{(1)}&=\frac{15}{539}-\frac{152 c_1 \zeta }{17017}-\frac{28492 c_2 \zeta ^2}{4946403}-\frac{8444 c_3 \zeta ^3}{1953875}
\,,\notag\\
\kappa_4^{(2)}&=\frac{6}{539}-\frac{115 c_1 \zeta }{34034}-\frac{11212 c_2 \zeta ^2}{4946403}-\frac{6709 c_3 \zeta ^3}{3907750}
\,,&
\kappa_4^{(3)}&=\frac{20}{539}-\frac{941 c_1 \zeta }{102102}-\frac{101648 c_2 \zeta ^2}{14839209}-\frac{63317 c_3 \zeta ^3}{11723250}
\,,\notag\\
\kappa_4^{(4)}&=\frac{8}{539}-\frac{25 c_1 \zeta }{7293}-\frac{39920 c_2 \zeta ^2}{14839209}-\frac{12571 c_3 \zeta ^3}{5861625}
\,,&
\kappa_4^{(5)}&=\frac{24}{539}-\frac{32 c_1 \zeta }{17017}+\frac{1880 c_2 \zeta ^2}{4946403}+\frac{247 c_3 \zeta ^3}{279125}
\,,\notag\\
\kappa_4^{(6)}&=\frac{32}{539}+\frac{257 c_1 \zeta }{51051}+\frac{56800 c_2 \zeta ^2}{14839209}+\frac{2561 c_3 \zeta ^3}{837375}
\,,&
\kappa_4^{(7)}&=\frac{128}{1617}+\frac{856 c_1 \zeta }{51051}+\frac{141440 c_2 \zeta ^2}{14839209}+\frac{5512 c_3 \zeta ^3}{837375}
\,,\notag\\
\kappa_4^{(8)}&=\frac{5}{198}-\frac{12569 c_1 \zeta }{1767150}-\frac{3838 c_2 \zeta ^2}{781011}-\frac{10267 c_3 \zeta ^3}{2713095}
\,,&
\kappa_4^{(9)}&=\frac{1}{99}-\frac{4889 c_1 \zeta }{1767150}-\frac{3055 c_2 \zeta ^2}{1562022}-\frac{8201 c_3 \zeta ^3}{5426190}
\,,\notag\\
\kappa_4^{(10)}&=\frac{1}{66}-\frac{2659 c_1 \zeta }{589050}-\frac{4687 c_2 \zeta ^2}{1562022}-\frac{2069 c_3 \zeta ^3}{904365}
\,,&
\kappa_4^{(11)}&=\frac{2}{99}-\frac{9481 c_1 \zeta }{1767150}-\frac{3022 c_2 \zeta ^2}{781011}-\frac{16321 c_3 \zeta ^3}{5426190}
\,,\notag\\
\kappa_4^{(12)}&=\frac{1}{231}-\frac{4429 c_1 \zeta }{3828825}-\frac{24515 c_2 \zeta ^2}{29678418}-\frac{14533 c_3 \zeta ^3}{22609125}
\,,&
\kappa_4^{(13)}&=\frac{4}{693}-\frac{16561 c_1 \zeta }{11486475}-\frac{16138 c_2 \zeta ^2}{14839209}-\frac{57901 c_3 \zeta ^3}{67827375}
\,,\notag\\
\kappa_4^{(14)}&=\frac{5}{462}-\frac{9523 c_1 \zeta }{2552550}-\frac{104122 c_2 \zeta ^2}{44517627}-\frac{5581 c_3 \zeta ^3}{3229875}
\,,&
\kappa_4^{(15)}&=\frac{1}{231}-\frac{3763 c_1 \zeta }{2552550}-\frac{83159 c_2 \zeta ^2}{89035254}-\frac{4463 c_3 \zeta ^3}{6459750}
\,,\notag\\
\kappa_4^{(16)}&=\frac{1}{154}-\frac{279 c_1 \zeta }{121550}-\frac{41893 c_2 \zeta ^2}{29678418}-\frac{7829 c_3 \zeta ^3}{7536375}
\,,&
\kappa_4^{(17)}&=\frac{2}{231}-\frac{1061 c_1 \zeta }{364650}-\frac{82862 c_2 \zeta ^2}{44517627}-\frac{62401 c_3 \zeta ^3}{45218250}
\,,\notag\\
\kappa_4^{(18)}&=\frac{1}{77}-\frac{1436 c_1 \zeta }{425425}-\frac{71675 c_2 \zeta ^2}{29678418}-\frac{14257 c_3 \zeta ^3}{7536375}
\,,&
\kappa_4^{(19)}&=\frac{4}{231}-\frac{1658 c_1 \zeta }{425425}-\frac{137806 c_2 \zeta ^2}{44517627}-\frac{56104 c_3 \zeta ^3}{22609125}
\,,\notag\\
\kappa_4^{(20)}&=\frac{20}{693}-\frac{944 c_1 \zeta }{1640925}+\frac{2180 c_2 \zeta ^2}{4946403}+\frac{89141 c_3 \zeta ^3}{135654750}
\,,&
\kappa_4^{(21)}&=\frac{8}{693}-\frac{239 c_1 \zeta }{1640925}+\frac{2770 c_2 \zeta ^2}{14839209}+\frac{17992 c_3 \zeta ^3}{67827375}
\,,\notag\\
\kappa_4^{(22)}&=\frac{4}{231}-\frac{334 c_1 \zeta }{546975}+\frac{3110 c_2 \zeta ^2}{14839209}+\frac{221 c_3 \zeta ^3}{587250}
\,,&
\kappa_4^{(23)}&=\frac{16}{693}-\frac{181 c_1 \zeta }{1640925}+\frac{6200 c_2 \zeta ^2}{14839209}+\frac{481 c_3 \zeta ^3}{880875}
\,,\notag\\
\kappa_4^{(24)}&=\frac{8}{231}+\frac{2434 c_1 \zeta }{546975}+\frac{15530 c_2 \zeta ^2}{4946403}+\frac{53443 c_3 \zeta ^3}{22609125}
\,,&
\kappa_4^{(25)}&=\frac{32}{693}+\frac{14356 c_1 \zeta }{1640925}+\frac{74440 c_2 \zeta ^2}{14839209}+\frac{237796 c_3 \zeta ^3}{67827375}
\,.\notag
\end{align}
\endgroup

One can continue this exercise up to an arbitrary $n$. 
For the purposes of this work, it suffices to run this procedure up to $n=5$: $F_5$ and $G_5$ 
are the highest order kernels necessary for the calculation of the two-loop power spectra, 
and therefore we may stop at the $\lambda^{(i)}_{5}, \kappa^{(i)}_{5}$ set of time coefficients
(all coefficients and corresponding kernels are given in the \textit{Mathematica} notebook, provided in the \textit{arXiv} source file of this paper).

\section{IR and UV limits of kernels} 
\label{sec: IR and UV limits}

In this appendix, we consider the IR and UV limits of the $\vec H_n$ kernels in the momenta configurations contributing to the one- and two-loop calculations. 
For IR limits, we have the following expansions
\eq{
\vec H_2 (\vec p, \vec k-\vec p) &\sim [\vec H_{2}]^{(1)}_{\rm IR} (\hat p) \frac{k}{p} +  [\vec H_{2}]^{(0)}_{\rm IR}(\hat p) 
+ \mathcal{O}(p^1)\, , \\
\vec H_3 (\vec k, \vec p,-\vec p) &\sim [\vec H_{3,{\rm I}}]^{(2)}_{\rm IR}(\hat p) \frac{k^2}{p^2} 
+  [\vec H_{3,{\rm I}}]^{(0)}_{\rm IR}(\hat p) 
+ \mathcal{O}(p^1) \, , \non\\
\vec H_{3}(\vec k - \vec q - \vec p, \vec q, \vec p) 
& \sim [\vec H_{3,{\rm II}}]^{(1)}_{\rm IR} (\hat p) \frac{k}{p} + [\vec H_{3,{\rm II}}]^{(0)}_{\rm IR} (\hat p) + \mathcal{O}(p^1) \, , \non\\
\vec H_{4}(\vec k -  \vec q , \vec q, \vec p, -\vec p) 
&\sim [\vec H_{4,{\rm I}}]^{(2)}_{\rm IR} (\hat p) \frac{k^2}{p^2}
+ [\vec H_{4,{\rm I}}]^{(0)}_{\rm IR} (\hat p) + \mathcal{O}(p^1) \, , 
\non \\
\vec H_{4}(\vec k -  \vec p , \vec p, \vec q, -\vec q) 
&\sim [\vec H_{4,{\rm II}}]^{(1)}_{\rm IR} (\hat p) \frac{k}{p}
+ [\vec H_{4,{\rm II}}]^{(0)}_{\rm IR} (\hat p) + \mathcal{O}(p^1)\, ,  
\non \\
\vec H_5(\vec k, \vec q, - \vec q, \vec p, -\vec p) 
&\sim [\vec H_{5}]^{(2)}_{\rm IR}(\hat p) \frac{k^2}{p^2} + [\vec H_{5}]^{(0)}_{\rm IR}(\hat p) + \mathcal{O}(p^1)\, , \non
}
as $p \to 0$. We can write explicitly the first few coefficients as 
$[\vec H_{2}]^{(1)}_{\rm IR} = \frac{\mu}{2}\lb1,1\rb$ and $[\vec H_{2}]^{(0)}_{\rm IR}  = \frac{1}{2}\lb1, 2\mu^2-1\rb$,
and $[\vec H_{3,{\rm I}}]^{(2)}_{\rm IR} = - \frac{\mu^2}{3}\lb1 , 1 , 0 , 0, 1, 1\rb$, 
$[\vec H_{3,{\rm I}}]^{(0)}_{\rm IR} = - \frac{1}{3}\lb \mu^2 , \mu^2 , \mu^2 - 2  , -\mu^2 , 2 - 6\mu^2 + 4\mu^4, 0 \rb$, where $\mu = \hat k \cdot \hat p$.
We shall not give the explicit form here for the
remaining coefficients. These are obtained in a similar manner and can be arrived at by expanding the full kernels.

In computing the loop contribution, it is crucial to isolate the leading product divergencies of the kernel products. Focusing on the IR limit at one-loop we have
\eq{
\vec h^{(2)}_{22,{\rm IR}} &= [\vec H_2]^{(1)}_{\rm IR} \otimes [\vec H_{2}]^{(1)}_{\rm IR} = \frac{\mu^2}{4} \mat{ 1 & 1 \\ 1 & 1}\, , \\
\vec h^{(1)}_{22,{\rm IR}} &= [\vec H_2]^{(1)}_{\rm IR} \otimes [\vec H_{2}]^{(0)}_{\rm IR} + [\vec H_{2}]^{(0)}_{\rm IR}\otimes [\vec H_{2}]^{(1)}_{\rm IR} = \frac{\mu}{2} \mat{ 1 & \mu^2 \\ \mu^2 & 2 \mu^2 - 1}\, , \non\\
\vec h^{(2)}_{13,{\rm IR}} &= \vec H_1(\vec k) \otimes [\vec H_{3,{\rm I}}]^{(2)}_{\rm IR} = - \frac{\mu^2}{3}\mat{ 1 & 1 & 0 & 0 & 1 & 1}\, , \non
}
while at two-loops one finds
\eq{
\vec h^{(2)}_{33,{\rm I,IR}} &= \vec H_3( \vec k, \vec q, -\vec q) \otimes [\vec H_{3,{\rm I}}]^{(2)}_{\rm IR} \, , \\
\vec h^{(2)}_{33,{\rm II,IR}} &=  [\vec H_{3,{\rm II}}]^{(1)}_{\rm IR} \otimes [\vec H_{3,{\rm II}}]^{(1)}_{\rm IR} \,, \non\\
\vec h^{(1)}_{33,{\rm II,IR}} &=  [\vec H_{3,{\rm II}}]^{(1)}_{\rm IR} \otimes [\vec H_{3,{\rm II}}]^{(0)}_{\rm IR} + [\vec H_{3,{\rm II}}]^{(0)}_{\rm IR} \otimes [\vec H_{3,{\rm II}}]^{(1)}_{\rm IR}\, , \non\\
\vec h^{(2)}_{24,{\rm I, IR}} &=  \vec H_2(\vec k - \vec q, \vec q ) \otimes [\vec H_{4,{\rm I}}]^{(2)}_{\rm IR}\, , \non\\
\vec h^{(2)}_{24,{\rm II, IR}} &= [\vec H_{2}]^{(1)}_{\rm IR} \otimes [\vec H_{4,{\rm II}}]^{(1)}_{\rm IR} , \non\\
\vec h^{(1)}_{24,{\rm II, IR}} &= [\vec H_{2}]_{\rm IR}^{(1)} \otimes [\vec H_{4,{\rm II}}]^{(0)}_{\rm IR} + [\vec H_{2}]_{\rm IR}^{(0)} \otimes [\vec H_{4,{\rm II}}]^{(1)}_{\rm IR}\, , \non\\
\vec h^{(2)}_{15,{\rm IR}} &= \vec H_1(\vec k) \otimes [\vec H_5]^{(2)}_{\rm IR}\, . \non
}
Similarly, one can derive the UV limit of the various terms, which are given by
\eq{
\vec H_2(\vec p, \vec k-\vec p) &\sim [\vec H_{2}]^{(2)}_{\rm UV} (\hat k) \frac{k^2}{p^2} + [\vec H_{2}]^{(3)}_{\rm UV} (\hat k) \frac{k^3}{p^3} + \mathcal{O}(k^4) \, , \\
\vec H_3( \vec k, \vec p,-\vec p) &\sim [\vec H_{3,{\rm I}}]^{(0)}_{\rm UV}(\hat k) + [\vec H_{3,{\rm I}}]^{(2)}_{\rm UV}(\hat k) \frac{k^2}{p^2}  + \mathcal{O}(k^4) \, , \non\\
\vec H_3(\vec k - \vec q - \vec p, \vec q, \vec p) &\sim [\vec H_{3,{\rm II}}]^{(1)}_{\rm UV} (\hat k) \frac{k}{p}
+ [\vec H_{3,{\rm II}}]^{(2)}_{\rm UV} (\hat k) \frac{k^2}{p^2} + [\vec H_{3,{\rm II}}]^{(3)}_{\rm UV} (\hat k) \frac{k^3}{p^3} + \mathcal{O}(k^4) \, , \non\\
\vec H_4( \vec k -  \vec q , \vec q, \vec p, -\vec p) &\sim
[\vec H_{4}]^{(2)}_{\rm UV} (\hat k) \frac{k^2}{p^2}  + [\vec H_{4}]^{(3)}_{\rm UV} (\hat k) \frac{k^3}{p^3}  
+ \mathcal{O}(k^4) \, , \non\\
\vec H_5(\vec k, \vec q, - \vec q, \vec p, -\vec p)
&\sim [\vec H_{5}]^{(0)}_{\rm UV}(\hat k)  +  [\vec H_{5}]^{(2)}_{\rm UV}(\hat k)  \frac{k^2}{p^2}  + \mathcal{O}(k^4) \, , \non
} 
as $k \to 0$. Again, the first few coefficients can be written as
$[\vec H_{2}]^{(2)}_{\rm UV} = \frac{1}{2} \lb 1 - 2\mu^2 , - 1 \rb$, 
$[\vec H_{2}]^{(3)}_{\rm UV} = \frac{\mu}{2} \lb 3 - 4\mu^2 , -1 \rb$,
$[\vec H_{3,{\rm I}}]^{(0)}_{\rm UV} = \frac{\mu^2}{3}\lb -1, -1, 1, 1, 0, 0 \rb$ 
and $[\vec H_{3,{\rm I}}]^{(2)}_{\rm UV} = - \frac{1}{3}\lb \mu^2, \mu^2, 2 (1-\mu^2)(2\mu^2-1), 0, 2 - \mu^2, \mu^2 \rb$. The kernel products relevant for the UV divergencies in the power spectra are, at one-loop power,
\eq{
\vec h^{(4)}_{22,{\rm UV}} &= [\vec H_{2}]^{(2)}_{\rm UV} \otimes [\vec H_{2}]^{(2)}_{\rm UV} = \frac{1}{4} \mat{ \lb 2 \mu^2 - 1 \rb^2 & 2 \mu^2 - 1 \\ 2 \mu^2 - 1 & 1}\, , \\
\vec h^{(0)}_{13,{\rm UV}} &= \vec H_1 \otimes [\vec H_{3,{\rm I}}]^{(0)}_{\rm UV} = - \frac{\mu^2}{3}\mat{ -1 & -1 & 1 & 1 & 0 & 0}\, , \non\\
\vec h^{(2)}_{13,{\rm UV}} &= \vec H_1 \otimes [\vec H_{3,{\rm I}}]^{(2)}_{\rm UV} = - \frac{1}{3} \mat{ \mu^2 & \mu^2 & 2 (1-\mu^2)(2\mu^2-1) & 0 & 2 - \mu^2 & \mu^2 }\, , \non
}
and, at two-loop,
\eq{
\vec h^{(0)}_{33,{\rm I,UV}} &= [\vec H_{3,{\rm I}}]^{(0)}_{\rm UV} \otimes [\vec H_{3,{\rm I}}]^{(0)}_{\rm UV} \, , \\
\vec h^{(2)}_{33,{\rm I,UV}} &= [\vec H_{3,{\rm I}}]^{(0)}_{\rm UV} \otimes [\vec H_{3,{\rm I}}]^{(2)}_{\rm UV} \, , ~~ {\rm not~symmetric~in~} \hat q {\rm ~and~} \hat p, \non\\
\vec h^{(4)}_{33,{\rm I,UV}} &= [\vec H_{3,{\rm I}}]^{(2)}_{\rm UV} \otimes [\vec H_{3,{\rm I}}]^{(2)}_{\rm UV} \, , \non\\
\vec h^{(2)}_{33,{\rm II,UV}}&=  [\vec H_{3,{\rm II}}]^{(1)}_{\rm UV} \otimes [\vec H_{3,{\rm II}}]^{(1)}_{\rm UV}\, , \non\\
\vec h^{(3)}_{33,{\rm II,UV}}&=  [\vec H_{3,{\rm II}}]^{(1)}_{\rm UV} \otimes [\vec H_{3,{\rm II}}]^{(2)}_{\rm UV} + [\vec H_{3,{\rm II}}]^{(2)}_{\rm UV} \otimes [\vec H_{3,{\rm II}}]^{(1)}_{\rm UV} \, , \non\\
\vec h^{(4)}_{33,{\rm II,UV}}&=  [\vec H_{3,{\rm II}}]^{(2)}_{\rm UV} \otimes [\vec H_{3,{\rm II}}]^{(2)}_{\rm UV} 
+ [\vec H_{3,{\rm II}}]^{(1)}_{\rm UV} \otimes [\vec H_{3,{\rm II}}]^{(3)}_{\rm UV} 
+ [\vec H_{3,{\rm II}}]^{(3)}_{\rm UV} \otimes [\vec H_{3,{\rm II}}]^{(1)}_{\rm UV} \, , \non\\
\vec h^{(4)}_{24,{\rm UV}} &= [\vec H_{2}]^{(2)}_{\rm UV} \otimes [\vec H_{4}]^{(2)}_{\rm UV} \, , \non\\
\vec h^{(0)}_{15,{\rm UV}} &= \vec H_1(\vec k) \otimes [\vec H_5]^{(0)}_{\rm UV} \, , \non \\
\vec h^{(2)}_{15,{\rm UV}} &= \vec H_1(\vec k) \otimes [\vec H_5]^{(2)}_{\rm UV} \, . \non
}

\section{Two-loop basis power spectra, IR and UV properties} 
\label{app:two-loops}

In this appendix, we look into the IR and UV properties
of the integrands given in Eq.~\eqref{eq:Inn_two_loop}. 
We first look at the $\vec I_{33}$ term, which has two  distinct contributions.
It is convenient to remap the contributions (see
\cite{Carrasco++:2013a}) as
\eq{
\vec I_{33,{\rm I}}
&= 9 P_{\rm lin}(\vec k)
\int_{\vec q, \vec p} \vec H_3(\vec k, - \vec q, \vec q) \otimes \vec H_3(\vec k, - \vec p, \vec p) P_{\rm lin}(\vec q) P_{\rm lin}(\vec p) \\
&= 9 P_{\rm lin}(\vec k)
\int_{\vec q, \vec p} \Big[ \vec H_3(\vec k, - \vec q, \vec q) \otimes \vec H_3(\vec k, - \vec p, \vec p) \Theta(q-p) + \vec q \leftrightarrow \vec p  \Big] P_{\rm lin}(\vec q) P_{\rm lin}(\vec p). \non
}
The IR and UV limits can be expressed as 
\eq{
\vec H_3(\vec k, - \vec q, \vec q) \otimes \vec H_3(\vec k, - \vec p, \vec p)
\sim 
\begin{cases}
\vec h^{(2)}_{33,{\rm I, IR}}  \lb \vec k, \vec q, \hat p \rb \frac{k^2}{p^2} + \mathcal{O}(p^0) \, ,
&{\rm as}~~ p \to 0 , \\[0.5em]
\begin{aligned} 
\vec h^{(0)}_{33,{\rm I,UV}}(\hat k, \vec q, \vec p) 
&+ \lb \vec h^{(2)}_{33,{\rm I,UV}} (\hat k, \vec q, \vec p) \tfrac{k^2}{q^2}  + \vec q \leftrightarrow \vec p \rb \\
&\hspace{2.0cm} + \vec h^{(4)}_{33,{\rm I,UV}}(\hat k, \vec q, \vec p) \tfrac{k^4}{p^2 q^2} + \mathcal{O}(k^5) \,
\end{aligned} , &{\rm as}~~ k \to 0 , 
\end{cases}
}
and we can thus write the regularized integral as
\eq{
\vec{\tilde I}_{33,{\rm I}}
&= 18 P_{\rm lin}(\vec k) \int_{\vec q, \vec p} \Bigg[
\vec H_{3}(\vec k, - \vec q, \vec q) \otimes \vec H_{3}(\vec k, - \vec p, \vec p) 
- \vec h^{(2)}_{33,{\rm I, IR}} (\vec k, \vec q, \hat p) \frac{k^2}{p^2} W^{\rm IR}_{33,{\rm I}} \\
&\hspace{1.0cm}
- \bigg( \vec h^{(0)}_{33,{\rm I,UV}}(\hat k, \vec q, \vec p) + \lb \vec h^{(2)}_{33,{\rm I,UV}} (\hat k, \vec q, \vec p) \frac{k^2}{q^2}  + \vec q \leftrightarrow \vec p \rb
+ \vec h^{(4)}_{33,{\rm I,UV}}(\hat k, \vec q, \vec p)  \frac{k^4}{p^2 q^2}  \bigg) W^{\rm UV}_{33,{\rm I}} \Bigg]  \Theta(q-p) P_{\rm lin}(\vec q)P_{\rm lin}(\vec p) . \non
}
Given that $\vec \lambda_3 \cdot [\vec H_{3,{\rm I}}]^{(0)}_{\rm UV} = \vec \kappa_3 \cdot [\vec H_{3,{\rm I}}]^{(0)}_{\rm UV} =0$,
both in EdS and $\Lambda$CDM case, the terms $\vec h^{(0)}_{33,{\rm I,UV}}$ and $\vec h^{(2)}_{33,{\rm I,UV}}$ are zero.
We thus have $\vec I_{33,{\rm I}} (k) = \vec{\tilde I}_{33,{\rm I}} (k) + [\vec I_{33,{\rm I}}]_{\rm IR} + [\vec I_{33,{\rm I}}]_{\rm UV}$,
where
\eq{
[\vec I_{33,{\rm I}}]_{\rm IR} 
&= 18 P_{\rm lin}(\vec k) \int_{\vec q, \vec p} \vec h^{(2)}_{33,{\rm I, IR}} (\vec k, \vec q, \hat p)  \frac{k^2}{p^2} W^{\rm IR}_{33,{\rm I}} \Theta(q - p) P_{\rm lin}(\vec q)P_{\rm lin}(\vec p) 
= \lb \vec h^{\rm IR}_{33,{\rm I}} W^{\rm IR}_{33,{\rm I}} \rb k^2 P_{\rm lin}(k) \, , \\
[\vec I_{33,{\rm I}}]_{\rm UV} 
&= 9 P_{\rm lin}(\vec k) \int_{\vec q, \vec p} \vec h^{(4)}_{33,{\rm I,UV}}(\hat k, \vec q, \vec p) \frac{k^4}{p^2q^2} W^{\rm UV}_{33,{\rm I}}P_{\rm lin}(\vec q)P_{\rm lin}(\vec p)
= \lb \vec h^{\rm UV}_{33,{\rm I}} W^{\rm UV}_{33,{\rm I}} \rb k^4 P_{\rm lin}(k). \non
}
In the last integral we have reverted back to the symmetric form of the integrand. We used the fact that 
\eeq{
\vec h^{\rm IR}_{33,{\rm I}}
= \int_{\vec q} \vec{\hat h}^{(2)}_{33,{\rm I, IR}} (\vec k, \vec q) \sigma^2_2(q) P_{\rm lin}(\vec q) \, , ~~{\rm and}~~
\vec h^{\rm UV}_{33,{\rm I}} 
= 2 \lb \vec h^{\rm UV}_{13} \otimes \vec h^{\rm UV}_{13} \rb  \lb \sigma^2_{2} \rb^2 \, ,
}
where we have introduced the coefficients $\vec{\hat h}^{(2)}_{33,{\rm I, IR}}(\vec k, \vec q) = 54 \int \frac{d \Omega_{\hat p}}{4 \pi}~ \vec h^{(2)}_{33,{\rm I, IR}} (\vec k, \vec q, \hat p)$,
integrating the $\vec h^{(2)}_{33,{\rm I, IR}}$ over $\hat p$. 
We have also introduced the variance due to the $p$ modes smaller than $q$ as 
\eeq{
\sigma^2_2(q) = \frac{1}{3}\int_{\vec p} \Theta(q - p)  P_{\rm lin}(\vec p)/p^2
= \frac{1}{3} \int_0^q \frac{dp}{2\pi^2} ~ P_{\rm lin}(p)\, .
\label{eq:variance_cut}
}
Note that the final IR term $\vec h^{\rm IR}_{33,{\rm I}}(k)$ is $k$ dependent.

Next we consider the $\vec I_{33,{\rm II}}$ integral
\eeq{
\vec I_{33,{\rm II}}
= 6 \int_{\vec q, \vec p} \vec H_{3}(\vec k - \vec q - \vec p, \vec q, \vec p) \otimes H_{3}(\vec k - \vec q - \vec p, \vec q, \vec p) 
P_{\rm lin} (\vec k - \vec q - \vec p) P_{\rm lin}(\vec q)P_{\rm lin}(\vec p) \, ,
}
which has leading divergencies when $\vec q$ and $\vec p$ go to zero, and when 
one of these momenta goes to zero while the other approaches $\vec k$.
The sub-leading divergencies arise when $\vec q$ or $\vec p$ go to zero, while the  other is finite, and when $\vec q +  \vec p \to \vec k $.
Since the integral can be symmetrized by introducing the delta function (see e.g. \cite{Carrasco++:2013a}), we can remap some of these divergencies into others by writing 
\eeq{
\vec I_{33,{\rm II}}
= 36 \int_{\vec q, \vec p} \vec H_{3}(\vec k - \vec q - \vec p, \vec q, \vec p) \otimes \vec H_{3}(\vec k - \vec q - \vec p, \vec q, \vec p) 
\Theta( q - p)  \Theta( |\vec k - \vec q - \vec p | - q ) 
P_{\rm lin} (\vec k - \vec q - \vec p) P_{\rm lin}(\vec q)P_{\rm lin}(\vec p) ,
}
where the leading divergencies appear only when $\vec q$ and $\vec p$ go to zero
and the subleading ones appear when $\vec q$ goes to zero for finite $\vec p$. The product of the kernels can be written as
\eeq{
\vec H_{3}(\vec k - \vec q - \vec p, \vec q, \vec p) \otimes  \vec H_{3}(\vec k - \vec q - \vec p, \vec q, \vec p) 
\sim 
\begin{cases}
\vec h^{(2)}_{33,{\rm II,IR}}(\vec k, \vec q, \hat p) \tfrac{k^2}{p^2} + \vec h^{(1)}_{33,{\rm II,IR}}(\vec k, \vec q, \hat p) \tfrac{k}{p} + \mathcal{O}(p^0) ,
~{\rm as}~~ p \to 0 \, , \\
\begin{aligned} 
\vec h^{(2)}_{33,{\rm II,UV}}(\hat k, \vec p, \vec q) \tfrac{k^2}{p^2} 
&+ \vec h^{(3)}_{33,{\rm II,UV}}(\hat k, \vec p, \vec q) \tfrac{k^3}{p^3} \\
&+ \vec h^{(4)}_{33,{\rm II,UV}}(\hat k, \vec p, \vec q) \tfrac{k^4}{p^4} + \mathcal{O}(k^5)
\end{aligned},
~{\rm as}~~ k \to 0 \, .
\end{cases}
}
We can thus write the regularized integral as
\eq{
\vec{\tilde I}_{33,{\rm II}}
&= 36 \int_{\vec q, \vec p} \Bigg[
\vec H_{3}(\vec k - \vec q - \vec p, \vec q, \vec p) \otimes  \vec H_{3}(\vec k - \vec q - \vec p, \vec q, \vec p) \Theta( |\vec k - \vec q - \vec p | - q ) P_{\rm lin} (\vec k - \vec q - \vec p) \\
&\hspace{0.5cm}
- \bigg(\vec h^{(2)}_{33,{\rm II,IR}}(\vec k, \vec q, \hat p) \frac{k^2}{p^2} + \vec h^{(1)}_{33,{\rm II,IR}}(\vec k, \vec q, \hat p) \frac{k}{p}  \bigg) 
W^{\rm IR}_{33,{\rm II}} \Theta( |\vec k - \vec q| - q ) P_{\rm lin} (\vec k - \vec q) \non\\
&\hspace{0.5cm}
- \bigg( \vec h^{(2)}_{33,{\rm II,UV}}(\hat k, \vec p, \vec q) \frac{k^2}{p^2} 
+ \vec h^{(3)}_{33,{\rm II,UV}}(\hat k, \vec p, \vec q) \frac{k^3}{p^3} 
+ \vec h^{(4)}_{33,{\rm II,UV}}(\hat k, \vec p, \vec q) \frac{k^4}{p^4}  \bigg) W^{\rm UV}_{33,{\rm II}} \Theta( |\vec q + \vec p | - q ) P_{\rm lin} (\vec q + \vec p)  \Bigg] \non\\
&\hspace{0.5cm} \times \Theta(q - p) P_{\rm lin}(\vec q)P_{\rm lin}(\vec p)\, . \non
}
After integrating over the $\hat p$ we see that the $\vec h^{(1)}_{33,{\rm II,IR}}$ term does not contribute. 
Similarly, the contribution of $\vec h^{(2)}_{33,{\rm II,UV}}$ and $\vec h^{(3)}_{33,{\rm II,UV}}$ vanish once contracted with $\vec \lambda_3$ and $\vec \kappa_3$. This is so since
$\vec \lambda_3 \cdot [\vec H_{3,{\rm II}}]^{(1)}_{\rm UV} = \vec \kappa_3 \cdot [\vec H_{3,{\rm II}}]^{(1)}_{\rm UV} =0$.
Note also that only the first term in $\vec h^{(4)}_{33,{\rm II,UV}}$ actually contributes.
To compute the full $\vec I_{33,{\rm II}}$ we have 
$\vec I_{33,{\rm II}}(k) = \vec{\tilde I}_{33,{\rm II}} + [\vec I_{33,{\rm II}}]_{\rm IR} + [\vec I_{33,{\rm II}}]_{\rm UV}$
where 
\eq{
[\vec I_{33,{\rm II}}]_{\rm IR}  
&= 36 \int_{\vec q, \vec p} \vec h^{(2)}_{33,{\rm II,IR}}(\vec k, \vec q, \hat p) \frac{k^2}{p^2} W^{\rm IR}_{33,{\rm II}} 
\Theta( |\vec k - \vec q| - q ) \Theta(q - p) P_{\rm lin} (\vec k - \vec q)  P_{\rm lin}(\vec q)P_{\rm lin}(\vec p)
= \lb \vec h^{\rm IR}_{33,{\rm II}} W^{\rm IR}_{33,{\rm II}}\rb k^2, \\
[\vec I_{33,{\rm II}}]_{\rm UV} 
&= 6 \int_{\vec q, \vec p} \vec h^{(4)}_{33,{\rm II,UV}}(\hat k, \vec p, \vec q) \frac{k^4}{p^4} W^{\rm UV}_{33,{\rm II}} 
P_{\rm lin} (\vec q + \vec p) P_{\rm lin}(\vec q)P_{\rm lin}(\vec p)
= \lb  \vec h^{\rm UV}_{33,{\rm II}} W^{\rm UV}_{33,{\rm II}} \rb k^4 . \non
}
The leading IR contribution gives
\eq{
\vec h^{\rm IR}_{33,{\rm II}}  
&= \int_{\vec q} \vec{\hat h}^{(2)}_{33,{\rm II, IR}}(\vec k, \vec q) \sigma^2_2(q)  \Theta( |\vec k - \vec q | - q ) P_{\rm lin} (\vec k - \vec q) P_{\rm lin}(\vec q) \, , \\
\vec h^{\rm UV}_{33,{\rm II}} 
&= \int_{\vec q, \vec p} \vec{\hat h}^{(4)}_{33,{\rm II,UV}}(\vec p, \vec q) P_{\rm lin} (\vec q + \vec p) P_{\rm lin}(\vec q)P_{\rm lin}(\vec p) \, ,
}
where we have introduced the coefficients $\vec{\hat h}^{(2)}_{33,{\rm II, IR}}(\vec k, \vec q) = 108 \int \frac{d \Omega_{\hat p}}{4 \pi}~ \vec h^{(2)}_{33,{\rm II, IR}} (\vec k, \vec q, \hat p)$. 
We also have the variance due to the $p$ modes smaller than $q$, defined in Eq.~\eqref{eq:variance_cut}.
We have defined the short scale noise $\vec h^{\rm UV}_{33,{\rm II}}$ contribution, which is scale independent 
given that $\vec{\hat h}^{(4)}_{33,{\rm II,UV}}(\vec p, \vec q) = 6 \int \frac{d \Omega_{\hat k}}{4\pi} \vec h^{(4)}_{33,{\rm II,UV}}/p^4$.

Next we look first at the $\vec I_{24}$ terms.
The integrals are of the form 
\eeq{
\vec I_{24} = 12 \int_{\vec q, \vec p} \vec H_2(\vec k -  \vec q, \vec q ) \otimes \vec H_4( \vec k -  \vec q , \vec q, \vec p, -\vec p) 
P_{\rm lin}( \vec k - \vec q ) P_{\rm lin}( \vec q ) P_{\rm lin}( \vec p ),
}
and the leading divergencies arise when $p \to 0$ \& $ q \to 0$, 
and $ p \to 0$ and $\vec q \to \vec k$. The latter divergence can be 
re-mapped again into $p \to 0$ \& $ q \to 0$ in the same way as was done for the $\vec I_{22}$ term. 
We have
\eq{
\vec I_{24} 
= \int_{\vec p, q < |\vec k - \vec q| } +  \int_{\vec p, q > |\vec k - \vec q| }
= 24 \int_{\vec p, \vec q} 
\vec H_2(\vec k -  \vec q, \vec q ) \otimes \vec H_4( \vec k -  \vec q , \vec q, \vec p, -\vec p) 
\Theta( |\vec k - \vec q | - q )
P_{\rm lin}( \vec k - \vec q ) P_{\rm lin}( \vec q ) P_{\rm lin}( \vec p ). \non
}
The remaining divergencies are now in $p \to 0$ and $ q \to 0$. 
Since the integral is not symmetric in these variables the two divergencies 
are distinct. However, we can symmetrise the integral first to get 
\eq{
\vec I_{24}
&= 12 \int_{\vec p, \vec q} 
\Big[
\vec H_2(\vec k -  \vec q, \vec q ) \otimes \vec H_4( \vec k -  \vec q , \vec q, \vec p, -\vec p) 
P_{\rm lin}( \vec k - \vec q ) \Theta( |\vec k - \vec q | - q )
+ \vec q \leftrightarrow \vec p
\Big]
P_{\rm lin}( \vec q ) P_{\rm lin}( \vec p ) .
}
We can remap the integral as
\eq{
\vec I_{24}
&= 24 \int_{\vec p, \vec q} 
\Big[
\vec H_2(\vec k - \vec q, \vec q ) \otimes \vec H_4( \vec k -  \vec q , \vec q, \vec p, -\vec p) 
P_{\rm lin}( \vec k - \vec q ) \Theta( |\vec k - \vec q | - q )
+ \vec q \leftrightarrow \vec p
\Big] \Theta( q - p)
P_{\rm lin}( \vec q ) P_{\rm lin}( \vec p ), 
}
where the $ q \to 0$ divergencies have been remapped into $p \to 0$ ones.
The two terms obviously give equal contributions to the leading 
divergence while in the sub-leading case they are different.
In the IR limit, we thus have
\eq{
\vec H_2(\vec k - \vec q, \vec q ) \otimes \vec H_4( \vec k -  \vec q , \vec q, \vec p, -\vec p) 
&\sim
\begin{cases}
\vec h^{(2)}_{24,{\rm I, IR}}(\vec k, \vec q, \hat p) \frac{k^2}{p^2}  + \mathcal{O}(p^0) , ~~{\rm as}~~ p \to 0 , \\[0.5em]
\vec h^{(4)}_{24,{\rm UV}} (\hat k, \vec q, \vec p) \frac{k^4}{p^4} + \mathcal{O}(k^5) , ~~{\rm as}~~ k \to 0 ,
\end{cases}  \\
\vec H_2(\vec k - \vec p, \vec p ) \otimes \vec H_4( \vec k -  \vec p , \vec p, \vec q, -\vec q) 
&\sim
\begin{cases}
 \vec h^{(2)}_{24,{\rm II, IR}} (\vec k, \vec q, \hat p) \frac{k^2}{p^2} + \vec h^{(1)}_{24,{\rm II, IR}} (\vec k, \vec q, \hat p) \frac{k}{p} + \mathcal{O}(p^0) ,~~{\rm as}~~ p \to 0 , \\[0.5em]
\vec h^{(4)}_{24,{\rm UV}} (\hat k, \vec p, \vec q) \frac{k^4}{q^4} + \mathcal{O}(k^5) , ~~{\rm as}~~ k \to 0.
\end{cases} \non
}
We thus have
\eq{
\vec{\tilde I}_{24}
&= 24 \int_{\vec p, \vec q} 
\bigg[
\bigg( \vec H_2(\vec k - \vec q, \vec q ) \otimes \vec H_4( \vec k -  \vec q , \vec q, \vec p, -\vec p) - \vec h^{(2)}_{24,{\rm I, IR}}(\vec k, \vec q, \hat p) \frac{k^2}{p^2} W^{\rm IR}_{24,{\rm I}} \bigg) P_{\rm lin}( \vec k - \vec q ) \Theta( |\vec k - \vec q | - q ) \\
&\hspace{0.5cm}
+ \vec H_2(\vec k - \vec p, \vec p ) \otimes \vec H_4( \vec k -  \vec p , \vec p, \vec q, -\vec q) P_{\rm lin}( \vec k - \vec p ) \Theta( |\vec k - \vec p | - p )  \non \\
&\hspace{0.5cm}
- \bigg( \vec h^{(2)}_{24,{\rm II, IR}} (\vec k, \vec q, \hat p) \frac{k^2}{p^2} + \vec h^{(1)}_{24,{\rm II, IR}} (\vec k, \vec q, \hat p) \frac{k}{p} \bigg) W^{\rm IR}_{24,{\rm II}} P_{\rm lin} (\vec k) \non\\
&\hspace{0.5cm}
- \bigg( \vec h^{(4)}_{24,{\rm UV}} (\hat k, \vec q, \vec p) \frac{k^4}{p^4} P_{\rm lin} (\vec q) + \vec q \leftrightarrow \vec p \bigg) W^{\rm UV}_{24}
\bigg]
P_{\rm lin}( \vec q ) P_{\rm lin}( \vec p ) \Theta( q - p). \non
}
To compute the full $\vec I$, we have $\vec I_{24} = \vec{\tilde I}_{24} (k) + [\vec I_{24,{\rm I}}]_{\rm IR} + [\vec I_{24,{\rm II}}]_{\rm IR} + [\vec I_{24}]_{\rm UV}$,
where 
\eq{
[\vec I_{24,{\rm I}}]_{\rm IR} 
&= 24 \int_{\vec p, \vec q} 
\vec h^{(2)}_{24,{\rm I, IR}}(\vec k, \vec q, \hat p) \frac{k^2}{p^2} W^{\rm IR}_{24,{\rm I}} \Theta( |\vec k - \vec q| - q )  \Theta( q - p)
P_{\rm lin} (\vec k - \vec q) P_{\rm lin}( \vec q ) P_{\rm lin}( \vec p )
= \lb \vec h^{\rm IR}_{24,{\rm I}} W^{\rm IR}_{24, {\rm I}} \rb k^2 \, , \\
[\vec I_{24,{\rm II}}]_{\rm IR} &= 24 P_{\rm lin} (\vec k)  \int_{\vec p, \vec q}
 \vec h^{(2)}_{24,{\rm II, IR}} (\vec k, \vec q, \hat p) \frac{k^2}{p^2} W^{\rm IR}_{24, {\rm II}}  \Theta( q - p) 
P_{\rm lin}( \vec q ) P_{\rm lin}( \vec p )
= \lb \vec h^{\rm IR}_{24,{\rm II}} W^{\rm IR}_{24, {\rm II}} \rb k^2 P_{\rm lin} (k) \, ,  \non
}
where 
\eq{
\vec h^{\rm IR}_{24,{\rm I}} &= \int_{\vec q} \vec{\hat h}^{(2)}_{24,{\rm I, IR}}(\vec k, \vec q) \sigma^2_2(q) \Theta( |\vec k - \vec q| - q ) P_{\rm lin} (\vec k - \vec q) P_{\rm lin}( \vec q ) \, ,
~~{\rm where}~~ \vec{\hat h}^{(2)}_{24,{\rm I, IR}} = 72 \int \frac{d \Omega_{\hat p}}{4 \pi}~ \vec h^{(2)}_{24,{\rm I, IR}}(\vec k, \vec q, \hat p) \, ,  \\
\vec h^{\rm IR}_{24,{\rm II}} &= \int_{\vec q} \vec{\hat h}^{(2)}_{24,{\rm II, IR}}(\vec k, \vec q) \sigma^2_2(q) P_{\rm lin}( \vec q ) \, , 
~~{\rm where}~~ \vec{\hat h}^{(2)}_{24,{\rm II, IR}} = 72 \int \frac{d \Omega_{\hat p}}{4 \pi}~ \vec h^{(2)}_{24,{\rm II, IR}} (\vec k, \vec q, \hat p) \, . \non
}
The UV components give
\eq{
[\vec I_{24}]_{\rm UV} &= 12 \int_{\vec p, \vec q} \vec h^{(4)}_{24,{\rm UV}} (\hat k, \vec q, \vec p) \frac{k^4}{p^4}W^{\rm UV}_{24} P_{\rm lin} (\vec q)^2 P_{\rm lin}( \vec p )
=  \lb \vec h^{\rm UV}_{24} W^{\rm UV}_{24} \rb k^4 \, ,
}
where 
\eeq{
\vec h^{\rm UV}_{24} = \int_{\vec p, \vec q} \vec{\hat h}^{(4)}_{24,{\rm UV}} (\vec q, \vec p) P_{\rm lin} (\vec q)^2 P_{\rm lin}( \vec p ), 
~~{\rm where}~~ \vec{\hat h}^{(4)}_{24,{\rm UV}} (\vec q, \vec p) = 12 \int \frac{d \Omega_{\hat k}}{4 \pi}~ \vec{\hat h}^{(4)}_{24,{\rm UV}} (\hat k, \vec q, \vec p).
}

At last, let us look at the $\vec I_{15}$ term. The integrals are of the form 
\eeq{
\vec I_{15} = 15 P_{\rm lin}(\vec k)  \int_{\vec q, \vec p} ~\vec H_1(\vec k) \otimes \vec H_5(\vec k, \vec q, - \vec q, \vec p, -\vec p) P_{\rm lin}( \vec q ) P_{\rm lin}( \vec p ). 
}
The leading divergencies arise when $p \to 0$ \& $ q \to 0$. 
We can re-map this so that we have
\eeq{
\vec I_{15} = 30 P_{\rm lin}(\vec k)  \int_{\vec q, \vec p} ~\vec H_1(\vec k) \otimes \vec H_5(\vec k, \vec q, - \vec q, \vec p, -\vec p) \theta(q-p)  P_{\rm lin}( \vec q ) P_{\rm lin}( \vec p ). 
}
The IR and UV limits are 
\eq{
\vec H_1(\vec k) \otimes \vec H_5(\vec k, \vec q, - \vec q, \vec p, -\vec p) 
&\sim 
\begin{cases}
\vec h^{(2)}_{15,{\rm IR}} (\vec k, \vec q, \hat p) \frac{k^2}{p^2}  + \mathcal{O}(p^0) ,
~~{\rm as}~~ p \to 0 , \\
\vec h^{(0)}_{15,{\rm UV}}(\hat k, \vec q, \vec p)
+ \vec h^{(2)}_{15,{\rm UV}} (\hat k, \vec q, \vec p) \frac{k^2}{p^2} + \mathcal{O}(p^0)
~~{\rm as}~~ k \to 0 . \non
\end{cases}
}
The regularised integrals are thus
\eq{
\vec{\tilde I}_{15} (k) 
&= 30 P_{\rm lin}(\vec k)  \int_{\vec q, \vec p} ~
\bigg[ \vec H_1(\vec k) \otimes \vec H_5(\vec k, \vec q, - \vec q, \vec p, -\vec p) 
 - \frac{k^2}{p^2} \vec h^{(2)}_{15,{\rm IR}} (\vec k, \vec q, \hat p)  W^{\rm IR}_{15} \\
&\hspace{5cm} - \lb \vec h^{(0)}_{15,{\rm UV}}(\hat k, \vec q, \vec p)
+ \vec h^{(2)}_{15,{\rm UV}} (\hat k, \vec q, \vec p) \frac{k^2}{p^2} \rb W^{\rm UV}_{15}
\bigg]
\theta(q-p)  P_{\rm lin}( \vec q ) P_{\rm lin}( \vec p ) \, .\non
}
The contributions from $\vec h^{(0)}_{15}$ vanish after contractions with the $\vec \lambda_5$ and $\vec \kappa_5$ coefficients.
To compute the full integral we have $\vec I_{15} (k) = \vec{\tilde I}_{15} (k) +  [\vec I_{15}]_{\rm IR} + [\vec I_{15}]_{\rm UV}$, 
with 
\eq{
[\vec I_{15}]_{\rm IR} 
&= 30  P_{\rm lin}(\vec k)  \int_{\vec q, \vec p} ~ \vec h^{(2)}_{15,{\rm IR}} (\vec k, \vec q, \hat p) \frac{k^2}{p^2} W^{\rm IR}_{15} \theta(q-p)  P_{\rm lin}( \vec q ) P_{\rm lin}( \vec p )
= \lb \vec h^{\rm IR}_{15}  W^{\rm IR}_{15} \rb k^2 P_{\rm lin} (\vec k)\, , \\
[\vec I_{15}]_{\rm UV} &= 15 P_{\rm lin}(\vec k)  \int_{\vec q, \vec p} ~ \vec h^{(2)}_{15,{\rm UV}}(\hat k, \vec q, \vec p) 
\frac{k^2}{p^2} W^{\rm UV}_{15}  P_{\rm lin}( \vec q ) P_{\rm lin}( \vec p )
= \lb \vec h^{\rm UV}_{15}  W^{\rm UV}_{15} \rb k^2 P_{\rm lin} (\vec k)\, .\non
}
We have also introduced 
\eq{
\vec h^{\rm IR}_{15} = \int_{\vec q} \vec{\hat h}^{(2)}_{15,{\rm IR}} (\vec k, \vec q) \sigma^2_2(q) P_{\rm lin}( \vec q )
~~{\rm and}~~
\vec h^{\rm UV}_{15} = \int_{\vec q} \vec{\hat h}^{(2)}_{15,{\rm UV}}(\vec q, \vec p) \frac{1}{p^2}  P_{\rm lin}( \vec q ) P_{\rm lin}( \vec p )\, ,
}
where $\vec{\hat h}^{(2)}_{15,{\rm IR}} (\vec k, \vec q) = 90 \int \frac{d \Omega_{\hat p}}{4 \pi}~\vec h^{(2)}_{15,{\rm IR}} (\vec k, \vec q, \hat p)$,
and $\vec{\hat h}^{(2)}_{15,{\rm UV}} (\vec q, \vec p) = 15 \int \frac{d \Omega_{\hat k}}{4 \pi} ~\vec h^{(2)}_{15,{\rm IR}} (\hat k, \vec q, \vec p)/p^2$.

\section{One- and two-loop results at redshift $z=1.0$.} 
\label{app:two-loops_higher_z}

\begin{figure}[t!]
\centering
\includegraphics[width=0.95\linewidth]{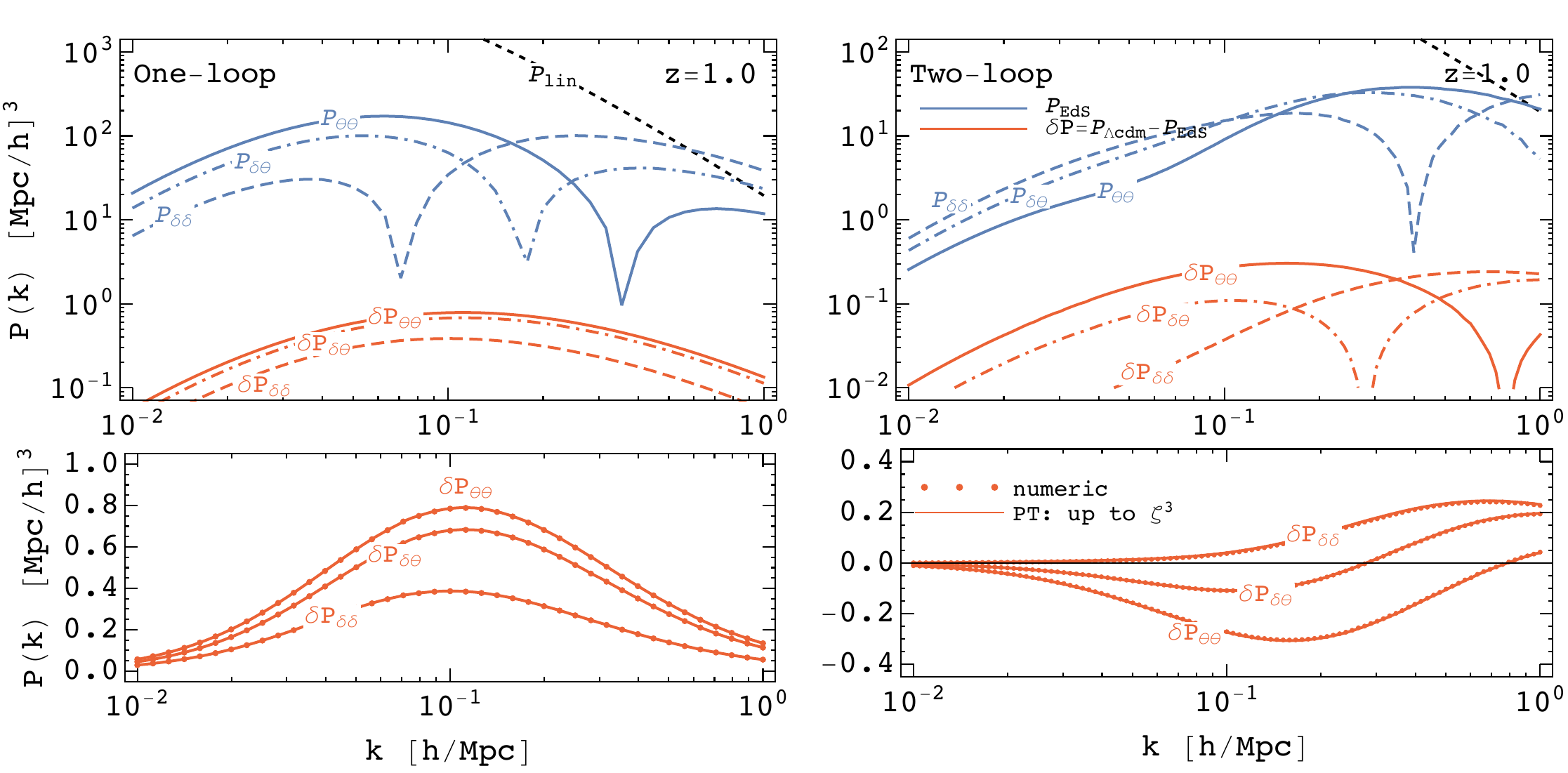}
\caption{Plotted above are the same quantities shown in Figure~\ref{fig:power_sepctra_I}, but this time at $z=1.0$. \textit{Upper panels} show the absolute contributions of EdS results (blue lines) compared to the $\Lambda$CDM corrections (red lines). We again see that the three different spectra $P_{\delta\delta}$ (dashed lines), $P_{\delta\theta}$ (dot-dashed lines) and $P_{\theta\theta}$ (solid lines) receive corrections of different size, whose relative importance is also a function of the scale dependence of the EdS terms. \textit{Lower panels} display the $\Lambda$CDM corrections $\delta P_{\delta\delta}$, $\delta P_{\delta\theta}$ and $\delta P_{\theta\theta}$ computed using the numerical evaluations of the $\lambda_n$ and $\kappa_n$ coefficients (show in dots). We also show the perturbative time dependence computation as described in Sec.~\ref{sec:PT_coeffs} up to the third order (solid lines).
}
\label{fig:power_sepctra_I_z10}
\end{figure}

In this appendix, we present the results of one- and two-loop contributions to the density-density, density-velocity and velocity-velocity power spectrum at redshift $z=1.0$, similar to what was presented in Sec.\ref{sec:one_two_loops} for the $z=0.0$ case. In analogy to the Figure~\ref{fig:power_sepctra_I}, the left panels of Figure~\ref{fig:power_sepctra_I_z10} show the one-loop contributions for power spectra of density and velocity fields at $z=1.0$: $P_{\delta\delta}$, $P_{\delta\theta}$ and  $P_{\theta\theta}$. We again show the EdS results and the corresponding $\Lambda$CDM correction. We see that the one-loop $\Lambda$CDM  corrections at $z=1.0$ are approximately two orders of magnitude smaller than the one-loop EdS contributions. Note that in the wavelength range where the EdS results exhibit the zero crossing the relative importance of $\Lambda$CDM correction is larger and similar in magnitude to the two-loop EdS results.

Analogous results hold for the two-loop order, shown on the right-hand side of Figure~\ref{fig:power_sepctra_I_z10}. One can observe that the corresponding $\Lambda$CDM corrections are approximately two orders of magnitude smaller than the EdS results (except in the $P_{\theta\theta}$ case for $k<0.1h/{\rm Mpc}$, where the corrections can reach up to 5\%). Here too, the exception is the wavenumber interval in the proximity of the k-value corresponding to the zero crossing of the EdS result. The correction due to such contributions can be partially mitigated once the UV counterterms are added, as was shown for the $z=0.0$ case in Figure~\ref{fig:power_sepctra_II}. In general, we see that these beyond-EdS corrections are overall smaller and less relevant at higher redshifts than is the case at $z=0.0$, as one would expect.

\section*{References}
\bibliography{References}

\end{document}